%% file: main.tex
\documentclass[11pt,a4paper]{article}

\usepackage[utf8]{inputenc}
\usepackage[T1]{fontenc}
\usepackage{graphicx}
\usepackage{amsmath}
\usepackage{amsfonts}
\usepackage{amssymb}
\usepackage{amstext}
\usepackage{natbib}
\usepackage{makecell}

\usepackage[hmargin=0.7in,vmargin={1.2in,1.1in}]{geometry}
\usepackage{csquotes}
\usepackage{kpfonts}
\usepackage{rotating}
\usepackage{booktabs}
\usepackage{csquotes}
\usepackage{amsbsy}
\usepackage{eqnarray}
\usepackage{lipsum}
\usepackage{dsfont}
\usepackage[affil-it]{authblk}
\usepackage{float}
\usepackage{setspace}
\usepackage{tabularx}
\usepackage{multirow}
\usepackage{color}
\usepackage{url}
\usepackage{soul}
\usepackage[official]{eurosym}
\usepackage[colorlinks]{hyperref}
\usepackage[dvipsnames,table,xcdraw,svgnames]{xcolor}
\usepackage{subfig}

\newtheorem{theorem}{Theorem}
\definecolor{deeppink}{rgb}{1.0, 0.08, 0.58}

\graphicspath{{Figures/}}

\makeatletter
\renewcommand\footnotesize{%
	\@setfontsize\footnotesize\@ixpt{11}%
	\abovedisplayskip 8\p@ \@plus2\p@ \@minus4\p@
	\abovedisplayshortskip \z@ \@plus\p@
	\belowdisplayshortskip 4\p@ \@plus2\p@ \@minus2\p@
	\def\@listi{\leftmargin\leftmargini
		\topsep 4\p@ \@plus2\p@ \@minus2\p@
		\parsep 2\p@ \@plus\p@ \@minus\p@
		\itemsep \parsep}%
	\belowdisplayskip \abovedisplayskip
}
\makeatother

\AtBeginDocument{%
\hypersetup{
citecolor=MidnightBlue,
linkcolor=Brown,   
urlcolor=MidnightBlue}
}

\title{Measuring productivity dispersion: a parametric approach using the L\'{e}vy alpha-stable distribution}
\date{}
\author[1,3]{Jangho Yang \thanks{We would like to acknowledge funding from Baillie Gifford, the Institute for New Economic Thinking at the Oxford Martin School, and the Rebuilding Macroeconomics project, which is funded by the ESRC. We are grateful to Jean-Philippe Bouchaud and Jos\'{e} Moran for very useful comments. \emph{Contacts:} j634yang@uwaterloo.ca (Corresponding Author); torsten.heinrich@wiwi.tu-chemnitz.de; julian.winkler@economics.ox.ac.uk; francois.lafond@inet.ox.ac.uk; pantelis.koutroumpis@oxfordmartin.ox.ac.uk; doyne.farmer@inet.ox.ac.uk}}
\author[2,3,4]{Torsten Heinrich}
\author[3,7]{Julian Winkler}
\author[3,5]{\\ Fran\c{c}ois Lafond}
\author[2,3]{Pantelis Koutroumpis}
\author[2,3,5,6]{J. Doyne Farmer}
\affil[1]{Management Sciences, Faculty of Engineering, University of Waterloo}
\affil[2]{Oxford Martin Programme on Technological and Economic Change, University of Oxford}
\affil[3]{Institute of New Economic Thinking at the Oxford Martin School, University of Oxford}
\affil[4]{Department of Economics and Business Administration, Chemnitz University of Technology}
\affil[5]{Mathematical Institute, University of Oxford}
\affil[6]{Santa Fe Institute}
\affil[7]{Department of Economics, University of Oxford}

\begin{document}

	\maketitle

	\begin{abstract}
		\onehalfspacing
It is well-known that value added per worker is extremely heterogeneous among firms, but relatively little has been done to characterize this heterogeneity more precisely. Here we show that the distribution of value-added per worker exhibits heavy tails, a very large support, and consistently features a proportion of negative values, which prevents log transformation. We propose to model the distribution of value added per worker using the four parameter L\'{e}vy stable distribution, a natural candidate deriving from the Generalised Central Limit Theorem, and we show that it is a better fit than key alternatives. Fitting a distribution allows us to capture dispersion through the tail exponent and scale parameters separately. We show that these parametric measures of dispersion are at least as useful as interquantile ratios, through case studies on the evolution of dispersion in recent years and the correlation between dispersion and intangible capital intensity.

		\bigskip
		
		
	\end{abstract}
	
%	\clearpage
	
%	\tableofcontents
	
%	\newpage

		\section{Introduction}
		\label{sec:introduction}

In the last two decades, the availability of micro-data has revealed the tremendous productivity differences between firms, even within sectors at a fairly detailed level \citep{bartelsman2000understanding,syverson2011what}. Measuring productivity dispersion is important in order to understand misallocation \citep{gopinath2017capital}, innovation and diffusion \citep{berlingieri2017great,andrews2016best}, business dynamism \citep{foster2018innovation}, and ultimately the aggregate productivity slowdown \citep{goldin2021productivity}.

\begin{figure}[!ht]
	\begin{center}
	\includegraphics[width=.7\textwidth]{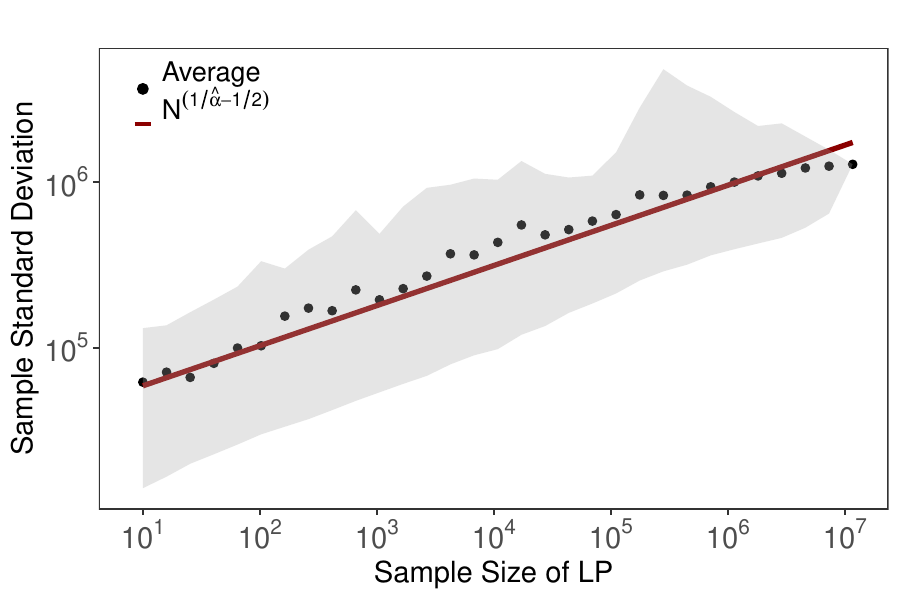}
	\caption{\textbf{Measured standard deviation in sub-samples of firm labour productivity.} We construct a distribution by pooling together the productivity levels of firms in France, Italy, Germany, and Spain (which have the same currency), for all years, expressed in thousands. Then, for each subsample size $N$, we compute the standard deviation of 100,000 subsamples, and report the average (black dots), and $5^{th}$ and $95^{th}$ percentiles. In Appendix~\ref{app:scaling}, we explain that on i.i.d. data with a power law tail, the theoretical scaling between the sample standard deviation and the sample size would be $N^{\frac{1}{\alpha}-\frac{1}{2}}$, where $\alpha$ is the tail exponent. The red line shows that this scaling holds empirically here (we use our estimate of the tail parameter $\hat{\alpha}=1.33$, and the intercept is calculated by simulating 100,000 i.i.d. random samples of size 10 from a L\'{e}vy alpha-stable with the same parameters as those we estimate on empirical data). Data from Orbis Europe.}
	\label{fig:std_17}
	\end{center}
    \end{figure}

In this paper, we focus on the distribution of value added (VA) per worker at the firm level, a common measure of productivity in the literature\footnote{
See for instance \citet{souma2009distribution,aoyama2015micro,andrews2016best,berlingieri2017great,gu2019frontier,ilzetzki2017measuring,campbell2019measuring,gouin2020productivity} and \citet{de2022firms}.
\citet{baily1996downsizing} note that it is a ``conceptually preferable measure of labour productivity'' to gross output per worker, even though they have to remove negative VA establishments in some of their analyses. \citet{oulton2000tale} discards roughy 1\% of his sample of UK firms, which have negative value added. \citet{aradanaz2017understanding} note that only a third of the UK firms with negative VA in 2010 had exited the market by 2015, which is only 10 percentage points higher than other firms. \citet{de2022firms} find a proportion of negative VA UK firms in both Orbis and administrative data.
While issues with VA-based measures of productivity have been noted \citep{campbell2019measuring,cunningham2021dispersion}, measures based on gross output are less comparable across different sectors, as intermediate consumption is naturally larger in some industries than in others, so that using gross output-based measures creates a need to use industry means to renormalize the values and compare them; this itself comes with its own issues, if only because classification systems are imperfect, lack granularity, and change over time.
}.
We employ a commercial dataset, \textit{Orbis Europe}. We have access to a comprehensive version, that includes around 23 million European firm-year observations for the period 2006-2017. The distribution of productivity levels, even within industry-country pairs, exhibits heavy tails, and therefore an infinite variance. In practice, this means that measuring the variance in a given sample is not meaningful, as the result is driven by the sample size, rather than reflecting a true moment of the data (see Fig.~\ref{fig:std_17}).

A common solution to this problem is to compute the variance of the \emph{logarithm} of productivity (instead of the variance of productivity). For variables that have positive support, such as revenue per worker, this is an acceptable solution, since the distribution of the log values likely exhibits finite variance. However, this solution is problematic for VA per worker, because firm-level datasets often contain a substantial proportion of firms that have \emph{negative} value added - typically of only a few percent, but up to 23\% for country-year pairs in our dataset (Table \ref{tab:obs_ctry_yr_neg}). To state the obvious, as far as measuring dispersion is concerned, removing the firms on the left tail is problematic. The prevalence of firms that stay alive despite very poor performance should be a key indicator in studies of misallocation, creative destruction, and business dynamism.

\paragraph{How can we measure dispersion in a dataset with negative values and heavy tails?} In this paper, we propose a transparent and straightforward approach: first, finding a good parametric model for the distribution of productivity, and second, using the fitted parameters to evaluate dispersion. For instance, if the data was normally distributed, an excellent way to evaluate how dispersion differs across time or sectors would have been to estimate the scale parameter of the normal distribution, that is, the variance. Here, however, we find that the data exhibits heavy tails and substantial asymmetry, so we turn to more complex models. We find that the L\'{e}vy alpha-stable distribution gives an excellent fit to the data, and we argue that its parameters - particularly the scale and the tail parameters - provide appropriate metrics of dispersion. 

The scale parameter of the L\'{e}vy distribution is similar to the standard deviation in a normal distribution. Roughly speaking, it provides information on the span of the support of the distribution where most of the data lies. In contrast, the tail exponent provides information on the prevalence of extremely productive or unproductive firms.

While the L\'{e}vy alpha-stable distribution elegantly addresses the issues of heavy tails and negative values, the problem with parametric measures is that the underlying model can be wrong. We address this in two ways. First, we show that the L\'{e}vy alpha-stable offers a surprisingly good fit to our productivity data, and that it is better than the main contender, the 5-parameter Subbotin (Asymmetric Exponential Power, AEP) distribution \citep{Bottazzi/Secchi11}. We add credibility to the model by using independent tests to confirm that the second moment is infinite. Second, we argue that the L\'{e}vy alpha-stable model is plausible \emph{a priori} simply because it is the result of the Generalized Central Limit Theorem (GCLT) \citep{NOLAN2020}. If VA per worker can be thought of as the aggregation of several micro-level variables such as the productivity of individual employees, contracts, tasks or routines, then under fairly general conditions the GCLT predicts that firm-level productivity will follow a L\'{e}vy alpha-stable distribution. 

Having established that the data is well fitted by a L\'{e}vy distribution, we can use the estimated parameters to evaluate dispersion. We propose the tail parameter as an indicator of dispersion in the tail, and the ratio of the scale to location parameters (similar to the coefficient of variation $(\sigma/\mu)$ for a normal distribution) as an indicator of dispersion in the body of the distribution. We compare these metrics with interquantile ratios, the most widely used non-parametric measures of dispersion, which can be designed to measure dispersion in body ($Q_{90}/Q_{10}$) or in the tail ($Q_{95}/Q_{50}$). 
While these metrics are all empirically correlated, the correlations are imperfect, so they capture fairly distinct aspects of dispersion.

We then proceed with two case studies: the evolution of dispersion in recent years, and the correlation between dispersion and intangible intensity, measured as intangible capital per worker.
We find that the quantile-based and L\'{e}vy-based metric give similar results. First, we find that for our period (2006-2017), there was no overwhelming evidence for a substantial, systematic increase in productivity dispersion in all countries. Second, we find a positive, but fairly noisy correlation between intangible intensity and dispersion. While the results are similar, we show that our parametric measures are superior because they make it possible to keep negative values, and avoid having to select specific quantiles. The results also appear somewhat less noisy with the L\'{e}vy metrics than with the quantile-based metrics.

\paragraph{Literature.}  Our results relate to three strands of literature. The first is the large body of literature on heavy tails in economics and finance \citep{Mitchell,Mandelbrot1960,Mandelbrot1963,Fama1965,samuelson_1967,Emberchts,fagiolo2008output,gabaix2011granular,acemoglu2017microeconomic, axtell2019dynamics}, more specifically on the distributions of firm sizes \citep{Ijiri/Simon77,Axtell01} and firm growth rates \citep{bottazzi2003common,Bottazzietal07,Bottazzi/Secchi11,schwarzkopf2010explanation,holly2013aggregate,moran}. Most related to our study is the work on estimating productivity distributions such as \citet{aoyama2010productivity} who estimate a Generalized Beta model for labour productivity levels\footnote{
Generalized Beta models are defined only over a positive support and thus are not suitable for our data, which contains negative productivity firms.
}, \citet{fujiwara2009distribution} who find a power law tail for labour productivity levels\footnote{
See also \citet{aoyama2009labour,mizuno2012power} and \citet{aoyama2015micro}. 
}, and \citet{gaffeo2008levy,gaffeo2011distribution} who estimates a L\'{e}vy alpha-stable distribution for the growth rates of sector-level total factor productivity. In the present paper, we focus on \emph{firm-level} data, productivity \emph{levels}, and productivity measured as \emph{value added per worker}. 

Second, our results on productivity levels are relevant to the literature that discusses how productivity dispersion may reflect the misallocation of factors of production \citep{hsieh2009misallocation,bartelsman2013cross,foster2018innovation,gopinath2017capital,haltiwanger2018misallocation}. In this paper, we critically discuss measures of dispersion in detail, showing how some statistics may be misleading and what parts of the distribution can affect dispersion. We also suggest that theoretical models should aim to derive a L\'{e}vy alpha-stable distribution, where ideally the parameters would be interpreted in terms of misallocation, or other sources\footnote{
There is an important theoretical literature on productivity dispersion, which recognizes that the right tail of the distribution of productivity levels is a power law, and explains this mostly as the result of innovation and imitation processes, see  \citet{ghiglino2012random,lucas2014knowledge,perla2014equilibrium} and \citet{konig2016innovation}. The presence of extreme values in empirical data also implies that data cleaning choices may affect substantive research results, as shown for the misallocation literature by \citet{rotemberg2021plant}.
}.

Third, our results relate to the work of statistical agencies around the world, who are developing procedures to make summary statistics of the micro data publicly available, and in particular industry-level productivity dispersion \citep{cunningham2021dispersion,berlingieri2017multiprod}. A key contribution of this paper is to make a suggestion regarding what statistics should be released, namely 5 quantiles necessary to perform \citet{McCulloch86}'s quantile estimation of the four L\'{e}vy alpha-stable parameters. Typically, readily available software packages use the $5^{th}, 25^{th}, 50^{th}, 75^{th}$ and $95^{th}$ quantiles, as in \citet{McCulloch86}'s original paper. We suggest these five quantiles are good candidates, and could be supplemented by more quantiles which can then be used for testing the quality of the fit.

\

The paper is organized as follows. Section~\ref{sec:data} describes the data sources and the basic patterns, including the presence of heavy tails. Section~\ref{sec:models} presents the L\'{e}vy alpha-stable distribution, the Generalized CLT, the main alternative parametric model, our fitting methods, and shows the quality of the fit. Section~\ref{sec:results} then compares the quantile-based and L\'{e}vy-based measures of dispersion in practice, showing the evolution of dispersion over time, and the correlation between industry-level dispersion and intangible capital intensity. Section~\ref{sec:conclusion} concludes.

	\section{Data and descriptive statistics}
	\label{sec:data}
	\subsection{Data sources}

In this section, we present the dataset and some basic patterns of the empirical data. We use data from the Orbis Europe database, compiled by Bureau van Dijk, which includes the balance sheets and profit-loss statements for 7 million unique firms across Europe, yielding approximately 23 million firm-year observations from 2006 to 2017. 

Unlike other widely used firm-level data such as Compustat and Worldscope, Orbis Europe records a large number of small and medium sized firms (SME) that are often not publicly traded. Table \ref{tab:obs_ctry_yr} in Appendix~\ref{sec:data_appendix} reports the number of observations per country-year.\footnote{
Orbis Europe includes many different types of firms, including Private limited company, Joint-stock company, Partnership, Cooperative, Consortium, Foundation, and Public agency/corporation.
For a more detailed discussions on the advantages and drawbacks of the Orbis Europe database, see \citet{KalemliOzcanetal15}.
}

To avoid double counting, we only use unconsolidated data\footnote{
For some corporate groups that consist of multiple firms, both consolidated and unconsolidated data is provided, i.e. data is listed for the entire group and again for each of the firms.
}. We then remove duplicated firm-year pairs, and we exclude self-employed firms, where the distinction between wage and profit is ambiguous. Finally, we regard negative values for total assets, fixed assets, sales, wage, and employment as missing values. For a detailed discussion on data cleaning, see Appendix \ref{sec:data_appendix}.

	\subsection{Construction of variables}	

For each firm $i$ in industry $j$ and year $t$, we define real value added ($Y_{i,t}$) as the sum of real labour income ($W_{i,t}$) and real capital income ($\Pi_{i,t}$), that is
\begin{align}
Y_{i,t} \equiv W_{i,t} + \Pi_{i,t} = \frac{\omega_{i,t}}{p^v_{j,c,t}} + \frac{\pi_{i,t}}{p^v_{j,c,t}}, 
\label{eq:VA} 
\end{align}
where $\omega$ and $\pi$ are nominal wage and nominal profit, and $p_{j,c,t}^v$ is the value added deflator of industry $j$ in country $c$ at time $t$. In Orbis, nominal wage and nominal profit variables are recorded in Cost of Employees (STAF) and Earnings Before Interest, Taxes, Depreciation \& Amortization (EBITDA)\footnote{
VA can also be calculated by subtracting intermediate costs (MATE in Orbis) from gross output (OPRE in Orbis). This is the output-based approach to computing VA. However, material costs are less frequently observed compared to wages and earnings in our sample. For example, only 52\% of firms in Germany report material costs. In some countries such as the UK and Denmark, no firms report material costs. Therefore, to get the largest possible coverage for VA, we use the income-based value added by adding wages and earnings. We note that the labour productivity variable constructed from the output based VA in our Orbis data tends to overestimate the labour productivity recorded in the national accounts. For example, the output-based productivity in France and Germany is almost twice as high as the labour productivity in the national accounts. Using the output-based labour productivity, only around 1\% of firms have negative productivity. \citet{gopinath2017capital}, who also used Orbis Europe data, construct output-based VA for Spanish firms, but this is justified in their case since around 90\% of firms in Spain report material costs, and for Spain the output-based productivity is very close to the productivity recorded in national accounts.
}. The firms' VA is deflated using the industry-level VA deflator from the EUKLEMS database ($VA\_P$ in \citet{Jaeger17}).  See Appendix~\ref{app:additionalcleaning} for more details.

Denoting the number of employees by $L_{i,t}$ (EMPL in Orbis), our measure of firm-level labour productivity ($LP_{i,t}$) is defined as the ratio of value added to the number of employees,
\begin{align}
LP_{i,t} &= \frac{Y_{i,t}}{L_{i,t}},
\label{eq:LP} 
\end{align}
and is expressed in units of currency of the country where the firm is located.

Researchers in productivity analysis typically proceed to a data transformation: they analyze the natural logarithm of productivity \citep{bartelsman2018measuring}. However, when output is measured as VA, firm-level labour productivity can be negative. If a firm's intermediate cost is  greater than revenues in a given fiscal year, the firm has negative VA and productivity. From the income perspective of the definition of VA (Wages + Earnings), the sum of earnings and wages can be negative since firms' losses (negative profits) can be greater than wages. 

Empirically, a sizeable share of firms has negative labour productivity. Table \ref{tab:obs_ctry_yr_neg} shows the proportion (\%) of negative observations per country-year in our data. Overall,  5\% of the firms in our sample have negative productivity. As a more extreme case, more than a fifth of UK firms had negative productivity in 2006-2008\footnote{
Other authors find smaller percentages for the UK, but there is systematically at least a few percents of negative VA firms. This is close to 10\% in \citet[Chart 17]{haldane2017productivity}. In employment weighted distributions, this is around 4-5\% \citep{ONS_ABS}. \citet{de2022firms} find only `a few' in Orbis, but remove many sectors and firms with less than 10 employees.
}.

\begin{table}[H]
	%\caption{}
	\footnotesize
    \begin{center}
    \caption{\textbf{Proportion (\%) of negative observations per country-year}.}
	\label{tab:obs_ctry_yr_neg}
\begin{tabular}{rllllllllllll}
  \hline
 & 2006 & 2007 & 2008 & 2009 & 2010 & 2011 & 2012 & 2013 & 2014 & 2015 & 2016 & 2017 \\ 
  \hline
Belgium &  2.58 &  1.33 &  1.56 &  1.62 &  1.49 &  1.44 &  1.55 &  1.59 &  1.53 &  1.49 &  1.53 &  1.58 \\ 
  Bulgaria &  0.00 &  9.33 &  4.02 &  5.79 &  5.86 &  7.45 &  7.60 &  7.02 &  6.35 &  6.10 &  5.47 &  5.13 \\ 
  Croatia &  6.91 &  4.46 &  4.60 &  5.66 &  6.18 &  5.27 &  5.79 &  4.54 &  4.00 &  3.55 &  2.86 &  0.00 \\ 
  Czech Republic &  7.10 &  4.89 &  5.32 &  6.49 &  6.26 &  6.35 &  6.33 &  6.19 &  5.48 &  4.52 &  4.30 &  3.69 \\ 
  Denmark &  0.00 &  0.00 &  0.00 &  0.00 &  0.00 &  0.00 &  2.51 &  2.98 &  2.83 &  2.51 &  2.91 &  3.04 \\ 
  Estonia &  0.00 &  2.90 &  4.09 &  6.09 &  4.65 &  3.25 &  3.11 &  3.12 &  3.37 &  3.38 &  3.30 &  2.75 \\ 
  Finland &  2.09 &  1.60 &  1.83 &  1.99 &  1.84 &  2.03 &  1.91 &  1.95 &  1.94 &  2.08 &  2.08 &  1.90 \\ 
  France &  1.19 &  1.10 &  1.31 &  1.27 &  1.13 &  1.16 &  1.32 &  1.46 &  1.43 &  1.60 &  1.81 &  1.89 \\ 
  Germany &  2.56 &  1.96 &  2.48 &  2.43 &  2.14 &  2.17 &  2.16 &  2.00 &  2.39 &  2.48 &  2.31 &  0.00 \\ 
  Hungary &  0.00 &  4.11 &  3.11 &  7.01 &  6.61 &  6.68 &  7.18 &  5.76 &  5.16 &  4.87 &  4.47 &  3.62 \\ 
  Italy &  4.64 &  3.08 &  3.64 &  4.16 &  3.68 &  3.78 &  5.00 &  4.74 &  4.32 &  3.84 &  3.16 &  2.79 \\ 
  Poland &  3.77 &  4.01 &  4.78 &  6.11 &  3.72 &  3.95 &  8.12 &  7.42 &  0.00 &  0.00 &  0.00 &  2.49 \\ 
  Portugal &  8.62 &  6.06 &  6.51 &  6.12 &  5.81 &  6.87 &  8.34 &  7.87 &  7.86 &  6.47 &  5.57 &  4.80 \\ 
  Romania &  0.00 &  7.58 &  8.52 & 10.73 & 11.73 & 11.46 & 12.31 & 12.47 & 10.40 &  7.34 &  6.39 &  0.00 \\ 
  Slovakia &  0.00 &  4.44 &  4.55 &  7.64 &  7.11 &  7.23 &  7.20 &  7.13 &  6.05 &  4.10 &  4.21 &  0.00 \\ 
  Slovenia &  0.00 &  0.00 &  0.00 &  1.50 &  1.69 &  1.81 &  1.88 &  1.56 &  1.44 &  1.35 &  1.25 &  1.15 \\ 
  Spain &  3.78 &  2.92 &  3.60 &  3.90 &  3.65 &  4.04 &  4.41 &  4.05 &  3.33 &  2.94 &  2.57 &  2.07 \\ 
  Sweden &  3.82 &  2.16 &  2.41 &  2.27 &  2.13 &  2.18 &  2.10 &  2.10 &  2.07 &  2.08 &  2.15 &  2.35 \\ 
  United Kingdom & 20.98 & 21.72 & 23.49 & 11.31 &  4.14 &  4.04 &  3.84 &  3.52 &  3.57 &  3.73 &  3.92 &  0.00 \\ 
   \hline
   \end{tabular}
	\end{center}
		 \footnotesize{This table records the proportion (\%) of negative observations of value added per country-year for all 19 countries. }
\end{table}

\subsection{Characteristic patterns of the distributions}
\label{sec:data:characteristics}

To motivate L\'{e}vy alpha-stable distributions as a model for $LP$,  we first show qualitatively that the distributions are heavy-tailed and asymmetric. 

\begin{figure}[H]
	\begin{center}
		\includegraphics[width=1\textwidth]{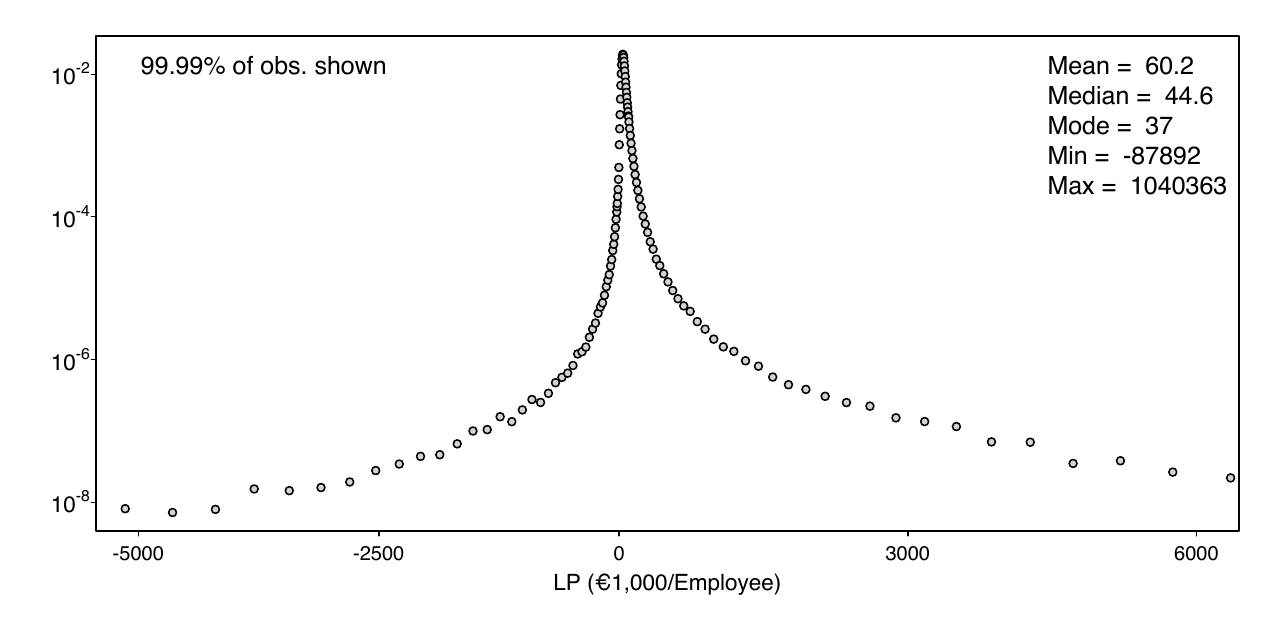}
		\caption{\textbf{Distribution of labour productivity in France, 2006-2015.} To display as many data points in the extreme tails as possible, we use log binning for the bottom and top 10\% productivity. The plot shows 99.99\% of the entire data.}
	\label{fig:levy_france}
	\end{center}
\end{figure}

Fig.~\ref{fig:levy_france} shows the distribution of labour productivity for France for the years covered by the data on a semi-log scale. An important observation is that even though the chart shows 99.99\% of observations, the tails appear relatively ``well behaved''. Once the bins mid-points are chosen appropriately, there are no large fluctuations and one could draw a fairly smooth line through the dots. This suggests that the common practice of windsorizing (removing e.g. 1\% of  observations in each tail) removes observations that are actually well-behaved statistically, rather than being unexpected or strange ``outliers'' (see e.g. \citet[p.196-197]{NOLAN2020} for a discussion).

We note the following five general characteristics of the empirical distribution of the labour productivity. First, the distribution appears unimodal. Second, the support of the distribution is very large. Third, the distribution is asymmetric, with a pronounced right skewness. Fourth, 
the distribution exhibits slowly decaying tails. And fifth, comparing the results for different years (not shown here), we find that the shape of the distribution is very persistent, as one would expect.

We now focus on one of the most important of these features: heavy tails.

\subsection{Testing for power law tails and infinite variance}
  \label{subsec: power_law}

We say that the tail of a distribution follows a power-law if its cumulative distribution function $F(x)$ takes the form
\begin{equation}
        1-F(x) \sim L(x)x^{-\alpha}, \quad x \rightarrow \infty,
\label{eq:tailbehaviour}
\end{equation}
where $L(x)$ is a slowly varying function, and $\alpha$ is the tail exponent. This tail exponent regulates the rate with which the size of extreme values decreases as we increase the sample of extreme values. Intuitively, the smaller the tail exponent, the slower the frequency of a large event decreases as we consider increasingly extreme values. Very extreme values will be relatively for frequent than a scenario with a higher tail exponent. In practice, it also determines what moments are finite; any moment greater than $\alpha$ will be infinite, while any moment less than $\alpha$ remains finite. Some authors use the term `heavy' tails for the case where $\alpha <2$, as in this case the variance is infinite.

Here we propose four methods to determine whether labour productivity distributions have heavy tails: (modified) QQ-plots, scaling of the sample standard deviation with sample size, direct estimates of the tail exponent based on extreme value distributions, and \citet{trapani2016testing}'s test for infinite moments.

\paragraph{QQ-plots}

\citet{resnick2007heavy} shows that QQ (Quantile-Quantile) plots for location-scale families can be used not only to provide visual evidence on the compatibility of a specific functional form with the data, but also to estimate the parameters. This is convenient because if $X$ is a random variable drawn from a Pareto distribution,
\begin{equation}
\label{eq:pareto_distribution}
    \text{P}(X > x) = \left( \frac{x}{k} \right)^{-\alpha}, \quad x > k,
\end{equation}
where $k$ is the threshold, and $\alpha$ is the tail exponent, the distribution of $\log X$ is a location-scale family. \citet{resnick2007heavy} shows that the values
\begin{equation}
\Big\{  \Big(-\log\Big(1- \frac{i}{N+1} \Big) ; \log X_{(i)} \Big), 1 \leq i  \leq N  \Big\},
\label{eq:scatter_pareto}
\end{equation}
where $X_{(i)}$ is the $i^{th}$ order statistic, would lie on a line with intercept $\log k$ and slope $1/\alpha$ under the null (Eq. \ref{eq:pareto_distribution}). To implement this procedure, we need to choose a cutoff $k$, or, alternatively, the sample size $N$. If we believe that the data has a Pareto tail but only as $x \to \infty$, we would prefer to choose the highest possible cutoff. However, this implies less data and thus more variance in the resulting estimates, so there is a trade-off.

\begin{figure}[!ht]
    \centering
    \includegraphics[width=.9\textwidth]{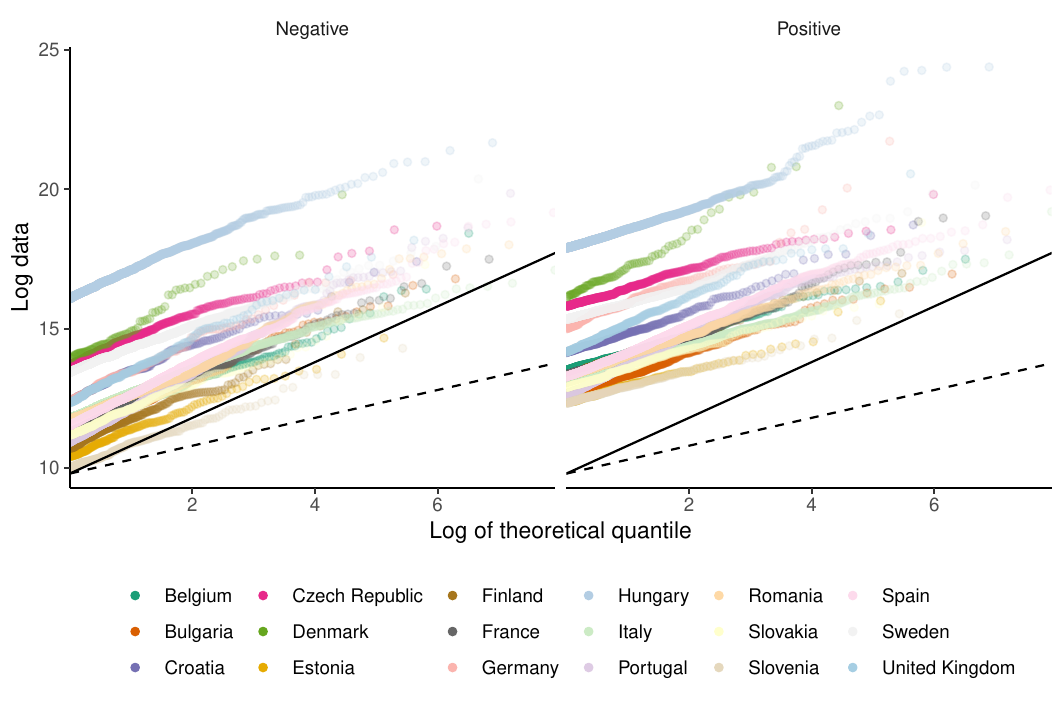}
    \caption{\textbf{QQ-plots for positive and negative labour productivity tails, 2015.} Quantiles of the labour productivity for the top and bottom 0.5\% of firms in 18 European countries are very close to the quantiles of a Pareto distribution. A Pareto tail with a exponent $\alpha$ would appear as a line with slope $a/\alpha$ \citep{resnick2007heavy}. A line with a slope of one (solid black), and a line with a slope of 1/2 (dashed) are plotted for visual comparison.}
        \label{fig:qq_plot}
\end{figure}

Fig~\ref{fig:qq_plot} shows the log of top and bottom 0.5\% observations of firm labour productivities. The datapoints by country roughly follow straight lines, which is good evidence that the tails are Pareto. Moreover, the slopes appear to lie between 1 and 1/2, suggesting $1<\alpha<2$.

One point to note is that the slopes for negative productivity appear slightly steeper than for positive productivity, suggesting a degree of tail asymmetry that is not in line with the L\'{e}vy hypothesis, where the tail exponent is the same on each side (Section \ref{section:levy_charac}). Indeed, as shown in Fig~\ref{fig:Levy-AEP-fits}, our fits may have a perceptible tendency to underestimate the left tail and overestimate the right tail. But overall, the data is substantially right-skewed, and this is well captured by the asymmetry parameter of the L\'{e}vy distribution, so the overall fit remains very good.

\paragraph{Scaling of the sample standard deviation with sample size}

If the distribution has power-law tail with $\alpha < 2$, the variance and higher-order moments are infinite. Due to this property, in finite samples, the larger the sample size, the higher the chance that an extreme event is drawn, leading to a larger sample standard deviation. More precisely, the sample standard deviation scales with sample size $N$ as $N^{\frac{1}{\alpha}-\frac{1}{2}}$ (see Appendix~\ref{app:scaling} for a heuristic derivation).

\begin{figure}[H]
	\begin{center}
	\begin{minipage}{.49\textwidth}
		\includegraphics[width=\textwidth]{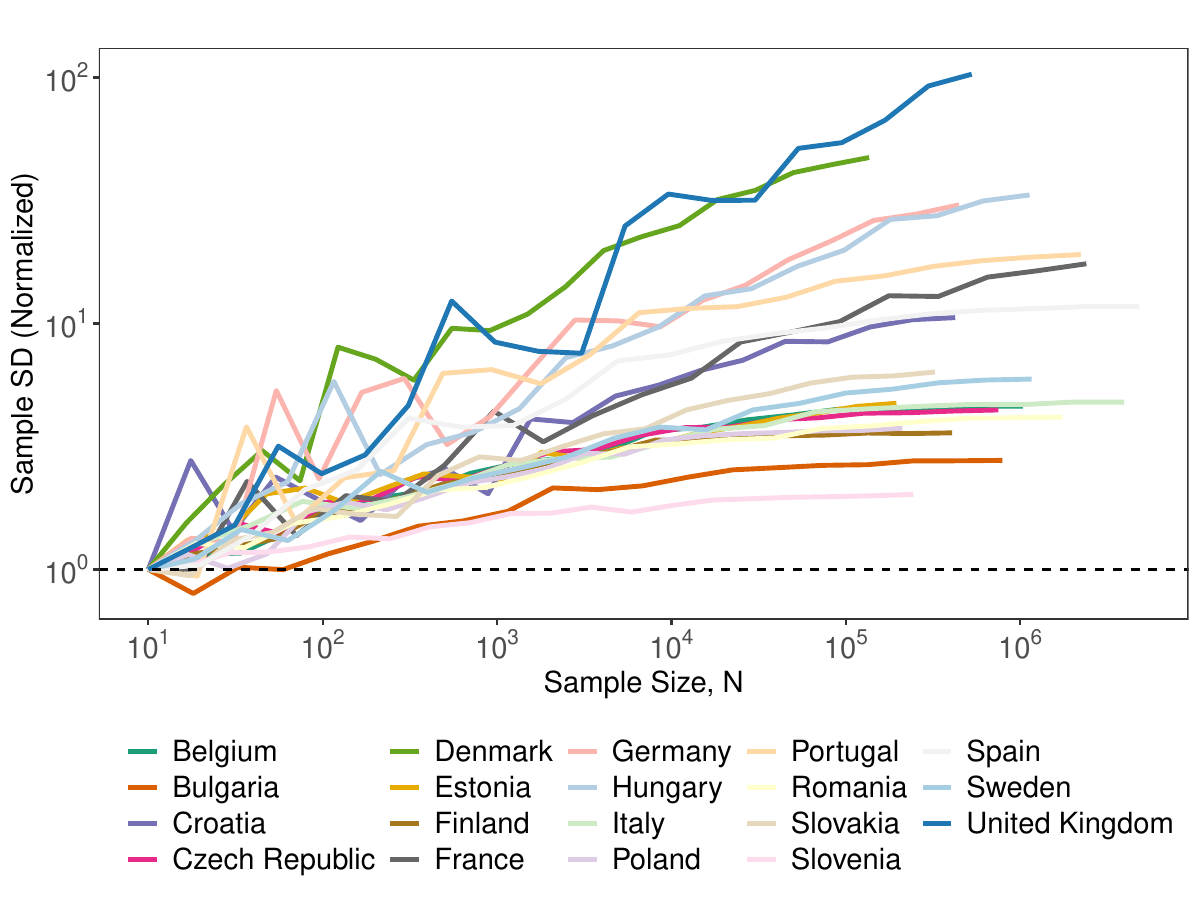}
	\end{minipage}
		\begin{minipage}{.49\textwidth}
		\includegraphics[width=\textwidth]{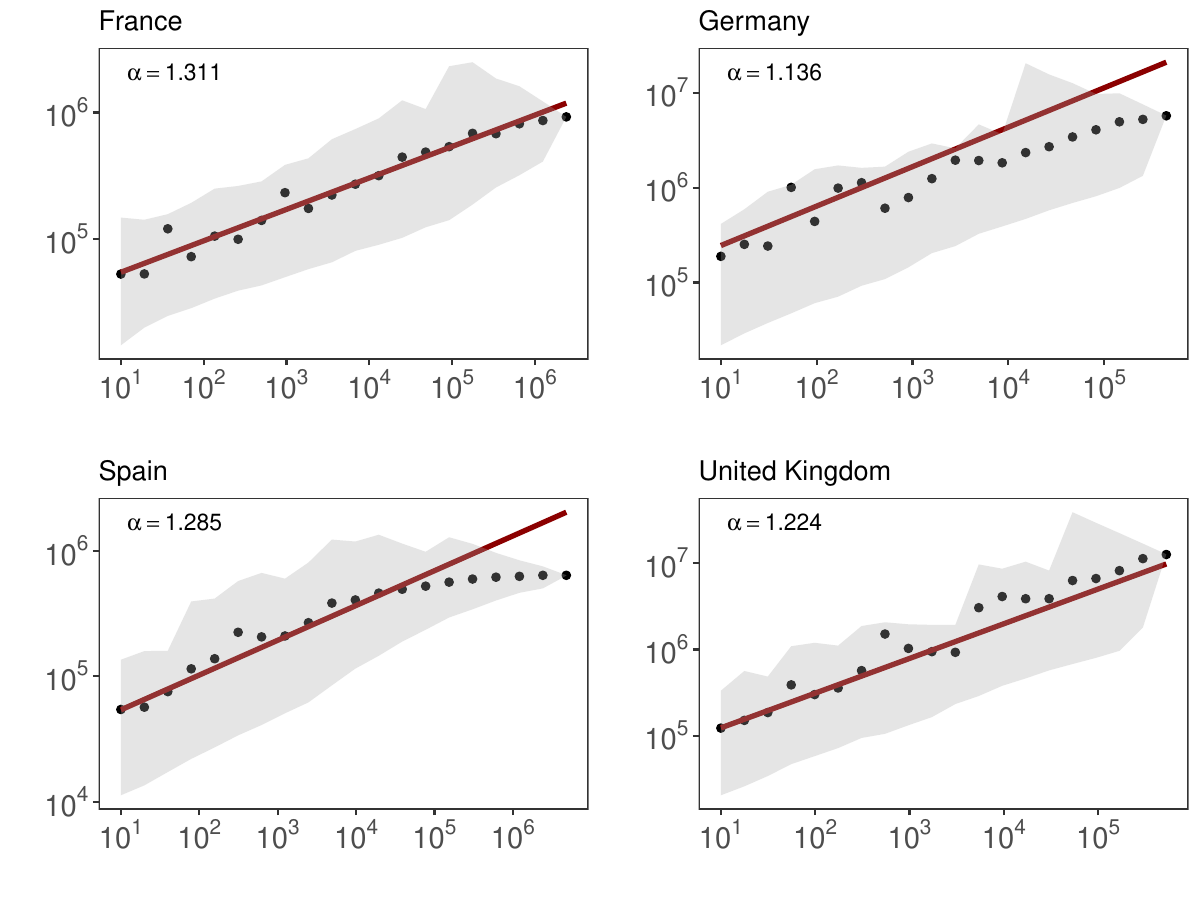}
	\end{minipage}
	\caption{\textbf{Measured standard deviation in sub-samples of firm labour productivity, country samples.} We construct a distribution by pooling together the productivity levels of firms in each of 19 countries, for all years. Then, for each subsample size $N$, we compute the standard deviation of each of 1,000 subsamples. The plot on the left shows the average of standard deviations for each of all 19 countries, while the plot on the right shows the same scaling for four selected countries, with the linear scaling calculated as $N^{\frac{1}{\hat{\alpha}}-\frac{1}{2}}$ (see Appendix~\ref{app:scaling}), where $\hat{\alpha}$ is estimated by fitting the L\'{e}vy distribution. The shaded area corresponds to the values of the sample standard deviation that fall between the $5^{th}$ and $95^{th}$ percentiles.}
	\label{fig:sd_scaling_all}
	\end{center}
  \end{figure}

Figure \ref{fig:sd_scaling_all} shows the scaling of the sample standard deviation of labour productivity for different sample sizes. It is clear that for most countries, the sample standard deviation gets larger as the sample size increases. For four selected countries, we show the theoretical scaling derived by estimating the power law exponent as the tail parameter of a L\'{e}vy alpha-stable distribution. The scaling of the sample standard deviation holds well empirically, although with a possible plateauing as $n \to \infty$ for some countries, such as Spain. Overall, this provides good evidence not only for power law tails, but also for the idea that the L\'{e}vy alpha-stable distribution can be a good model to retrieve the value of the tail index.

\paragraph{Estimating tail exponents}

Another way to test for infinite second moment is to estimate the tail exponent and see if it is lower than 2. Typically, one considers data from the tail only; that is, order statistics of up to order $k$, which makes it possible to estimate tail behaviour and determine the finiteness of moments, independently of the behaviour of the rest of the distribution. An important issue is that one has to chose a value $k^*$ which determines which data is used to estimate the tail parameter. Usually, the tails are influenced in a non-negligible way by the slowly varying function $L(x)$ (see Eq. \ref{eq:tailbehaviour}), and therefore the choice of $k^*$ may be difficult and lead to biased estimates of $\alpha$.

An early and popular method for estimating tail exponents is the Hill estimator, but as noted in \citet{resnick2007heavy}, the Hill estimator provides very different estimates of the tail depending on which value of $k^*$ we chose (``Hill Horror Plots''). Because of these well-known issues for regularly varying distributions (Eq. \ref{eq:tailbehaviour}) that are somewhat far from pure power laws, several estimators of the tail exponents have been developed, and tested for cases where the slowly varying function is non-negligible. Here we use the estimators described and implemented by \citet{voitalov2019scale}. \citet{voitalov2019scale}'s implementation includes an automatic double bootstrapping procedure for picking $k^*$. We use their package with default values. In addition, we also use the popular Hill estimator described in \citet{clauset2009power}, which finds $k^*$ by minimizing the Kolmogorov-Smirnov statistic (computed assuming that the true model is a \emph{pure} power law). We add the estimate of $\alpha$ based on the L\'{e}vy distribution as well.

\begin{figure}[H]
	\begin{center}
		\includegraphics[width=0.4\textwidth]{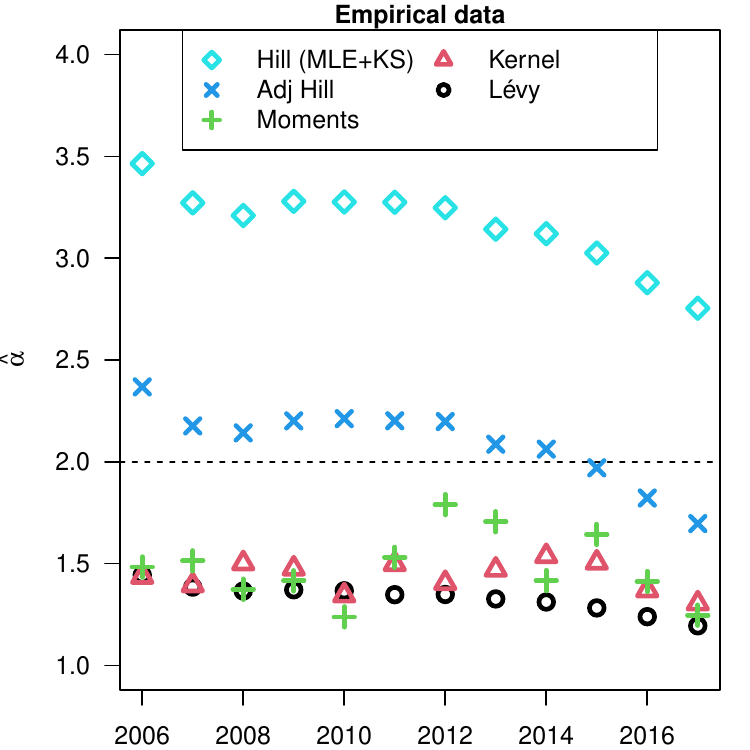}
		\includegraphics[width=0.4\textwidth]{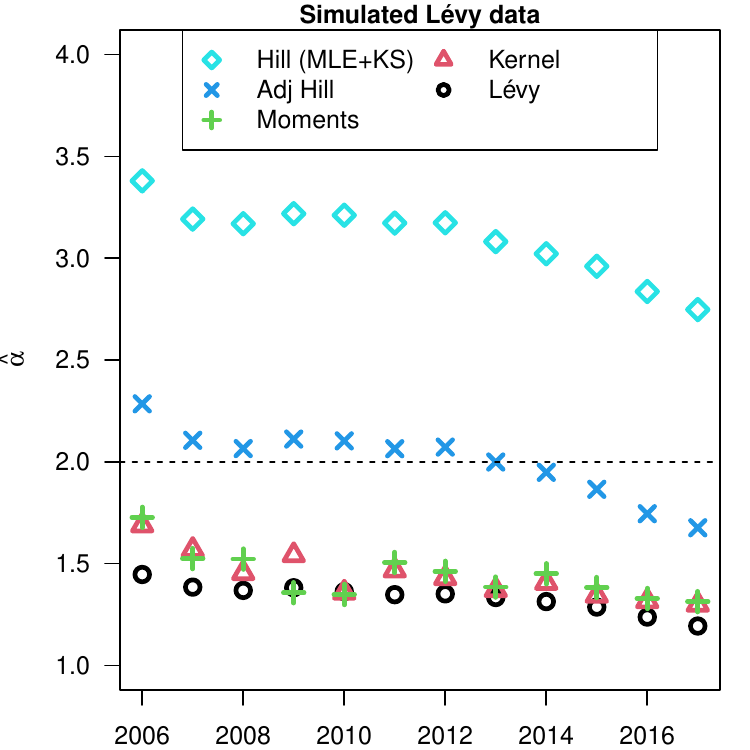}
	\caption{\textbf{Estimated values of the tail exponent $\hat{\alpha}$}, using pooled data from France for each year separately, and using 5 different estimators. The left chart shows $\hat{\alpha}$ for the empirical data. The right chart shows $\hat{\alpha}$ for data drawn from a L\'{e}vy-stable distribution with parameters equal to those estimated on the empirical data. The left chart shows that 3 estimators suggest infinite variance ($\hat{\alpha}<2$), while the two Hill-based estimators suggest finite variance on empirical data, but would also fail to capture infinite variance if the data were indeed L\'{e}vy-distributed (right). We conclude that labour productivity has indeed a power law tail with $\alpha<2$. }
	\label{fig:tail_exp}
	\end{center}
  \end{figure}

Figure \ref{fig:tail_exp} (left) shows the results using data from France for each year separately, and strongly suggests that $\alpha<2$. Three out of 5 estimators suggest similar values of $\alpha \approx 1.4$ (dots for 2010 are missing for two estimators as the routines sometimes fail). The two estimators that suggest $\alpha>2$ are Hill-based estimators, which are known to produce poor results for L\'{e}vy distributed data \citep{resnick2007heavy}. To further confirm the plausibility of these estimates, for each year, we simulate one sample of L\'{e}vy distributed data using the empirical sample size, and using parameters for the L\'{e}vy distribution estimated from the empirical data. In this case (right panel), the 5 estimators behave almost exactly as in the empirical data, suggesting that the L\'{e}vy distribution is a plausible model, and reinforcing the previous conclusion that the second moment is infinite.

The left panel of Figure \ref{fig:tail_exp} also helps to make an important point: if the data is \emph{not} L\'{e}vy distributed, but is a regularly varying distribution with a tail exponent that is correctly estimated by the Kernel or Moments estimators, then estimating the exponent using simply 5 quantiles and assuming a L\'{e}vy distribution gives relatively good results, in the case of our data. This is important, because statistical offices holding detailed micro-data may not want to release the order statistics that are necessary to run the Kernel or Moments estimators, but they may be willing to release 5 quantiles, particularly since the lowest/highest quantiles ($5^{th}$ and $95^{th}$) are likely to be uninformative about the situation of specific firms (in contrast to, say, the $99.9^{th}$ quantile).

\paragraph{Trapani's test}

As a final check, we implemented a more formal statistical test due to \citet{trapani2016testing}. The testing procedure exploits the fact that non-finite sample moments diverge with sample size. A full overview of the testing procedure and the detailed results are available in Appendix \ref{app:finite_moment_test}.

In the vast majority of both the country-year and country-industry samples, we cannot reject the hypothesis that the second moment of labour productivity is infinite. For example, for 84\% of country-year samples, Trapani's finite moment test failed to reject the null hypothesis of the infinite second moment at the 5\% confidence level. 

We also perform Trapani’s test for country-year-industry samples at the level of two-digit NACE codes. For 55\% of country-industry samples, Trapani's finite moment test failed to reject the null hypothesis of the infinite second moment at the 5\% confidence level. For the largest 5 European countries (France, Germany, Italy, Spain, and the UK), the same result holds for nearly 70\% of samples\footnote{
We also estimated the tail exponent using \citet{voitalov2019scale}'s estimators at the country-sector-year level and found similar results. While the majority of cases have a tail exponent lower than 2, there are many cases having an exponent between 2 and 3, and many cases having too few values to give reliable estimates. In Appendix \ref{app:scaling_example}, we show examples of the scaling (or lack thereof) of the empirical standard deviation with sample size at the country-industry-year level.
Of course, part of the dispersion on the pooled distributions is due to pooling heterogenous subsamples, and we acknowledge that we cannot validate the infinite moment hypothesis (or, for that matter, the L\'{e}vy hypothesis) on detailed subsamples as clearly as on the pooled distribution. 
}.

In Appendix \ref{app:finite_moment_test}, we evaluate the performance of the test, and find that it tends to over-reject the null, sometimes substantially; this is unsurprising since we make a conservative choice for the parameters of the test. As a result, despite somewhat mixed results on the fine-grained sub-samples, we conclude that in a large majority of cases, the second moment of the distribution of VA per worker is likely to be infinite.

\section{Models and estimation } 
\label{sec:models}

This section provides a detailed discussion of the key properties of the L\'evy alpha-stable distribution, and introduces the estimation methods. For details on the competing distributional models (the Asymmetric Exponential Power (AEP) with 4 or 5 parameters), see Appendix \ref{app:comparisonAEP} and \citet{Bottazzi/Secchi11}.

\subsection{The L\'{e}vy alpha-stable distribution}
\label{sec:levy}

The L\'evy alpha-stable distribution is a natural candidate for many distributions exhibiting heavy tails, as it emerges from Generalized Central Limit Theorem (GCLT). We first describe the key properties of the L\'evy alpha-stable distribution, and then discuss the GCLT.

\subsubsection{Characteristics of the L\'evy alpha-stable distribution}
\label{section:levy_charac}

The L\'evy alpha-stable distribution is a four-parameter distribution with parameters $\alpha$ (tail exponent), $\beta$ (skewness), $\gamma$ (scale), and $\delta$ (location). The density function exists in closed form only in a few special cases such as $\alpha=2 $ (Gaussian), $\{\alpha=1, \beta=0\}$ (Cauchy), and $\{\alpha=1/2, \beta=1\}$ (standard L\'{e}vy). Out of several alternatives, we use \citeauthor{NOLAN1998187}'s \citeyearpar{NOLAN1998187} $\textbf{S}_0$ parametrization, which has the  characteristic function
\begin{eqnarray}
 \varphi(t)\equiv\operatorname{E}[\exp{(itX)}] ={\begin{cases}
	\exp{ \Big( -|\gamma t|^\alpha \Big[ 1+i\beta \text{tan}\left({\tfrac {\pi \alpha }{2}}\right) \operatorname {sgn}(t)\left(|\gamma t|^{1-\alpha}-1 \right) \Big] + i\delta t \Big)}
	&\alpha \neq 1,
	\\\exp{\Big( -\gamma|t| \Big[1+i\beta{\tfrac {2}{\pi}} \operatorname {sgn}(t)\log(\gamma|t|) \Big] + i\delta t \Big)}&\alpha =1,
	\end{cases}}
\end{eqnarray} 
where $i$ is the imaginary unit, $t \in R$ is the argument of the characteristic function, and $ \operatorname {sgn}()$ is the sign function.
This particular parametrization is useful for numerical work and statistical inference since the characteristic function is continuous in all four parameters.

  \begin{figure}[H]
	\begin{center}
		\includegraphics[width=\textwidth]{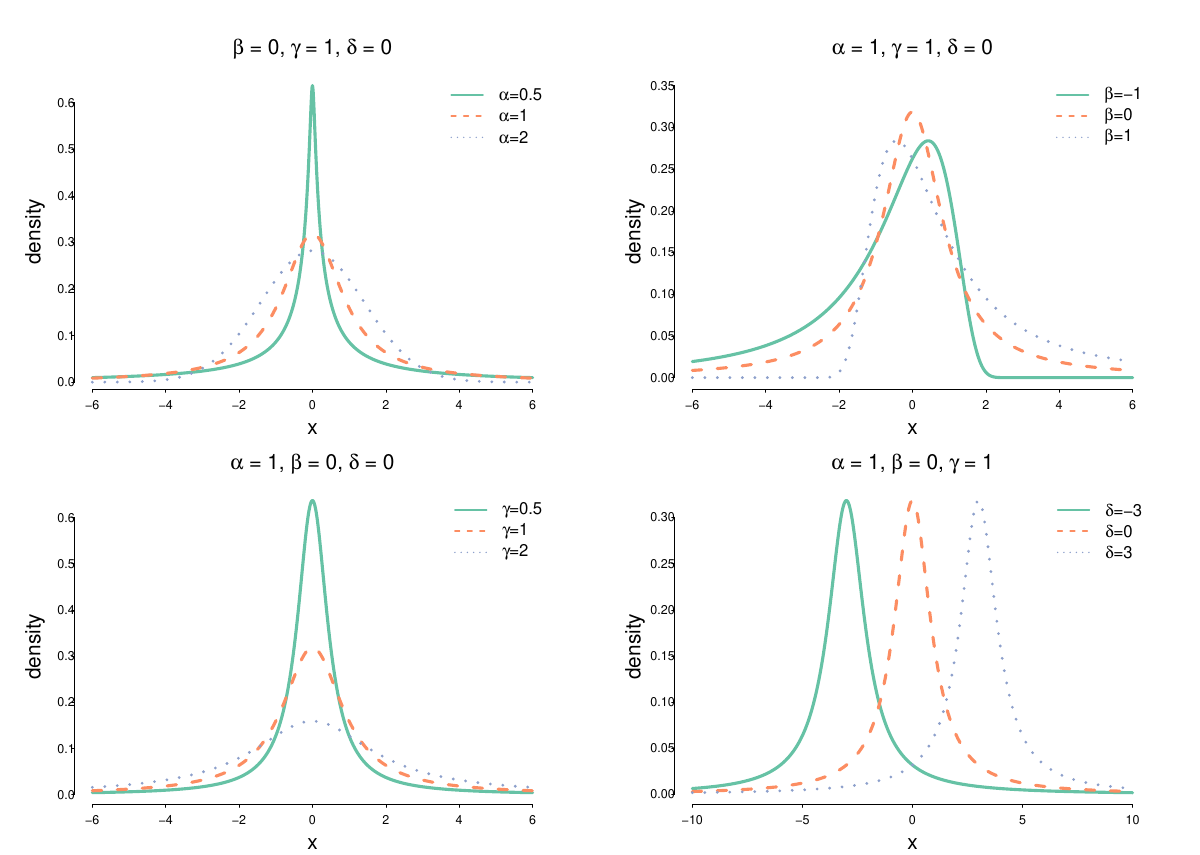}
		\caption{\textbf{Probability densities of L\'evy alpha-stable distributions for different parameter values.} The top left plot varies the $\alpha$ parameter (tail exponent), which determines how heavy the tail is. The top right plot varies the $\beta$ parameter (skew parameter). The bottom left plot varies the $\gamma$ parameter, which determines the `scale', or `width', of the distribution. The last plot varies the $\delta$ parameter, which shifts the location of the modal value of the distribution.}
	\label{fig:levy_pdf}
	\end{center}
\end{figure}

Fig \ref{fig:levy_pdf} shows the probability density function of the  L\'evy alpha-stable distribution, with three different values for each of the four parameters. Starting from the top left, the parameter  $\alpha  \in (0,2]$ is also called the tail exponent because the tail of the distribution decays as a power law with exponent $\alpha$, that is, $P(X>x) \approx Cx^{-\alpha}$. This implies that scaling the value of $x$ by a factor $h$ just scales the tail probability by $h^{-\alpha}$, since $h^{-\alpha}P(X>x)\approx C(hx)^{-\alpha}$. For example, if $\alpha=1$, values of $x$ that are twice as large (i.e., $h=2$) as some reference $\tilde{x}$ will be half as common, no less. As a result, with power law tails extreme values are very common and can dominate certain moments of the distribution. Thus, the tail parameter $\alpha$ is an important indicator of dispersion. The lower $\alpha$, the thicker the tail (the more ``tail dispersion'' there is). However, note that while a smaller $\alpha$ indicates more dispersion because of a heavier tail, a smaller $\alpha$ also leads to a higher concentration of values close to the mode, and thus, in this sense, to a lower dispersion in the body of the distribution.

The top right panel shows the effects of the ``skew'' or ``asymmetry'' parameter $\beta \in [-1,1]$. For $\beta=0$, the distribution is symmetric, for $\beta>0$ right-skewed, and for $\beta<0$ it is left-skewed (except when $\alpha=2$, as the skew parameter $\beta$ vanishes in that case). In practice, we will find that productivity distributions are right skewed.

The bottom-left panel shows the effect of the scale parameter $\gamma \in [0, +\infty]$. The higher the scale parameter $\gamma$, the wider the body of the distribution. The parameter $\gamma$ is not dimensionless; it is expressed in the same units as the data. We can get more intuition on the role of $\gamma$ by considering special cases \citep[p.168]{NOLAN2020}. First, $\gamma$ is highly related to the (25-75) Inter Quartile Range; in fact, when $\beta$ is small and $\alpha$ is large (say, $\beta=0.5$ and $\alpha =1.3$), The interquartile range divided by 2 is a good estimator of $\gamma$. Second, for large enough $\alpha$, the re-centred data (so that $\delta=0$) can be used to estimate $\gamma$ as the median of the absolute values.

Finally, the bottom right panel shows the effect of the location parameter $\delta \in (-\infty, +\infty)$, which is fairly intuitive even though it is not in general exactly related to key quantities such as the mode, mean, or median. For symmetric stable distributions, the median is equal to the location, and for asymmetric distributions the median remains close to the location as long as $|\beta|$ is not too large and $\alpha$ is not too small \citep[p.168]{NOLAN2020}. When $\alpha>1$, the sample mean is actually a consistent (but slowly converging) estimator of $\delta$ \citep[p.174]{NOLAN2020}. \citet[p.92]{NOLAN2020} defines an $\textbf{S}_2$ parametrization specifically so that $\delta$ is the mode.

We can gain further intuition by considering special cases: For $\alpha = 2$, the distribution becomes a Gaussian with variance $2\gamma^2$ and mean $\delta$. For $\alpha = 1, \beta = 0$ the distribution becomes Cauchy with scale parameter $\gamma$ and location parameter $\delta$. The role of $\delta$ and $\gamma$ as location and scale parameters can be seen clearly by writing the Cauchy density as $f(x) \propto \Big[ 1 + \Big( \frac{x-\delta}{\gamma} \Big)^2\Big]^{-1}$. More generally, if $X \sim \textbf{S}_0(\alpha,\beta,\gamma,\delta)$, then $(X-\delta)/\gamma \sim \textbf{S}_0(\alpha,\beta,1,0)$.

One of the key characteristics of the L\'{e}vy alpha-stable distribution is that $\gamma$ and $\alpha$ measure qualitatively different aspects of dispersion, providing a richer perspective on observed empirical patterns. In discussing the estimation results in the following sections, we will compare the estimated $\alpha$ and $\gamma$ with other conventional measures of dispersion, to demonstrate the strength of parametric measures of dispersion based on the L\'{e}vy alpha-stable distribution in the case of productivity. But before, let us explain that alpha stable distributions are (in general) likely to arise, due to the GCLT.

\subsubsection{The Generalized Central Limit Theorem}
 \label{sec:gclt}
 
Consider the sum of i.i.d. variables (with a zero mean, without loss of generality). If they have a finite variance, then this sum divided by $N^{1/2}$ tends in distribution to a Gaussian distribution - this is the CLT. If we are unwilling to assume that the i.i.d. variables have finite variance, the Generalized CLT states that the sum should be normalized $N^{1/\alpha}$, where $\alpha$ is the tail exponent of the i.i.d. variables (or $\alpha=2$ if they have finite variance), and it tends to the L\'evy alpha-stable distribution (which is the Normal distribution if $\alpha=2$).

We take a more formal presentation from \citet{NOLAN2020}, Theorem 3.12, and state it assuming $1<\alpha<2$, which is the relevant case here, and simplifies exposition\footnote{
For $\alpha$>2, the rescaled sum converges to a normal distribution. For $0<\alpha\leq 1$, the mean of $X$ does not exist so the expressions for centering the sum are different, but the sum still converges to an alpha stable distribution. Note also that we state the Theorem in terms of convergence to Nolan's $\boldsymbol{S_1}$ parametrization, but our empirical work is conducted with the $\boldsymbol{S_0}$ parametrization.
}.

\begin{theorem}
Let $X_1, X_2 \dots$ be i.i.d. copies of $X$, where $X$ has characteristic function $\phi_X(u)$ and satisfies the tail conditions
\[
x^\alpha F(-x) \to c^- \text{ and } x^\alpha (1-F(x)) \to c^{+} \text { as } x \to \infty,
\]
where $F(x)=P(X \leq x)$ is the cumulative distribution function of $X$, $c^{-}\geq 0$, $c^{+} \geq 0$, $0<c^{-}+c^{+}<\infty$, and $1<\alpha<2$. Then
\[
a_N \Big(X_1+X_2+\dots+ X_N - N E[X] \Big) \xrightarrow{d} Z \sim \textbf{S}_1(\alpha,\beta,1,0),
\]
where
\[
\beta=\frac{c^{+}-c^{-}}{c^{+}+c^{-}}
\]
and 
\[
a_n=\Big(\frac{2 \Gamma(\alpha) \sin (\pi \alpha/2) }{\pi (c^{+}+c^{-})} \Big)^{1/\alpha} N^{-1/\alpha}.
\]
\end{theorem}
In plain English, the sum can be re-centered using the first moment (thanks to $1<\alpha<2$), the rescaling factor is proportional to $N^{1/\alpha}$  (and a function of $\alpha$ and the tail balance parameters $c^{+}$ and $c^{-}$), and the asymmetry parameter $\beta$ depends on the relative difference between tail balance parameters $c^{+}$ and $c^{-}$.

This theorem is remarkably general. For instance, assume that 
employees draw their productivity from a heavy tail distribution\footnote{
There exist many mechanisms generating power laws, see \citet{mitzenmacher2004brief} and \citet{gabaix2009power}. Although not focusing on tail dispersion, \citet{ilzetzki2017measuring} provide evidence of substantial dispersion in vote counting productivity across polling stations.
}. Firms' productivity being an average of the productivity of their employees, sufficiently large firms will have L\'evy distributed productivity.

A legitimate concern with this justification for the alpha stable distribution is that employees are unlikely to draw their productivity from the same distribution, and even if they did, these draws are unlikely to be independent: \- the two ``i''s of the i.i.d. assumption are violated. 

In the case of finite variance variables, the CLT has been extended to cover independent but heterogeneous variables, non-independent but identically distributed variables, and non-independent, heterogenous variables, see \citet{white2000asymptotic}.
There are similar attempts for infinite variance variables, but less is known and the conditions are typically more difficult to state, less intuitive, and harder to check empirically.

Starting with dependent variables, there is growing literature but it is generally focused on time series, rather than cross-sectional correlations\footnote{
In spatial and network econometrics, it is well-known that spatial correlations are qualitatively different from time series correlations, as reflected for instance in the fact that one needs to decide whether to use ``infill'' or ``increasing domain'' asymptotics \citep{kelejian2017spatial}.
}. That said, it is known that under ``mild'' time series dependence (e.g. $m$-dependent sequences), the GCLT still applies (see \citet{bartkiewicz2011stable} for precise conditions and a history of this literature).

Regarding heterogeneous variables, it is important to realize that thanks to the property of stability, a sum of independent but heterogenous stable distributions (with the same $\alpha$, also called ``index of stability'') is still a stable distribution, with parameters being an explicit function of the weights and parameters of underlying distributions \citep[Proposition 1.3]{NOLAN2020}. As a result, if we have a fixed number of heterogeneous distributions (say, two types of employees), we can let the number of copies go to infinity for each kind of distributions, obtain a stable distribution for each of the sums, and then sum up all the stable variables, which will give another stable distribution. A more precise statement appears in \citet{shintani2018super}.

While this discussion suggests that the basic GCLT has a broader scope than i.i.d. variables, there remain a number of limitations, such as a fixed $\alpha$ for heterogeneous variables\footnote{
Generically, the smaller $\alpha$ dominates (see e.g. \citet{gabaix2009power}, Eq. 3), but convergence might be slow so that these theoretical results do not always work well in samples of even moderate size. See also \citet{cohen2020heavy} for the case of heterogenous $\alpha$ (but focusing on $0<\alpha<1$).
}, the relative lack of results on GCLT for cross-sectional dependence (but see \citet{cohen2020heavy} for progress on this and on heterogenous variables), and limited results on large deviations. As a result, we refrain from attempting to provide a specific data-generating process (DGP) here. It is likely that several plausible DGPs can deliver a stable distribution, but in our view, evaluating which one is more plausible would require fitting its parameters, and simulating it using \emph{finite} samples of sizes similar to the empirical data, as well as checking additional predictions of each model. 
We thus leave this for further research, and in this paper, we focus on establishing that the distribution of productivity is alpha stable, and discuss the values of the parameters for subsamples, such as different points in time, countries, and industries.

\subsection{Distribution fit and model comparison}

We use two members of the Asymmetric Exponential Power (AEP) family of distributions as benchmark. Like the L\'{e}vy alpha-stable distribution, AEP distributions generalize the Gaussian distribution to allow for skewness and more probability density in the tails. In contrast to the L\'{e}vy distribution, however, the AEP does not feature power law tails, so all its moments are finite. Various formulations of AEP distributions have been introduced since the 1980s, notably the 4-parameter AEP by \citet{Delicado/Goria08} and the 5-parameter AEP by \citet{Bottazzi/Secchi11}. They have been suggested as distributional models for quantities that are known to have heavier tails than Gaussians, such as changes in currency exchange rates \citep{Ayebo/Kozubowski03}, and (logarithmic) firm growth rates \citep{Bottazzietal07}. 

The AEP distributions provide a good benchmark because (i) they have been applied to firm-level data related to productivity (namely, sales growth rates \citep{Bottazzietal07}), (ii)  they are broadly able to fit the body of the empirical distribution, but (iii) they differ radically from the L\'{e}vy alpha-stable distribution in their tail behavior.

Among the approaches to fitting AEP distributions are applications of maximum likelihood with interval-constrained optimization \citep{Bottazzi14} and the method of L-moments \citep{ASQUITH2014955}. While the former returns a fit that is faithful to the likelihood of the empirical density, the latter will match the overall shape of the distribution better at the expense of exactness around the modal value, given the role of the third and fourth moment in the fitting procedure\footnote{
See \citet{moran} for a discussion of the inability of the Laplace distribution to fit growth rates due to the sharp change of behaviour around the mode.
}. Appendix~\ref{app:comparisonAEP} gives the functional forms of the 4-parameter and 5-parameter AEP models, the details of the fitting procedure and goodness comparisons to the L\'{e}vy alpha-stable model fit.

\begin{figure}[H]
\centering
 \subfloat{\includegraphics[width=.85\textwidth]{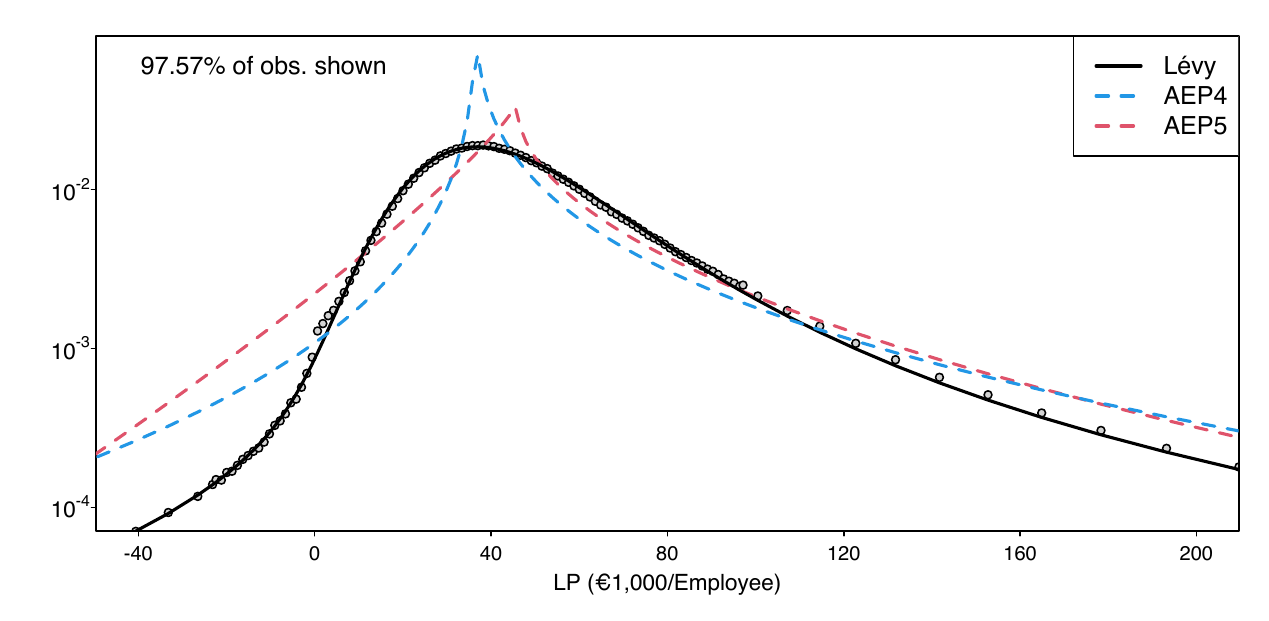}}\\
 \vspace{-1em}

 \subfloat{\includegraphics[width=.85\textwidth]{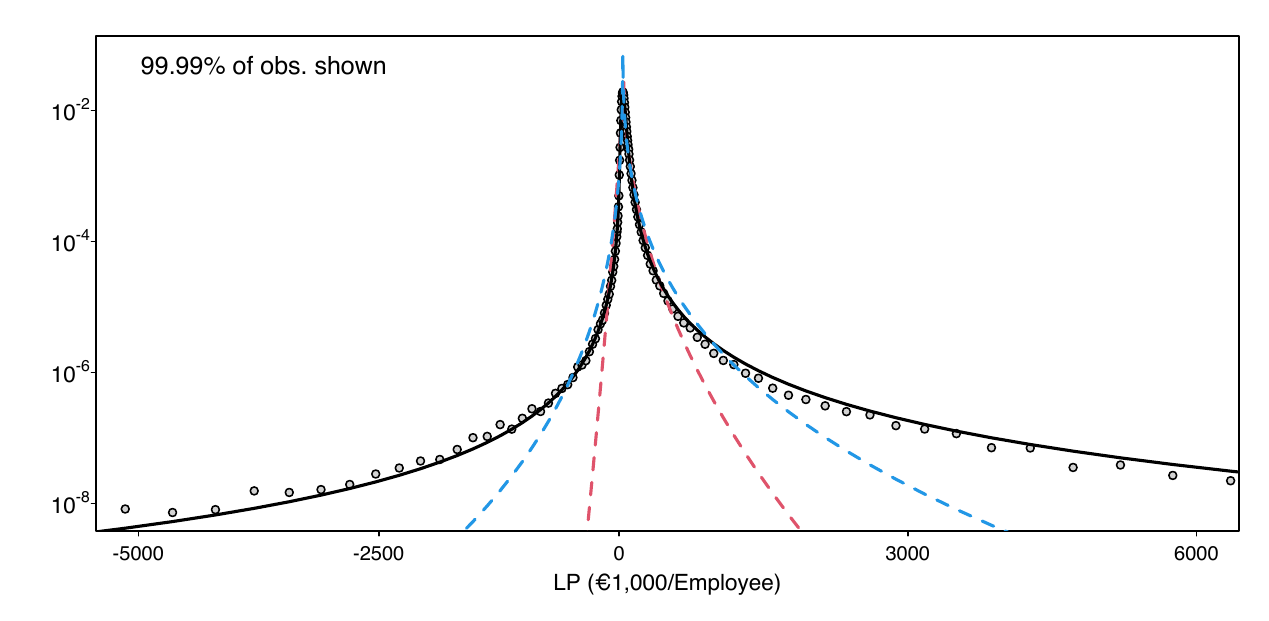}}\\
  \caption{\textbf{Distribution of labour productivity with L\'evy alpha-stable and AEP fits.} The solid black line indicates the L\'evy alpha-stable, The solid red line indicates the 4-parameter AEP fit obtained with the method of L-moments and the dashed blue line indicates the 5-parameter AEP fit obtained with interval-constrained likelihood optimization.}
  \label{fig:Levy-AEP-fits}
\end{figure}

\begin{figure}[h!]
	\begin{center}
	\begin{minipage}{.49\textwidth}
		\includegraphics[width=\textwidth]{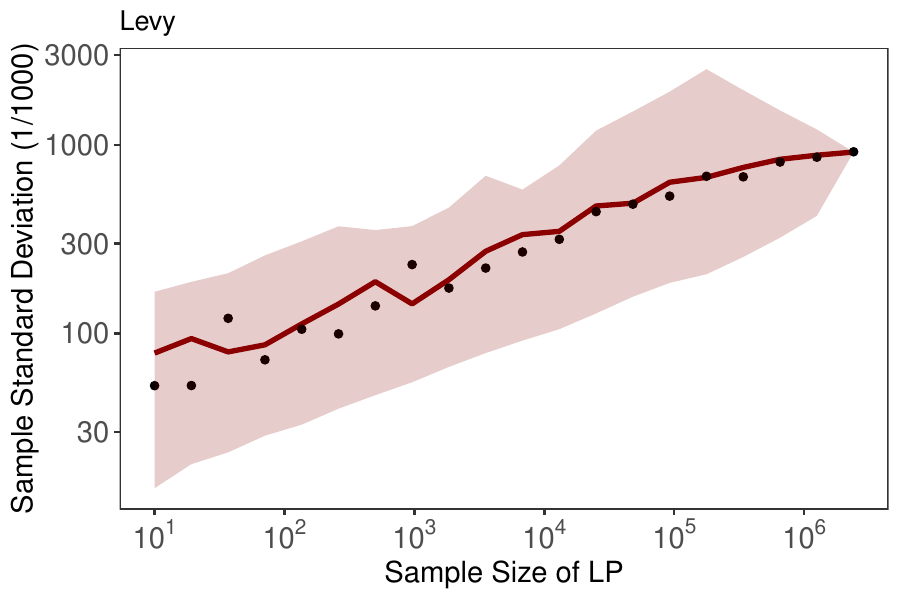}
	\end{minipage}
		\begin{minipage}{.49\textwidth}
		\includegraphics[width=\textwidth]{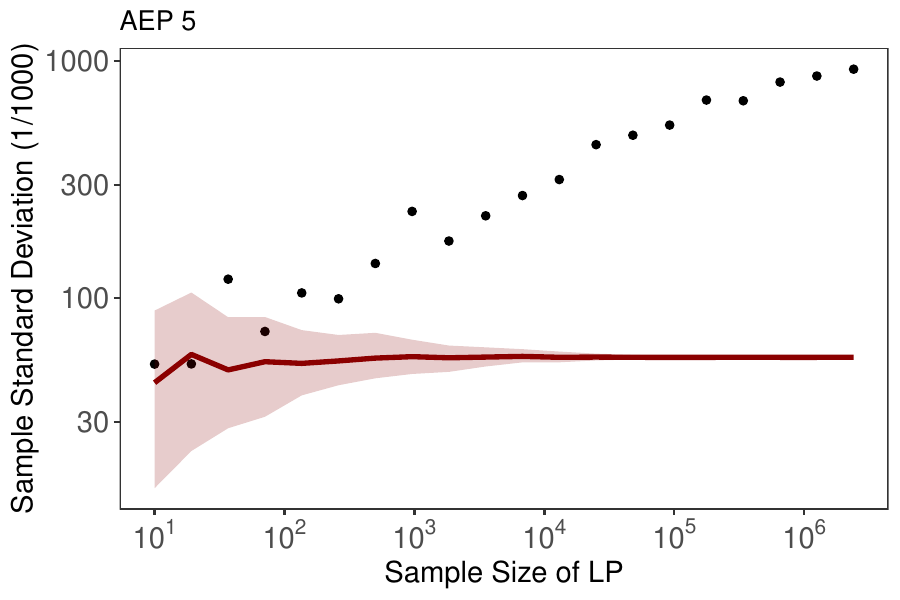}
	\end{minipage}
	\caption{\textbf{Simulation of the scaling of the sample standard deviation with sample size for L\'{e}vy alpha-stable and AEP distributed data.} We generate surrogate datasets that follow either the L\'{e}vy alpha-stable (left) and for the 5-parameter AEP distribution (right), using the same sample size as the empirical data for France ($2.4$ million), and the parameters estimated on the empirical data. We then estimate the standard deviation in subsamples of the surrogate datasets, using many subsamples for each subsample size. The figures show mean and 5\%-95\% quantiles for estimates of the standard deviation. The black dots show the mean (across subsamples of given size) of the sample standard deviation using the empirical dataset.}
	\label{fig:scaling-simulation}
	\end{center}
  \end{figure}

Fig~\ref{fig:Levy-AEP-fits} shows the empirical densities and the fits for the L\'{e}vy alpha-stable, AEP distributions fitted with interval-constrained likelihood optimization and again with L-moments\footnote{
We employ \citet{Bottazzi14}'s Subbotools package for the interval-constrained likelihood optimization of the 5-parameter AEP and the R-package lmomco \citep{ASQUITH2014955} for the L-moments fitting of the 4-parameter AEP. L-moments of the 5-parameter AEP distribution have to our knowledge not yet been derived.
} for the France sample. The L\'{e}vy alpha-stable distribution is able to fit the entire domain of the empirical distribution extremely well, except for a small irregularity in the data at zero, and possibly a slightly imperfect ability to fit deal with asymmetric tail behavior (as noted in Section \ref{subsec: power_law}). 
In contrast, the AEP distributions fail to fit the tails and have an unsatisfactory fit for the body of the distribution. The same result holds for the other country samples, as shown in  Table~\ref{tab:Model_Comparison_main} where the log-likelihood of each model is compared.

As a further check, Fig~\ref{fig:scaling-simulation} compares the scaling of the estimated standard deviation with sample size in the L\'{e}vy alpha-stable model and the 5-parameter AEP model (using a simulated sample of 2.4 million observations, which is the size of our empirical data for France). Both distributions are generated with the parameters we estimated for the respective model with our empirical data for France. Clearly, an AEP random sample fails to reproduce the pattern which we have shown holds for both the L\'{e}vy distribution and the empirical data (Fig~\ref{fig:std_17}, and, for the case of France, Fig~\ref{fig:sd_scaling_all}).

\section{Patterns of productivity dispersion}
\label{sec:results}

In this section, we compare our indicators of productivity dispersion with other metrics, using two case studies: dispersion over time, and correlation of dispersion with intangible capital intensity. For the case studies, we use the most disaggregated samples at the country-year-industry level, which leads to nearly 3,500 sub-groups\footnote{
In principle, we would have preferred to estimate and test goodness of fit at the disaggregated level, and then discuss whether a mixture model is more appropriate when data is pooled. We acknowledge that the problem of estimating mixture models remains, and will be an important way forward to understand the extent to which country-level heterogeneity is due to within- vs between-sector heterogeneity. However, the sample sizes at a detailed level are fairly small, so here we can establish the quality of the L\'{e}vy hypothesis only at a pooled level. We use country-year-industry samples to provide an example of how our metrics could be used in practice when one has access to the larger samples typically available in administrative data.
}
. We start by explaining our choice of dispersion metrics.

\subsection{Comparing measures of dispersion}
\label{sec:comment_log}

To sum up from the earlier discussion, we cannot measure productivity dispersion using the empirical standard deviation because it is likely to be a measurement of a moment that is infinite, so its measured value depends on sample size and is not meaningful. We also cannot measure dispersion as the standard deviation of the log of VA/worker, because too many values are negative, and would need to be removed when taking the log. As a result, we are left with only one serious contender: Inter-quantile ratios. 

\paragraph{Computing Inter-Quantile Ratios}
Inter-Quantile Ratios (IQRs) are very popular in part because they are easy to interpret. Let us denote the value of the $p^{th}$ quantile for VA per worker by $Q_p(V)$. Considering for instance firms at the $90^{th}$ and $10^{th}$ quantiles, we define the IQR as the ratio
\begin{equation}
\text{IQR}_{90/10}=\frac{Q_{90}(V)}{Q_{10}(V)},
\label{eq:IQR}
\end{equation}
which tells us how many times more productive the firm at the $90^{th}$ percentile is compared to the firm at the $10^{th}$ percentile. Usually, the variables are initially log transformed. If all values are positive, the firm sitting at the $p^{th}$ quantile of VA per worker is also the firm sitting at the $p^{th}$ quantile of the log(VA per worker), so we have $\log(Q_p(V))=Q_p(\log(V))$, and therefore Eq. \ref{eq:IQR} can be rewritten

\[
\text{IQR}_{90/10}=\exp \Big( Q_{90}(\log V) - Q_{10}(\log V) \Big).
\]
In other words, the Inter-Quantile Ratio is the exponential of the Inter Quantile \emph{range} of the log-transformed values. While one usually log transforms the data before computing an Inter-Quantile range, this is by no means necessary because one could just directly compute an Inter-Quantile Ratio.

This makes clear a key advantage of IQRs: even when there are negative values, IQRs can still be computed, as long as the bottom quantile is positive. For example, Spain has roughly 3-4\% of negative labour productivity observations (Table~\ref{tab:obs_ctry_yr_neg}), so we do not need to drop the negative values if we want to compare, say, the bottom 5\% to the top 5\%. Unfortunately, the practice of automatically taking log as a first step of the analysis has sometimes led to removing non-positive values before computing IQRs, leading to a bias in the measurement of dispersion.\footnote{
If one wants to take the log but keep all the observations (to keep the ranks, at least for the positive values \citep{campbell2019measuring}), one can code the log of negative values as some value strictly lower than the lowest value of the log values of the positive values available in the sample. Note that coding log labour productivity as 0 would be incorrect as there could be values of VA per worker that are between 0 and 1 (particularly if expressed in thousands or millions of currency units), leading to log values that are negative.
}

\begin{figure}[H]
	\begin{center}
	    	\begin{minipage}{.55 \textwidth}
    		\includegraphics[width=\textwidth]{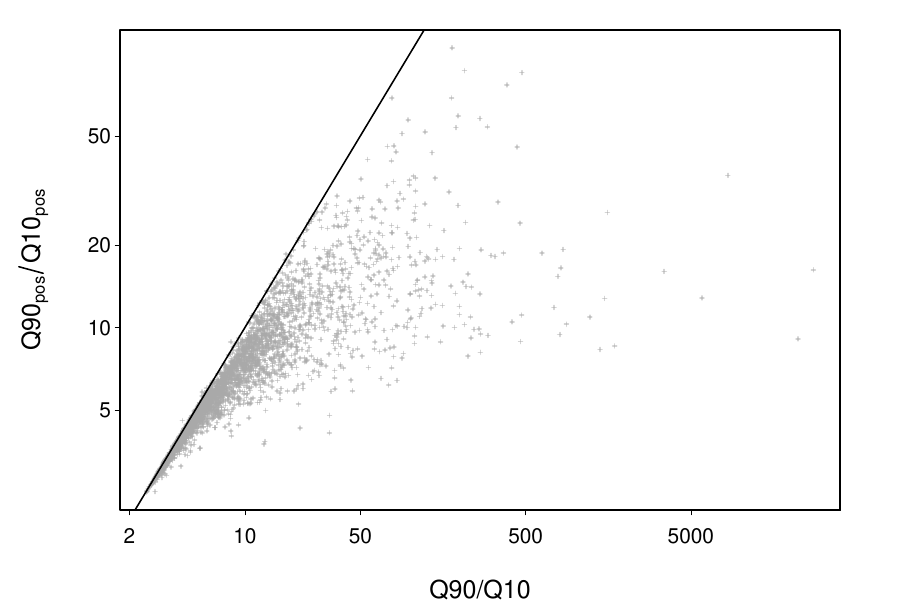}
    	\end{minipage}
    	\caption{\textbf{Scatterplot of quantiles ratios, with and without removing non-positive values.} 
    	Both axes are on the log scale. The black line is a 45 degree line.}
\label{fig:cut_effect}
	\end{center}
\end{figure}

Fig~\ref{fig:cut_effect} checks how substantial this bias can be in practice, by comparing the ratio of two quantiles before and after removing non-positive values, Q90/Q10 and Q90$_{\text{pos}}$/Q10$_{\text{pos}}$. We use the country-year-industry samples at a 2-digit industry level. We exclude the samples where the bottom 10\% quantile of the labour productivity is negative. By definition, the log quantile ratio based only on positive productivity values (y-axis) is always smaller than the actual log ratio, so all the points are below the 45 degree line. It appears that the larger the actual dispersion is, the higher is the degree of distortion. While in the vast majority of cases, the distortion is likely to be minor, in extreme cases the biased IQRs is two orders of magnitude smaller than the true IQRs. This is likely to happen when the lower quantile is close to zero.

\paragraph{Choosing IQRs}
We will study two inter-quantile ratios as non-parametric measures of dispersion:  $\log(Q_{90}/Q_{10})$, which measures dispersion in the body of the distribution, and $\log(Q_{95}/Q_{50})$, which measures the extent to which super-star firms deviate from a ``typical'' firm. Our choice is driven by previous work that has identified an increase in productivity dispersion mostly due to the divergence of superstar firms \citep{andrews2016best,de2022firms}. We compute the quantiles on the non-log data, but we remove observations of IQRs that are negative (i.e., when $Q_{10}$ or $Q_{50}$ is negative).

\paragraph{Measuring dispersion using L\'{e}vy parameters.} Because the IQRs are relative, we do not want to compare them directly with $\gamma$, which is expressed in monetary units. To express dispersion in the body of the distribution in relative terms, we propose to use $\gamma/\delta$. Using a ratio of a scale over a location parameter is fairly common - for instance the standard deviation over the mean defines the well-known coefficient of variation, which is the inverse of the well-know Sharpe ratio in finance. Of course, $\delta$ can be close to zero or even negative, but the issue also exists and is indeed worse with IQRs: the lower quantile of a (body dispersion) IQR is more likely than $\delta$ to be very close to zero or negative, since $\delta$ is usually close to the median.

Second, we consider $\alpha$, the tail parameter. We consider that a fatter tail ($\alpha$ closer to zero) indicates more dispersion, since fatter tails indicate that there is a higher mass of super-productive firms (in almost all cases, $\beta$ is substantially above 0, so we consider $\alpha$ as an indicator of the fatness of the right tail.). We often use $-\alpha$ instead of $\alpha$, so that this indicator, like all the others, indicates more dispersion when it increases.

We compute these four indicators at the level of industry (2 digit)-country-year. The samples are fairly small, so we remove samples with size $N<50$, and we use maximum likelihood to estimate the L\'{e}vy parameters when $50 \leq N <200$. We remove values of $\delta$ that are negative and a few cases where the MLE fails or the quantile methods gives corner solutions. This leaves us with more than 3700 country-industry-year triples with estimated values of the parameters, out of an ideal maximum that would have been 19 countries $\times$ 20 industries $\times$ 12 years = 4560 triples. Appendix \ref{sec:fittingmethod} justifies these choices based on a Monte Carlo simulation, and Fig. \ref{fig:hist_country_industry_year} shows the histograms of the estimated parameters.

We also checked that the scaling of the sample standard deviation with sample size persists in some country-year and country-industry subsamples (Appendix~\ref{app:scaling_example}).

\begin{figure}[H]
	\begin{center}
		\includegraphics[width=.8\textwidth]{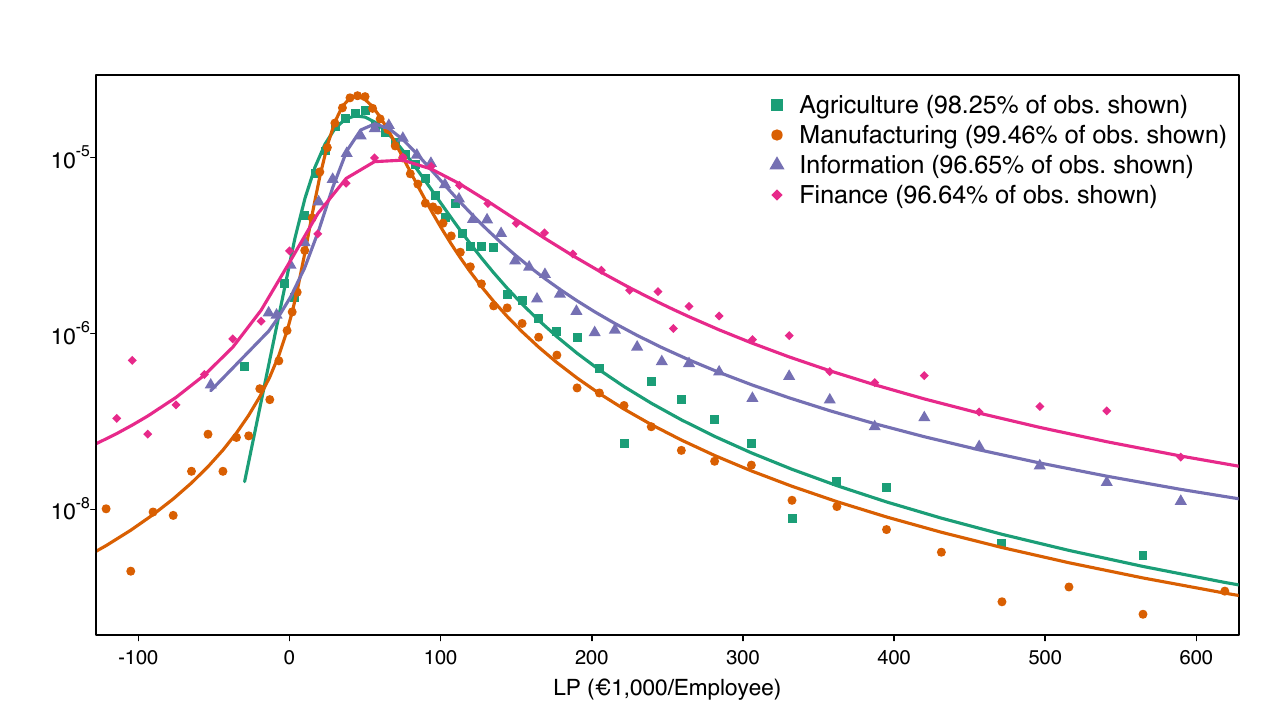}
		\caption{\textbf{Distribution of French labour productivity by industry for 2015 with L\'evy alpha-stable fits.} Solid lines show the fitted L\'evy alpha-stable distributions.}
		           	\label{fig:lp_ctry_ind_year_levy_fit}
	\end{center}
\end{figure}

As representative examples, Fig~\ref{fig:lp_ctry_ind_year_levy_fit} shows the  distributions with the fitted lines for the French sample in 2015, looking at 4 industries. The L\'evy alpha-stable model generally gives a very good fit. In particular, the L\'{e}vy alpha-stable model captures the power-law decay of the tails well. Further, the L\'{e}vy alpha-stable model outperforms the AEP 5 in terms of the log-likelihood for 88\% of the subsamples.

\paragraph{Correlations between metrics: Body vs Tail as two concepts of dispersion}
\input{Tables/intan_correlations.tex}

Table \ref{table:intan_correlations} shows the correlation between these four metrics. It confirms that $\text{IQR}_{90/10}$ and $\gamma/\delta$ capture similar aspects of dispersion (dispersion in the body), and $\text{IQR}_{95/50}$ and $-\alpha$ capture similar aspect of dispersion (dispersion from a fatter tail). The correlation between the body- and tail- measures is fairly high for the quantile-based metrics (0.72), but much lower for the L\'{e}vy-based measures (0.37). This shows that the L\'{e}vy-based measures are potentially better able to capture distinct aspects of dispersion.

We are now ready to address two case studies: productivity dispersion over time, and the correlation between dispersion and intangible intensity.

\subsection{Has productivity dispersion increased?}
\label{sec:results_1}

Time series patterns of productivity dispersion have been a central concern of the current literature on productivity, including reports by the OECD for advanced economies \citep{andrews2016best, berlingieri2017great}, \citet{haldane2017productivity} and \citet{de2022firms} for the UK, \citet{gopinath2017capital} for European economies, and \citet{cette2018firm} for France. Our goal here is to propose alternative measures of dispersion and compare them with the existing literature. 

\begin{figure}[H]
	\begin{center}
    		\includegraphics[width=\textwidth]{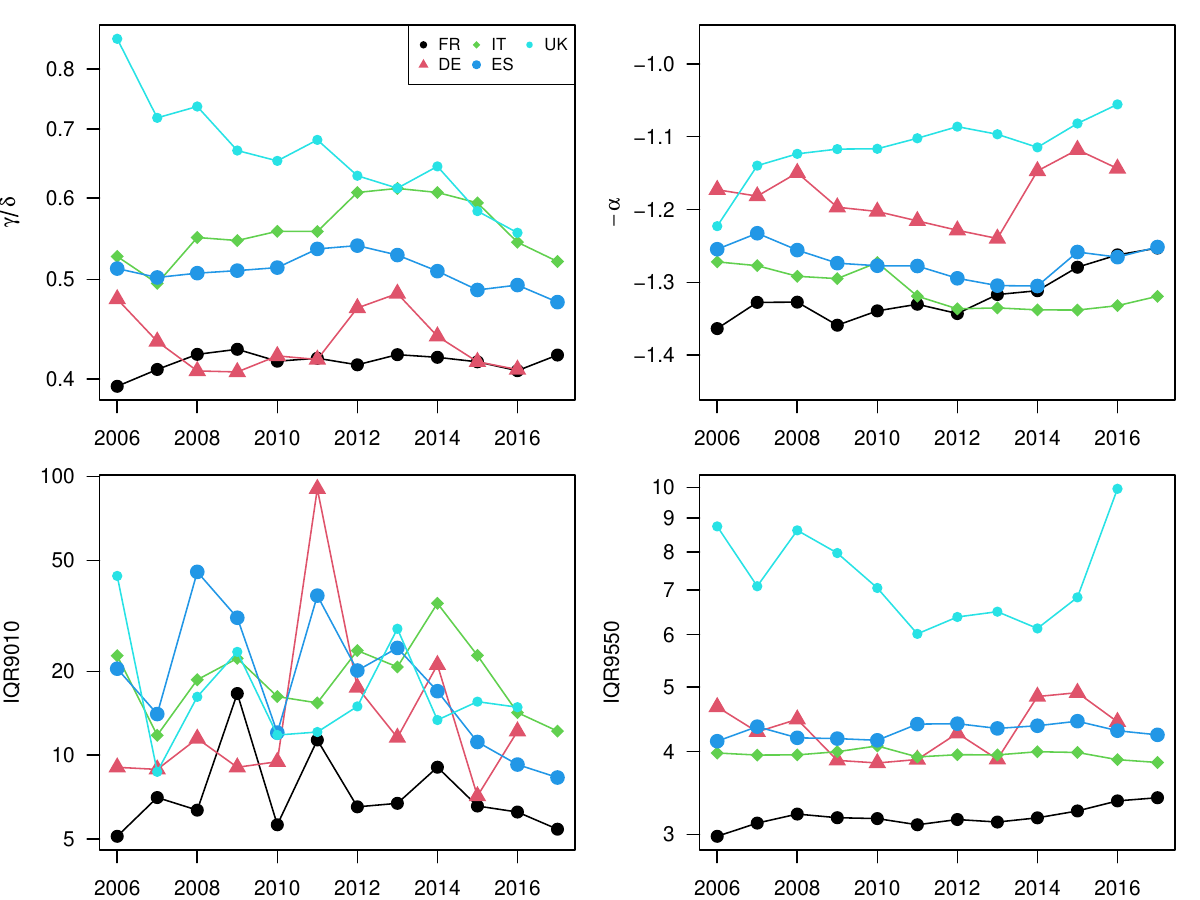}
    	\caption{\textbf{Dispersion metrics parameters for selected countries over time}. Each dot is an (unweighted) average over all industries for which a value is available. We show $-\alpha$ so that all 4 metrics should go up if dispersion increases.} 
    	\label{fig:evolution_dispersion}
    \end{center}
\end{figure}

Fig~\ref{fig:evolution_dispersion} shows the time evolution of the dispersion metrics. For each country-year, we use unweighted averages of the industry-level estimates\footnote{%
This procedure implies that different years will not include the same set of industries, as some industries fail to have the required sample size only in some years, or have a negative $\delta$ or a negative $10^{th}$ percentile only in some years.
}

There is no evidence of a generalized increase in productivity dispersion in our sample. The only country where both body and tail dispersion have increased is France. The time trends of estimated parameters are quite heterogeneous across countries (See also the country-year L\'{e}vy parameters in Fig~\ref{fig:lp_ctry_yr_levy_par} in Appendix~\ref{app:est_para} for all 19 countries).

This is not necessarily at odd with previous work. Some have found that the increase in productivity dispersion took place mostly before 2006 (typically with data starting in 1996), with productivity dispersion only weakly increasing or possibly stagnating after 2006 (\citet[Fig.2]{andrews2016best}, \citet[Figs 3 and 4b]{de2022firms} for the UK); \citet[Fig 4a]{de2022firms} find increasing dispersion mostly after 2006 using Orbis, but note that using administrative data, the patterns are weaker. \citet[Figs 8 and 10]{oliveira2021business} find broadly stable dispersion for the UK and OECD countries. Various studies use different datasets, different subsets (i.e. industries), different concepts (employment weighted or not), cleaning strategies (in particular, include or exclude negative value added),  periods and base years, and ways to present the data.

Putting aside the substantive results on the evolution of dispersion, however, we see that the year-on-year volatility of body dispersion is higher using quantiles rather than L\'{e}vy-based parameters, suggesting an advantage of L\'{e}vy-based metrics at capturing signal over noise. We think that part of the reason for this is the denominator, which is more stable for our L\'{e}vy-based metrics than for $\text{IQR}_{90/10}$. This can be seen by comparing the range of the y-axis in Fig.~\ref{fig:evolution_dispersion} (top-left vs bottom-left), where we see that IQRs have a numerator that is up to 100 times higher than the denominator, while $\gamma/\delta$ is always less than 1. IQRs are also more likely to have a negative denominator, in which case these are simply dropped from our sample, which, aside from the loss of information, may increase volatility. Turning to the comparison of tail dispersion metrics, we find that L\'{e}vy-based and quantile-based metrics have relatively similar tendencies and overall volatility, with perhaps the exception of the UK, where tail behaviour clearly suggests increasing dispersion but $\text{IQR}_{95/50}$ are volatile without clear tendency.

\subsection{Productivity dispersion and intangible capital intensity}
\label{sec:intangibles}

A prominent hypothesis in contemporary macroeconomics is that market concentration, markups and productivity dispersion have increased, and that this may be partly due to the rise of intangible capital (see e.g. \citet{goldin2021productivity} or \citet{de2022firms} for a review). Intangible capital - software and databases, knowledge capital developed from investment in R\&D, branding, and employee skills - is typically a large fixed cost that also makes marginal cost lower. This results in an increase in economies of scale, which makes firms able to invest even more productive, diverging from the rest, and able to charge higher markups and capture more market shares.
A recent paper \citep{corrado2021new} found that industries with high intangible capital per worker also tend to have a higher dispersion of productivity. Here we revisit this finding to compare traditional parametric and non-parametric measures of dispersion.

We gather data on industry-level intangibles from the latest version of EU-KLEMS. Our measure of intangible intensity is a volume measure of capital services per hours worked, for each letter-level industry in 19 countries, between 2006 and 2017\footnote{
See \url{https://euklems-intanprod-llee.luiss.it}. We use the February 2022 release of EU KLEMS 2021, downloaded in March 2022 \citep{Bontadini2021}. We divide the variable \texttt{Kq\_Intang} (Total Capital stock intangibles, volume 2015 ref.prices, NAC mn) by the variable \texttt{H\_EMP} (from the ``statistical'' national accounts). We obtained similar results when using GFCF rather than capital stocks.
}.

Since we are interested only in correlations (rather than causation), we regress intangible intensity on measures of dispersion rather than the other way around as \citet{corrado2021new} do. This has several advantages: first, we can compare the goodness of fit obtained using different metrics: putting intangible intensity on the left-hand side ensures that the $R^2$ and the standard error of the regression are comparable. Second, we can see whether including two separate metrics in the same regression provides additional explanatory power. And third, because $\alpha$ is bounded between 0 and 2, this avoids having to consider an appropriate estimator for bounded dependent variables.

\input{Tables/intan_reg_iqr.tex}

\input{Tables/intan_reg_levy.tex}

We broadly follow \citet{corrado2021new}: we include a control for the central values of the distribution, we provide a specification with country-by-industry and time-by-industry dummies, and we provide conservative standard errors, clustered at the industry-country level.

Tables \ref{table:intan_reg_iqr} and \ref{table:intan_reg_levy} show the results. The first and perhaps most important remark is that the number of observations is generally higher when we use the L\'{e}vy-based metrics, particularly compared to regressions that use the $10^{th}$ quantile, because the $10^{th}$ quantile is negative in more than 10\% of the cases\footnote{
\citet{campbell2019measuring} note that in Australian administrative data, ``At the 4-digit ANZSIC level, (...) only four industries (around 1.5 per cent of all 4-digit industries) [are] excluded for the [Inter-Quartile Range] analysis but 139 industries (around 51 per cent) [are] excluded for the 90-10 differential analysis. Nearly all 4-digit industries in Retail Trade and Wholesale Trade have a negative value added in the $10^{th}$ percentile.''
}.

Considering regressions that do not include fixed effects, we find similar results using parametric and non parametric measures. Taken independently, both body and tail metrics have the expected sign and are significant, in both the parametric and the non parametric cases. The estimated parameters suggest an important economic effect. The elasticity of intangible capital to the 90/10 IQR is only slightly weaker than in \citet[Table 3]{corrado2021new} (0.16-0.24). The elasticity for the tail dispersion IQR is more than three times higher. The elasticity for $\gamma/\delta$, close to 0.75, is substantial. For the tail parameter $\alpha$, an increase of $-\alpha$ by 0.1 (e.g. a fattening of the tail from say 1.4 to 1.3, roughly like the evolution observed for France in Fig. \ref{fig:evolution_dispersion}) is associated with an increase in intangible intensity of around 10\%. 

The goodness of fit is similar across the IQR and L\'{e}vy tables, because they are driven by the large explanatory power of the statistics for central location (median, or $\delta$). There is a sensibly higher $R^2$ for the L\'{e}vy variables when body and tail dispersion are included together.

\input{Tables/intan_lasso.tex}

When including fixed effects (the last three columns of each Table), in all cases the coefficients drop to levels where they are no longer significant at the 5\% level. Including all these dummies leads to a loss of statistical power (degrees of freedom fall from roughly 2800 to 2300), and clear overfitting ($R^2>0.99$). As a final check, we repeat the regression of the last column of Tables \ref{table:intan_reg_iqr} and \ref{table:intan_reg_levy}, but using a standard penalized likelihood method. Specifically, we use a standard Lasso procedure with 10-fold cross validation as implemented in \citet{friedman2017}. Table \ref{table:intan_lasso} shows that once insignificant dummies are selected out, the coefficients are zero or very close to zero for the IQR metrics, but of the expected sign for the L\'{e}vy variables, although substantially smaller than when no fixed effects are included at all. If we attempt to regress intangible intensity on all metrics of dispersion, we find inconclusive evidence, which we interpret as showing that the correlations among the regressors lead to unstable results.

Overall, these results indicate that our metrics provide similar or slightly superior results than traditional non-parametric measures. To be sure, we are confident that our L\'{e}vy-based dispersion metrics are superior \emph{in principle}: they make it possible to keep negative values, avoid having to select specific quantiles, and while they may be slightly more difficult to interpret than quantile-based measures, they provide a link to the large body of literature in economics and finance that has shown the relevance of heavy tails and the importance of estimating precisely the tail exponent. However, we also acknowledge that in specific applications such as the one explored here, a number of methodological choices\footnote{
For instance, using investment instead of capital, changing the threshold for the minimum sample size, changing the estimators for the L\'{e}vy-based parameters, changing the quantiles of the quantile-based metrics, and adding or removing controls.
} could affect the conclusions regarding which metric provides a better fit.

\section{Discussion and conclusion}
\label{sec:conclusion}

The distributions of firm-level VA per worker have extremely large support, are asymmetric, and are heavy-tailed. A major consequence of this is that measuring dispersion is not straightforward: standard deviations are poor metrics because second moments do not exist, and log transformations are not recommended due to the proportion of the negative values.

We propose the L\'{e}vy alpha-stable distribution as a sensible distributional model for labour productivity, motivated by empirical evidence of an infinite variance, and by the fact that the L\'{e}vy distribution should be fairly common, as it emerges from the generalized Central Limit Theorem. We show that the L\'{e}vy distribution provides a better fit to the data than its main competitor, the five-parameter AEP distribution.

Good distributional models make it possible to offer a richer picture of dispersion. While the scale parameter captures the overall width of the distribution, the tail parameter captures the occurrence of the extreme events. These are qualitatively distinct aspects of dispersion. While existing research does attempt to distinguish between body and tail dispersion using quantile ratios, we argue that parametric measures make this distinction clearer and more objective, as there is no need to log-transform the data or to choose specific quantiles.

We have provided two case studies, comparing our parametric measures with non parametric measures of dispersion. We find that they give similar results: a change in dispersion (2006-2017) that differs across countries, and a positive but noisy correlation between intangible intensity and dispersion. While the results are similar, our parametric measures limit the need to discard data when measuring body dispersion, and limit the need to make arbitrary choices of quantiles.

To go further, the research on productivity dispersion needs administrative data, which typically cannot be made available. A limitation of our results is that country-industry-year samples, even at the 2-digit level, remain fairly small so it is difficult not only to estimate the parameters of the L\'{e}vy distribution, but also to test the quality of the fit. Our findings have implications for statistical offices releasing moments of micro-data. Because the L\'{e}vy alpha-stable distribution can be estimated using 5 quantiles (more conveniently the $5^{th},25^{th},50^{th},75^{th}$ and $95^{th}$), if statistical offices release more than these 5 quantiles, then the L\'{e}vy alpha-stable can be estimated on 5 quantiles and the fit can tested on the other quantiles. If the fit is good, researchers can then compute any other statistic about the distribution almost as if they had access to the raw data.

Yet, our paper establishes a number of facts that are useful for the burgeoning literature on misallocation and productivity dispersion, suggesting that models should be able to reproduce L\'{e}vy-distributed VA per worker, with tail parameters $\alpha \in [1,1.5]$ (although possibly with a broader range when considering country-industry-year samples), substantially positive skewness ($\beta \in [0.5,1]$), and a ``body'' dispersion $\gamma/\delta \in [0.2,1]$.

\small
\bibliography{main}
\bibliographystyle{agsm}
\normalsize

\pagebreak
\appendix

\section{Data appendix}
\label{sec:data_appendix}

\subsection{Raw data} 
	
We use the Orbis Europe firm level database, which is part of the Orbis data provided by Bureau van Dijk. The database encompasses more than 21 million firms of all sizes from 44 different European countries. Around 23 million firm-year observations for 7 million unique firms in 19 countries are used in the present analysis; see Table~\ref{tab:obs_ctry_yr}. The variables used are listed in Table~\ref{tab:variables}.

\begin{table}[H]
\centering
\begin{tabular}{l l p{10cm}}
 \hline
Orbis Code & Notation & Description \\ 
 \hline
IDNR&  & Firm's identification number\\
EBITDA& $\pi$ & Nominal Earnings Before Interest, Taxes, Depreciation \& Amortization \\
STAF& $\omega$ & Nominal Wages (staff costs)\\
EMPL& $L$ & Employment \\
NACE\_PRIM\_CODE&  & Industrial classification code (NACE Rev. 2) \\
CLOSDATE\_year&  & Year\\
 \hline
\end{tabular}
\caption{Variables from Orbis Europe used for the analysis. See Eq. \ref{eq:VA} for the construction of VA and Labour productivity.} 
\label{tab:variables}
\end{table}

\subsection{Data cleaning} 
\label{app:additionalcleaning}
	
The first step in processing the data is to ensure that no observations have missing values for their reporting IDNR or year. We only use unconsolidated in order to avoid double counting. Since some accounts provide a default year for various reasons, observations where the reported year is not between 2006 and 2017 are removed. Negative total asset, fixed asset, sales, wages, and employment observations are regarded as missing.
	
Furthermore, we remove firm-year observations where employment or value added are missing. In addition, we discard country-year samples without at least 10,000 observations for five years or more. This reduces the list of countries considered from 44 to 19 countries (See Table \ref{tab:obs_ctry_yr}). \footnote{
For 25 countries, there was insufficient data: Albania, Austria, Belarus,  Bosnia and Herzegovina,  Cyprus,  Greece, Iceland, Ireland, Kosovo, Latvia, Liechtenstein, Lithuania, Luxembourg, Malta, Monaco, Montenegro, Netherlands, Norway, North Macedonia,  Moldova, Russian Federation, Serbia, Switzerland, Turkey, Ukraine.
}
	
\subsection{Deflation} 
	
The KLEMS database\footnote{
\url{http://www.euklems.eu}. This is version that was available when we started the project, so it differs from the EU-KLEMS data used in Section \ref{sec:intangibles}.
} provides the most comprehensive data on deflation for the countries and industries covered by the Orbis Europe sample. In particular, tables \verb|Statistical_National-Accounts.rds| and \verb|Statistical_Capital.rds| provide value added and gross output for two-digit NACE Rev.2 industries for the countries in the Orbis Europe sample. Depending on its location, industry, and year of reporting, we deflate firms' value added using the value added deflator

% latex table generated in R 4.1.1 by xtable 1.8-4 package
% Fri Jan 21 21:55:29 2022
\begin{table}[H]
	\scriptsize
    \begin{center}
\begin{tabular}{|c|llllllllllll|}
  \hline
 & 2006 & 2007 & 2008 & 2009 & 2010 & 2011 & 2012 & 2013 & 2014 & 2015 & 2016 & 2017 \\ 
  \hline
Belgium &  24,823 &  88,877 &  94,373 &  94,899 &  95,631 &  96,810 &  97,357 &  97,093 &  97,324 &  97,629 &  96,909 &  60,423 \\ 
  Bulgaria &       0 &  12,231 &  38,205 &  38,136 &  39,190 &  81,215 &  93,304 &  94,817 &  95,577 &  99,594 &  99,453 & 102,942 \\ 
  Croatia &  16,451 &  38,875 &  39,957 &  42,387 &  40,536 &  39,360 &  40,340 &  40,472 &  41,208 &  42,434 &  44,062 &       0 \\ 
	\makecell{Czech \\Republic} &  24,328 &  52,697 &  46,049 &  70,294 &  73,284 &  77,457 &  76,924 &  78,795 &  75,475 &  77,139 &  60,192 &  40,554 \\ 
  Denmark &       0 &       0 &       0 &       0 &       0 &       0 &  10,153 &  13,116 &  12,649 &  14,232 &  41,054 &  45,520 \\ 
  Estonia &       0 &  14,505 &  16,095 &  13,939 &  15,932 &  16,909 &  18,088 &  19,017 &  20,185 &  20,511 &  20,926 &  19,876 \\ 
  Finland &  14,585 &  31,271 &  30,235 &  32,541 &  32,344 &  38,403 &  38,749 &  39,827 &  40,825 &  37,529 &  35,797 &  36,760 \\ 
  France & 188,682 & 248,892 & 235,860 & 249,724 & 259,352 & 233,105 & 199,456 & 207,906 & 207,671 & 155,439 & 128,023 &  82,804 \\ 
  Germany &  19,537 &  36,256 &  37,810 &  38,440 &  39,498 &  42,164 &  50,222 &  73,045 &  40,261 &  37,871 &  33,155 &       0 \\ 
  Hungary &       0 &  65,589 &  16,220 & 109,059 &  91,033 &  88,773 & 120,705 & 121,833 & 129,449 & 132,779 & 133,472 & 129,107 \\ 
  Italy &  80,270 & 205,580 & 300,766 & 255,141 & 201,252 & 406,372 & 425,845 & 418,763 & 422,876 & 437,585 & 435,280 & 380,695 \\ 
  Poland &  13,548 &  25,968 &  31,798 &  54,341 &  18,129 &  14,024 &  15,311 &  13,514 &       0 &       0 &       0 &  25,350 \\ 
  Portugal &  97,015 & 202,418 & 208,151 & 201,864 & 197,862 & 192,167 & 187,614 & 184,607 & 187,208 & 191,268 & 193,563 & 193,080 \\ 
  Romania &       0 & 167,761 & 175,666 & 161,536 & 162,342 & 173,394 & 179,953 & 182,637 & 178,535 & 183,662 & 191,385 &       0 \\ 
  Slovakia &       0 &  16,158 &  22,700 &  33,137 &  32,910 &  35,012 &  33,821 &  39,242 &  43,622 &  34,241 &  34,956 &       0 \\ 
  Slovenia &       0 &       0 &       0 &   9,659 &  29,250 &  29,453 &  27,599 &  27,442 &  28,932 &  29,746 &  30,959 &  31,813 \\ 
  Spain & 270,505 & 424,877 & 468,998 & 461,257 & 436,631 & 424,314 & 404,536 & 390,900 & 392,478 & 403,412 & 405,003 & 345,054 \\ 
  Sweden &  18,283 &  83,593 & 100,106 & 102,341 & 104,274 & 106,361 & 107,975 & 108,888 & 110,179 & 112,690 & 114,515 & 100,681 \\ 
  \makecell{United \\ Kingdom}&  16,128 &  49,873 &  53,686 &  51,906 &  50,259 &  49,457 &  49,832 &  51,090 &  52,126 &  52,793 &  52,818 &       0 \\ 
   \hline
   \end{tabular}
	\end{center}
\caption{Number of observations per country-year (after cleaning)}. 
\label{tab:obs_ctry_yr}
\end{table}

\subsection{Industry classification} 
\label{app:nace2_cat}

Each firm reported in Orbis Europe includes its four-digit NACE Rev. 2 industrial classification. We further aggregate the industry classes using the highest aggregation level in the original NACE Rev. 2 industrial classification. This yields 19 industries. Table \ref{tab:nace2_key} summarizes the industry category and descriptions. 
	
    	% latex table generated in R 3.4.1 by xtable 1.8-2 package
        % Fri Jan 04 13:52:26 2019
        \begin{table}[H]
        \centering
        \scriptsize
        \begin{tabular}{cl}
          \hline
       Broad Category &  Original NACE Rev. 2 Category \\ 
          \hline
       \multirow{2}{*}{Agriculture }  & Agriculture, forestry and fishing (Agr) \\ 
           & Mining and quarrying (Mine) \\ 
          \midrule
          \multirow{1}{*}{Manufacturing }  &  Manufacturing (Manu) \\ 
          \midrule
          \multirow{2}{*}{Energy } & Electricity, gas, steam and air conditioning supply (Elec)  \\ 
      & Water supply; sewerage, waste management and remediation activities (Water) \\ 
          \midrule
          \multirow{1}{*}{Construction } & Construction (Cons) \\ 
          \midrule
          \multirow{7}{*}{ \makecell{Non-Financial \\ Service }} & Wholesale and retail trade; repair of motor vehicles and motorcycles (Whole) \\ 
         & Transportation and storage (Trans) \\ 
     & Accommodation and food service activities (Accom)\\ 
        & Administrative and support service activities (Adm-S) \\ 
        & Arts, entertainment and recreation (Art) \\ 
       & Other service activities (O-Serv) \\ 
          \midrule
          \multirow{1}{*}{Information} &  Information and communication (Info) \\ 
          \midrule
          \multirow{2}{*}{\makecell{Finance, Insurance, \\Real Estate }} & Financial and insurance activities (F\&I) \\ 
     & Real estate activities (Real)\\ 
          & Professional, scientific and technical activities (Prof-S) \\ 
          \midrule
           \multirow{3}{*}{\makecell{Non-Market \\ Service }} & Public administration and defence; compulsory social security (Pub-S) \\ 
         & Education (Edu) \\ 
     &  Human health and social work activities (Health) \\
         \bottomrule
        \end{tabular}
                \caption{Industry Category Codes and Descriptions} 

        \label{tab:nace2_key}
        \end{table}

\section{Fitting procedure}

\subsection{Fitting methods}
\label{sec:fittingmethod}

\citet[Chap. 4]{NOLAN2020} provides a recent review and evaluation of the various methods for estimating the parameters of stable distributions. For asymmetric distributions, there are three main contenders: maximum likelihood (MLE), the quantile-based estimator of \citet{McCulloch86}, and a method based on empirical characteristic functions. Nolan evaluates these methods for 10 different points in the parameter space. The most relevant to us is likely to be his case 8 ($\alpha=1.5$, $\beta=0.5$). In this case and for sample sizes of 100, the MLE is the best method, followed by the empirical characteristic function, and the quantile method. As can be seen in his Fig. 4.22 p.196, while the differences in performance are not very high in general, they are fairly substantial for $\alpha$, with a RMSE of 0.156 for the MLE and 0.227 for the quantile estimator. This means that a roughly constructed 95\% interval ($\alpha \pm 2 \text{ RMSE}$), which assumes unbiasedness as suggested by Nolan's Fig.4.22, would be [1.19-1.81] for the MLE, and [1.05-1.95] for the quantile estimator, so the difference is noticeable.

Unfortunately, the MLE is computationally highly prohibitive even for moderate sample sizes, so we want to use it only when absolutely necessary. To decide on the minimal sample size needed for using the quantile estimator, we run a Monte Carlo simulation as follows. We set $\alpha=1.3, \beta=0.75, \gamma=10, \delta=20$, and we investigate sample sizes $N=20, 50, 100, 200, 500$. For each sample size we draw 200 samples and estimate the parameters using the MLE and the quantiles method. We then compute the mean and the $10^{th}$ and $90^{th}$ quantile of these 200 observations. We discard non convergent MLE results, but we keep ``corner'' solutions from the quantile estimator ($\alpha=0.5$, which is the lower bound for this estimator \citep[p.168]{NOLAN2020}).

  \begin{figure}[H]
	\begin{center}
		\includegraphics[width=0.8\textwidth]{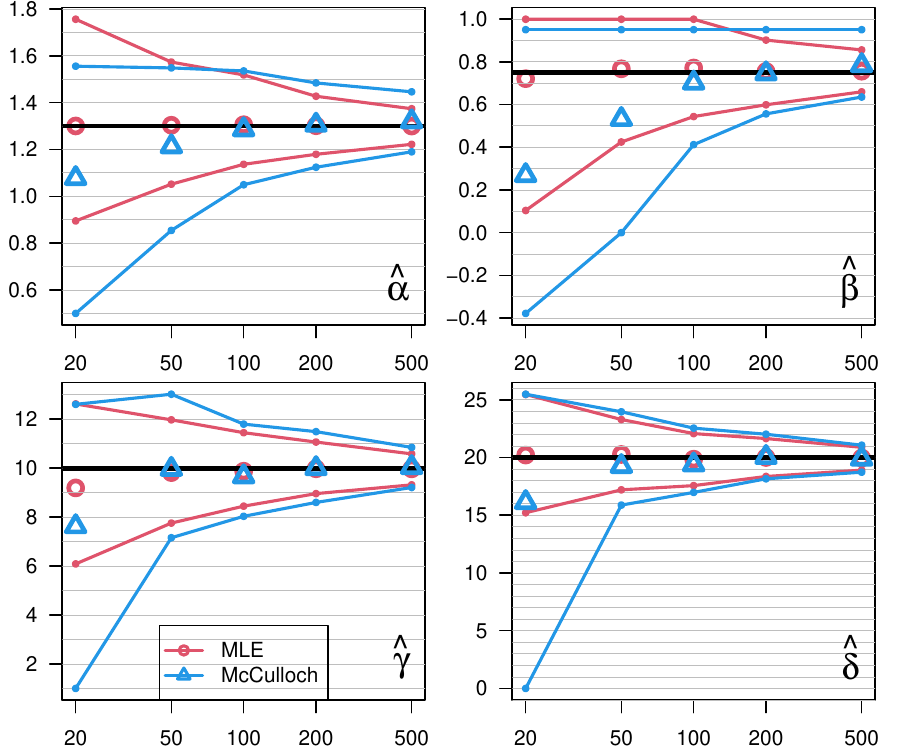}
	\end{center}
	\caption{\textbf{Performance of the MLE and Quantile estimators in small samples.} For each sample size, we sample 200 sets of i.i.d. values. The black lines show the true values of the parameters. The dots show the average of the estimated parameters over the 200 repetitions, and the lines show the $10^{th}$ and $90^{th}$ percentiles.}
	\label{fig:montecarlo}
  \end{figure}

Fig.~\ref{fig:montecarlo} shows the results. As expected, the MLE generally outperforms the quantile estimator. A important point is that the quantile estimator gives \emph{biased} estimates of all parameters for small sample sizes. However, the MLE remains approximately unbiased even for small samples. While the estimates are quite imprecise, it is still clear that valuable information can be extracted from a small sample, particularly if using the MLE on samples of size greater than 50. We choose $N>50$ as our criteria for the minimal sample size in the paper. When $N$ reaches 200, both estimators appear unbiased, and the precision advantage of the MLE is less obvious (at least in absolute value, note that the y-axes are not in log scale). Given our computational constraints, we set $N=200$ as the minimum sample size for using the quantile estimator rather than the MLE.

  \begin{figure}[H]
	\begin{center}
		\includegraphics[width=0.9\textwidth]{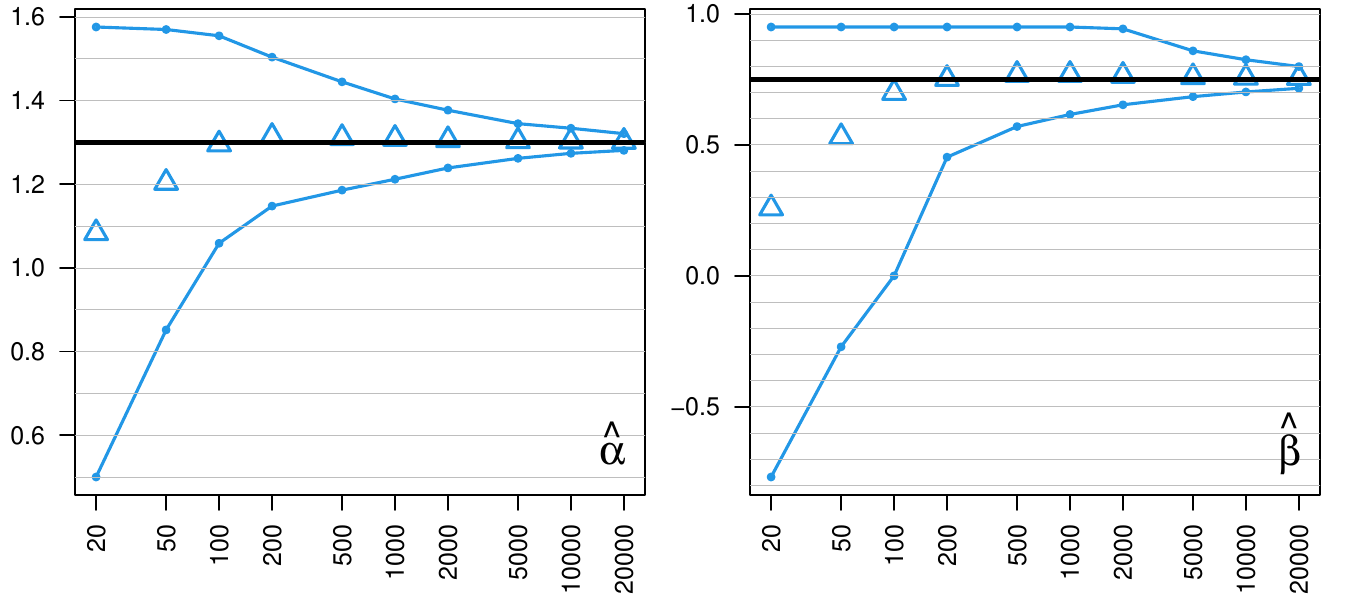}
	\end{center}
	\caption{\textbf{Performance of the Quantile estimators in moderate sample sizes.}. For each sample size, we sample 10,000 sets of i.i.d. values. The black lines shows the true values of the parameters. The dots show the average of the estimated parameters over the 200 repetitions, and the lines show the $2.5^{th}$ and $97.5^{th}$ percentiles.}
	\label{fig:montecarlo2}
  \end{figure}

One remaining concern with the quantile estimator is the estimate of $\beta$. The implementation we use reports a maximum $\hat \beta=0.95$ rather than $\hat \beta=1$, and this maximum is still reached quite often for moderate sample sizes $N=200,500$ when $\beta=0.75$. Similarly, the minimum value that the estimator of $\alpha$ can return is 0.5, so the value 0.5 for the $10^{th}$ percentile at $N=20$ is not very informative. To investigate this issue we run another set of simulations, using the same values of the parameters but 10,000 replications, and sample sizes up to 20,000. We also change the reported percentiles from [0.1,0.9] to [0.025,0.0975]. Fig. \ref{fig:montecarlo2} reports the results for $\hat{\beta}$, which is our main interest, and $\hat{\alpha}$, to provide an idea of the precision of the quantile estimator for this key parameter for large sample size. The 95\% intervals for $\hat{\beta}$ do still include 0.95 when $N=2,000$, but not when $N=5,000$. The Figure also complements Fig. \ref{fig:montecarlo} to make clear the bias and lack of precision for $N<200$.

\subsection{Estimated parameters}
\label{app:est_para}

\begin{figure}[H]
	\begin{center}
    	\begin{minipage}{.8\textwidth}
    		\includegraphics[width=\textwidth]{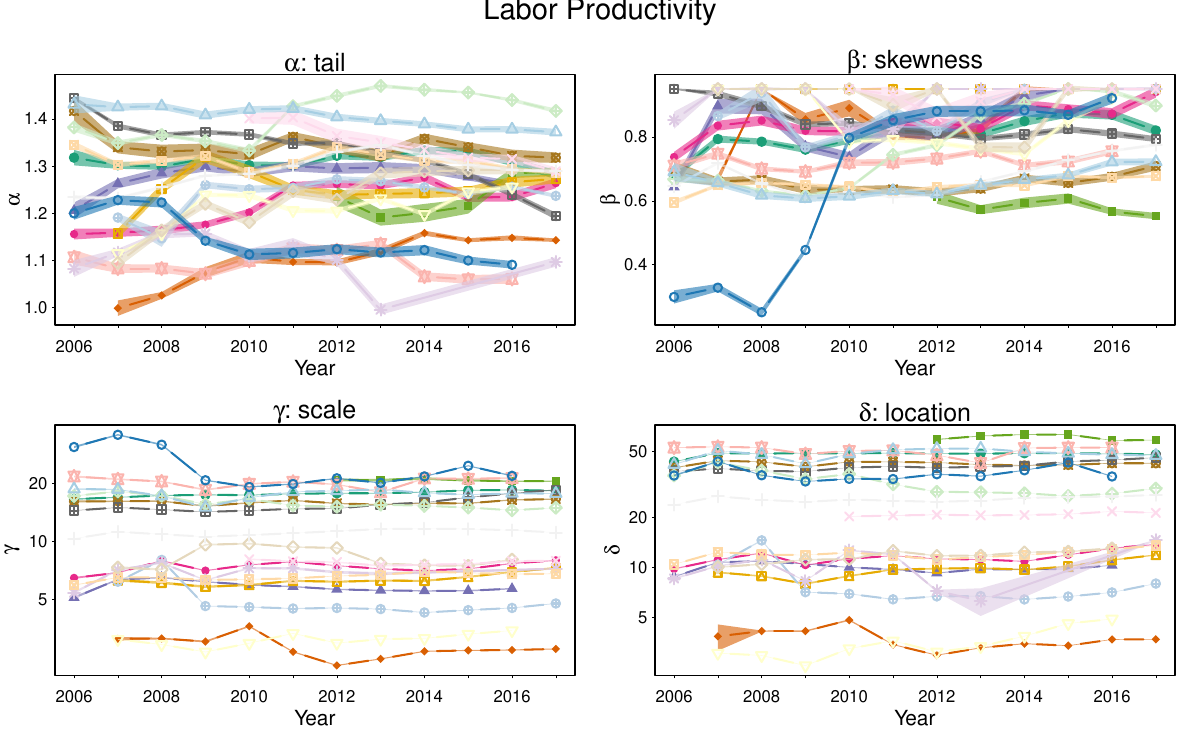}
    	\end{minipage}
    		\begin{minipage}{.9\textwidth}
    		\includegraphics[width=\textwidth]{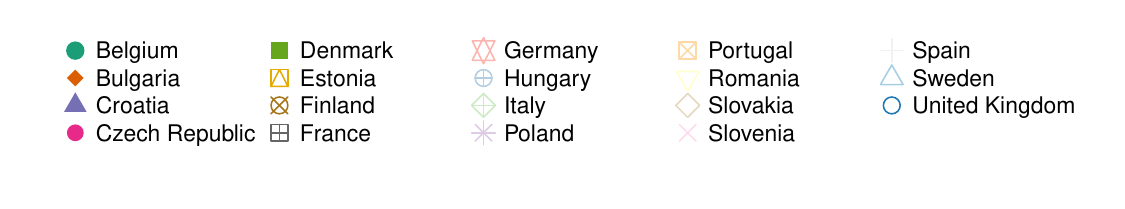}
    	\end{minipage}
    	\caption{\textbf{Estimated parameters for country-year samples.} The four parameters of the fitted L\'evy alpha-stable distributions for labour productivity levels (top two rows) and change (bottom two rows) are plotted by year (2006-2017 for levels, 2007-2017 for change). $\gamma$ and $\delta$ are denominated in \euro{1,000}/employee and in log scale, $\alpha$ and $\beta$ are dimensionless. Note that $\gamma$ and $\delta$ parameters in non-euro countries are converted to Euro using the average exchange rates for the country's sample time period. The exchange rate data is obtained from Eurostat\protect\footnotemark. The shaded area represents the range of $\pm1$ standard errors. The standard errors are calculated by bootstrapping with 1000 replications. Each replication is obtained by sampling from the data \textit{with replacement}. We use R package boot \citep{boot}. } 
    	\label{fig:lp_ctry_yr_levy_par}
    \end{center}

\end{figure}
\footnotetext{\url{https://ec.europa.eu/eurostat/databrowser/view/ert_bil_eur_a/default/table?lang=en}}

\begin{figure}[H]
	\begin{center}
\includegraphics[width=\textwidth]{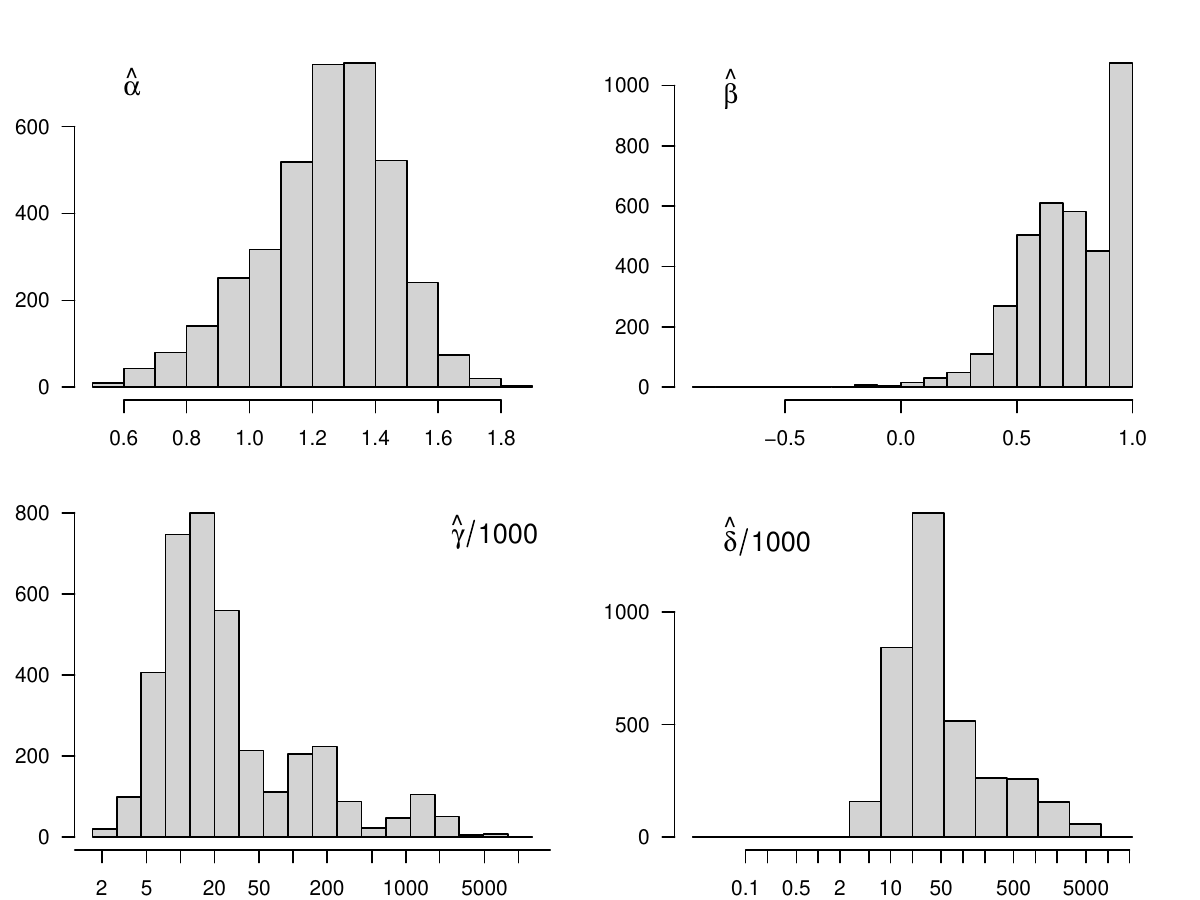}
\caption{Distribution of the estimated parameters for the country-industry-year samples. The parameters are not estimated for sample sizes $N<50$, are estimated by MLE for sample sizes $50 \leq N <200$ and are estimated by the quantile method for samples $N \geq 200$. We removed a few cases where $\hat \delta<0$ or where we had both $\hat \alpha=0.5$ and $\hat \gamma =1$ (a sign of failure of the quantile method).}
\label{fig:hist_country_industry_year}
\end{center}
\end{figure}

\section{Reference model and comparison of the fit}
\label{app:comparisonAEP}

As a reference model, we use two variants of the Asymmetric Exponential Power (AEP) distribution - the 4-parameter AEP by \citet{Delicado/Goria08} and the 5-parameter AEP by \citet{Bottazzi/Secchi11} - that are estimated with two different fitting procedures with radically different characteristics. In the following, we give the functional forms, details of the fitting procedures, and a comparison of the goodness of the results.

\subsection{Functional forms of the AEP models and fitting procedure}

\subsubsection{4-parameter AEP}

The probability density function of the distribution is 
\[
f(x) = \frac{\alpha \kappa}{\sigma (1+\kappa^2)\Gamma\left(\frac{1}{\alpha}\right)} \exp \Big( -\left(\kappa^{\operatorname {sign}(x-\theta)}\left|\frac{x-\theta}{\sigma}\right|\right)^{\alpha} \Big),
\]
where the parameters are location ($\theta$), scale ($\sigma$), skewness ($\alpha$) and tails ($\kappa$). We use the method from \citep{ASQUITH2014955} to estimate these parameters. It used uses L-moments, the 3rd and 4th moments specifically, and will therefore achieve a better representation of the shape of the empirical data in the tails at the expense of a more exact fit in the body. For the computation, we use the R-package \textbf{lmomco} \citep{lmonco}.

\subsubsection{5-parameter AEP}

The probability density function of the 5-parameter AEP distribution is \citep{Bottazzi/Secchi11}

$$
f(x)=
\begin{cases}
  \frac{1}{A} \exp \Big( -\frac{1}{b_l}\left|\frac{x-\theta}{a_l}\right|^{b_l} \Big) \quad\quad\quad \textup{for } x<\theta
  \\
  \frac{1}{A} \exp \Big( -\frac{1}{b_r}\left|\frac{x-\theta}{a_r}\right|^{b_r} \Big) \quad\quad\quad \textup{for } x>\theta
\end{cases}
$$

with 

$$A=a_lb_l^{\frac{1}{b_l}}\Gamma\left(1+\frac{1}{b_l}\right) + a_rb_r^{\frac{1}{b_r}}\Gamma\left(1+\frac{1}{b_r}\right).$$

$\theta$ remains the location parameter; the scale parameter ($\sigma$ in the 4-parameter version) is now separate for the left and right side of the distribution $x<\theta$ and $x>\theta$ as scale parameters $a_l$ and $a_r$. The skewness and tail parameters of the 4-parameter version are rearranged into a left-shape and a right-shape parameter $b_l$ and $b_r$. The distribution recovers the normal distribution $\mathcal{N}(\mu, \sigma)$ for parameter values $\theta=\mu, a_l=a_r=\sigma, b_l=b_r=2$.

The distribution can be fitted with interval-constrained likelihood optimization, which is, however, computationally expensive. We used the Subbotools package \citep{Bottazzi14}. To speed up the computation, we first perform a 3-parameter exponential power\footnote{This distribution is the standard Subbotin distribution and is obtained for $a_l=a_r$ and $b_l=b_r$ for the 5-parameter AEP.} fit and then a 5-parameter AEP fit that is initialized with the estimates from the 3-parameter exponential power fit.

The L-moments have to our knowledge not been derived, therefore the alternative fitting method of L-moments \citep{ASQUITH2014955} is not available in this case.

\subsection{Comparison of the goodness of fit}

Table~\ref{tab:Model_Comparison_main} compares the average log-likelihoods of the L\'{e}vy alpha-stable and the AEP distributions for all country samples.  The log-likelihood of L\'{e}vy alpha-stable model is always higher.

\begin{table}[H]

 	\begin{center}
\begin{tabular}{lrrrr}
  \hline
Country & Obs & AEP4 Log Lik & AEP5 Log Lik & L\'{e}vy Log Lik \\ 
  \hline
Belgium & 1,042,148 & -5.20 & -5.11 & -5.04 \\ 
  Bulgaria & 794,664 & -4.18 & -4.31 & -4.03 \\ 
  Croatia & 426,082 & -6.09 & -6.05 & -5.97 \\ 
  Czech Republic & 753188 & -7.70 & -7.69 & -7.51 \\ 
  Denmark & 136,724 & -7.65 & -7.34 & -7.25 \\ 
  Estonia & 195,983 & -4.17 & -4.05 & -4.02 \\ 
  Finland & 408,866 & -5.10 & -5.02 & -4.91 \\ 
  France & 2,396,914 & -5.04 & -4.95 & -4.86 \\ 
  Germany & 448,259 & -5.81 & -5.46 & -5.36 \\ 
  Hungary & 1,138,019 & -9.67 & -9.50 & -9.45 \\ 
  Italy & 3,970,425 & -5.00 & -4.95 & -4.87 \\ 
  Poland & 211,983 & -5.88 & -5.75 & -5.67 \\ 
  Portugal & 2,236,817 & -4.36 & -4.16 & -4.07 \\ 
  Romania & 1,756,871 & -4.96 & -4.87 & -4.84 \\ 
  Slovakia & 325,799 & -4.51 & -4.41 & -4.35 \\ 
  Slovenia & 244,853 & -4.26 & -4.21 & -4.19 \\ 
  Spain & 4,827,965 & -4.86 &  & -4.61 \\ 
  Sweden & 1,169,886 & -7.30 & -7.29 & -7.20 \\ 
  United Kingdom & 529,968 & -5.72 & -5.41 & -5.32 \\ 
   \hline
\end{tabular}
	\caption{\textbf{Model Comparison: L\'{e}vy alpha-stable vs. AEP distributions}. The table shows the average log-likelihood of the L\'{e}vy alpha-stable model and the AEP models. The log-likelihood of  L\'{e}vy alpha-stable model is generally higher than that of the AEP4 and AEP5 models. Note that some countries have a few observations that 
	have a zero probability for the AEP5 density function.  The probability is not actually zero but is so close to it and is recorded as zero in the numerical software. We exclude those observations when calculating the log likelihood for the AEP5. The number of observations that are excluded from the country sample is the following: Bulgaria (3), Czech Republic (1), Finland (2), France (8), and Sweden (2). The AEP5 result for Spain is omitted because its estimation fails to be completed within a reasonable amount of time.}
\label{tab:Model_Comparison_main}
 	\end{center}
\end{table}

\section{Derivation of the scaling of the sample standard deviation with sample size}
\label{app:scaling}

Here we provide a highly stylized, heuristic derivation for the scaling of the sample standard deviation with sample size in the case of a L\'{e}vy alpha-stable distributed random variable. The key to this phenomenon is that because the theoretical moment is infinite, the larger the sample size, the higher is the chance that an extreme event is drawn. These extreme events are so extreme that they dominate the sum of squares from which the variance is computed. Thus, the larger the sample size, the larger the sample variance.

\cite{sornette2006critical} and \cite{bouchaud2003theory}, for instance, provide more precise statements. Here we expose the argument in the simplest, albeit non rigorous way. We discuss the maximum, but symmetric arguments apply to the minimum.

First, we note that for $N$ large enough, the sample maximum will be dictated by the tail. A key characteristic of the L\'{e}vy alpha-stable distribution is that it has power law tails, that is, for large $x$ \citep{NOLAN2020}, 
\begin{equation}
P(X>x) \sim x^{-\alpha}.
\label{eq:tailprob}
\end{equation}

Now, in a sample of size $N$, we would hardly expect to see an extreme value that has chances of occurring less than $1/N$.
Thus, we may define the ``typical'' value of the maximum as the value $X_{\text{max}}$ such that $1/N=P(X>X_{\text{max}})$. Using Eq. \ref{eq:tailprob},
we have\footnote{
\cite{sornette2006critical} shows that this is the value of the maximum that is not exceeded with probability $1/e\approx 37\%$. Generalizing to the value of the maximum that is not exceeded with probability $p$ implies the condition $\frac{\ln(1/p)}{N}=P(X>X_{\text{max}})$, which does not change the scaling of $X_{\text{max}}$ with $N$. \citet{newman2005power} shows that the expected value of the maximum also scales as $N^{\frac{1}{\alpha}}$.
}
$1/N \sim X_{\text{max}}^{-\alpha}$, and solving for $X_{\text{max}}$ gives
\begin{equation}
    X_{\text{max}} \sim N^{\frac{1}{\alpha}},
\label{eq:Xmax}
\end{equation}
For simplicity, let us assume a mean of zero\footnote{
This is without loss of generality, see (2.8) in \citet{cohen2020heavy}.
}, so that the sample variance is just the average squared value,
\begin{equation}
    \text{Var}(X) \sim \frac{1}{N} \sum^N_{i = 1} X_i^2.
\label{eq:vardef}
\end{equation}

In a L\'{e}vy alpha-stable distribution, the square of the maximum (or minimum, if larger in absolute value) is so large that it dominates the entire sum of squares, such that we may approximate
\begin{equation}
    \sum^N_{i = 1} X_i^2 \approx X_{\text{max}}^2 \sim N^{\frac{2}{\alpha}},
\label{eq:sumequalmax}
\end{equation}
where the last step uses Eq. \ref{eq:Xmax}. Now, inserting Eq. \ref{eq:sumequalmax} into \ref{eq:vardef}, and taking square root, we find that the standard deviation depends on the sample size as 
\[
\sqrt{\text{Var}(X)} \sim N^{\frac{1}{\alpha}-\frac{1}{2}}.
\]
Note that when $\alpha=2$, so that the distribution is Gaussian, the sample standard deviation does \emph{not} increase with sample size, as one expects.

See also the derivations leading to (4.52) in \cite{sornette2006critical} and the heuristic proof based on the rank-size rule in \citet[Proof of Prop. 2]{gabaix2009power}. A rigorous proof
\citep{gabaix2011granular,cohen2020heavy} simply uses the fact that since $X_i$ is (by assumption) regularly varying with index $\alpha$, it follows that $X_i^2$ is regularly varying with index $\alpha/2$ \citep[e.g. Eq.8]{gabaix2009power}, and we can apply the GCLT to the sum of $X_i^2$ to obtain the right scaling. The rigorous proof makes it clear that we expect the sample standard deviation to exhibit substantial variations, since the convergence is in distribution to a L\'{e}vy stable variable.

\section{The scaling of the estimated sample standard deviation with sample size}
\label{app:scaling_example}

\begin{figure}[H]
	\begin{center}
	\begin{minipage}{.49\textwidth}
		\includegraphics[width=\textwidth]{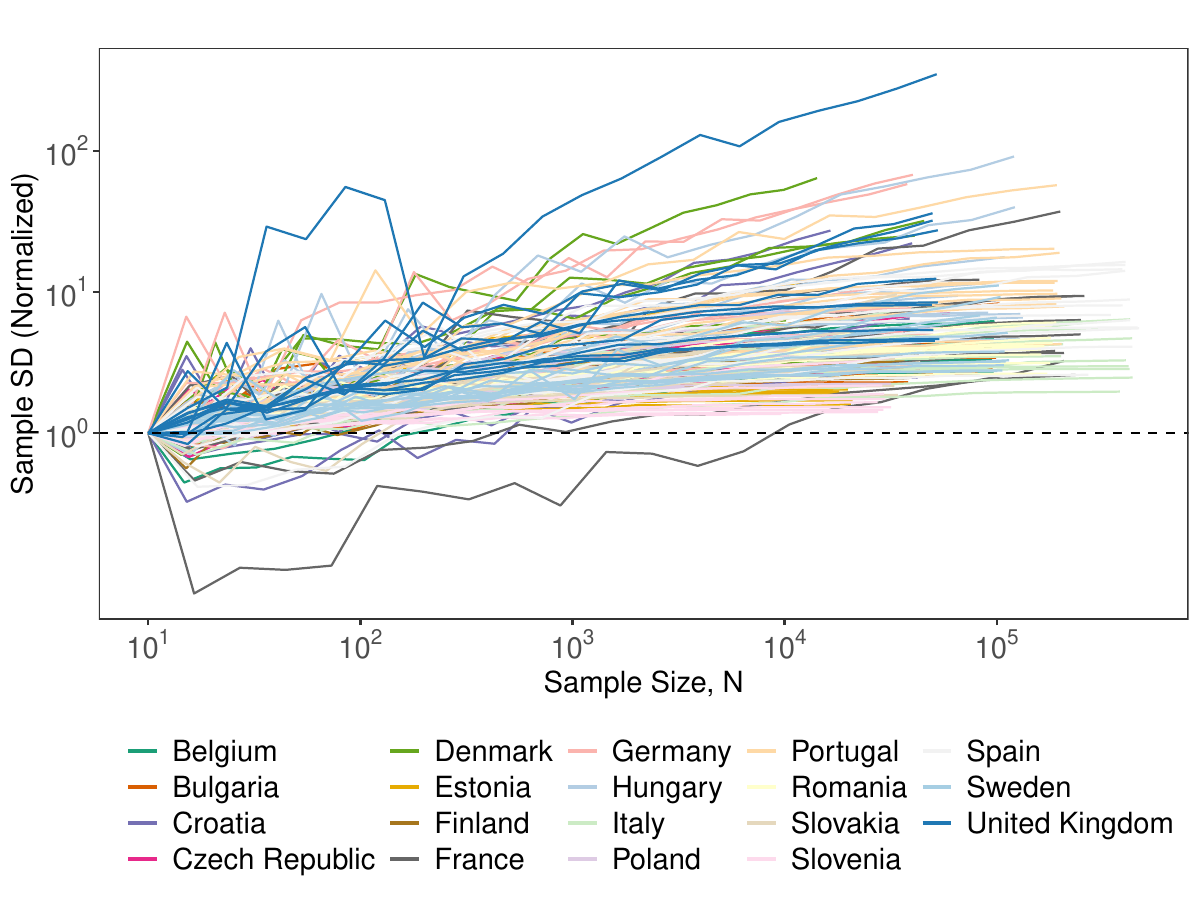}
	\end{minipage}
		\begin{minipage}{.49\textwidth}
		\includegraphics[width=\textwidth]{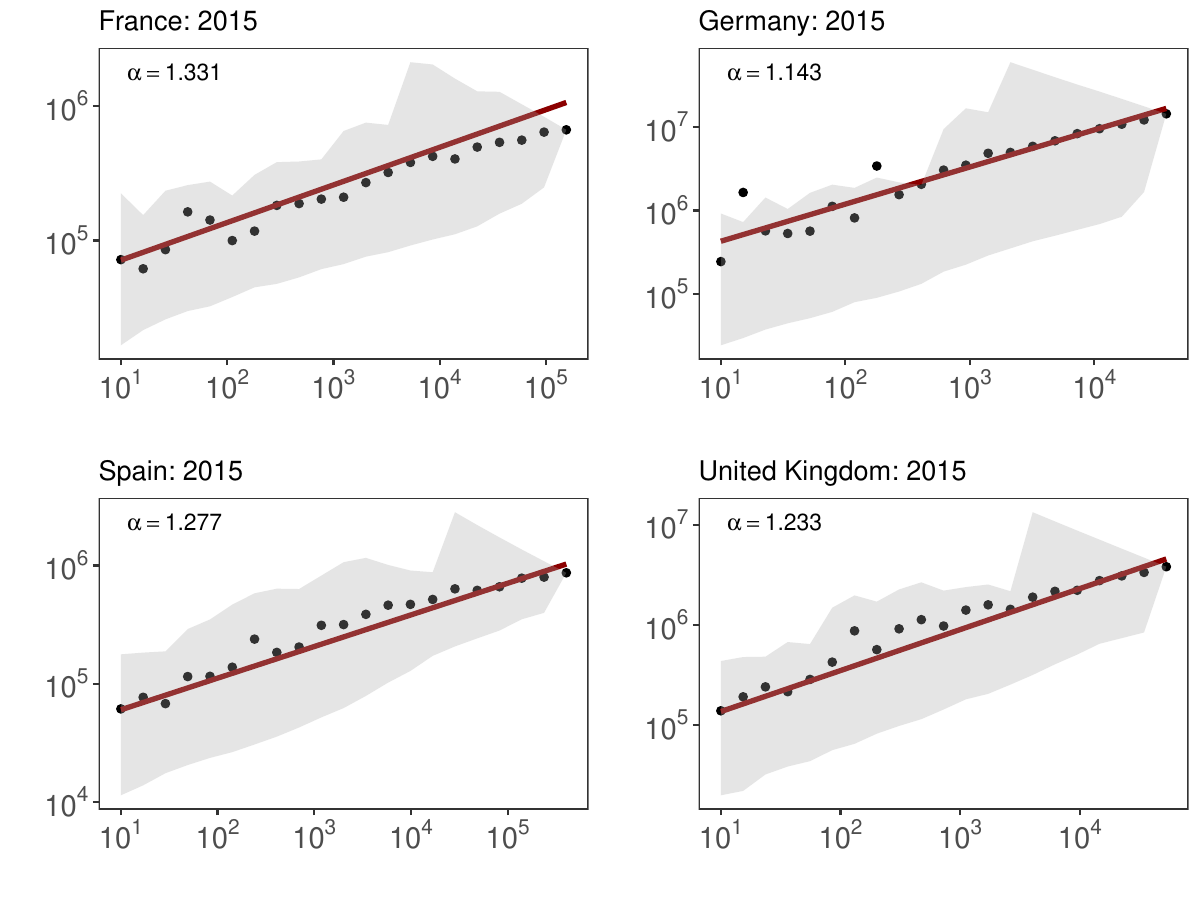}
	\end{minipage}
	\caption{\textbf{Measured standard deviation in sub-samples of firm labour productivity, country-year sample.} We construct a distribution by pooling together the productivity levels of firms in each of 19 countries for each year. Then, for each subsample size $N$, we compute the standard deviation of each of 1,000 subsamples. The plot on the left shows the average of standard deviations (dots) for each of all country-year samples, while the plot on the right shows the same scaling for four selective countries, France, Germany, Spain and the UK, for Year 2015 with the linear scaling line calculated by $N^{\frac{1}{\alpha}-\frac{1}{2}}$ from Appendix~\ref{app:scaling} (we use our estimate of the tail parameter $\alpha$). The shaded area is the $5^{th}$ and $95^{th}$ percentiles of the sample standard deviation. }

		\label{fig:sd_scaling_year}
	\end{center}
  \end{figure}

\begin{figure}[H]
	\begin{center}
	\begin{minipage}{.49\textwidth}
		\includegraphics[width=\textwidth]{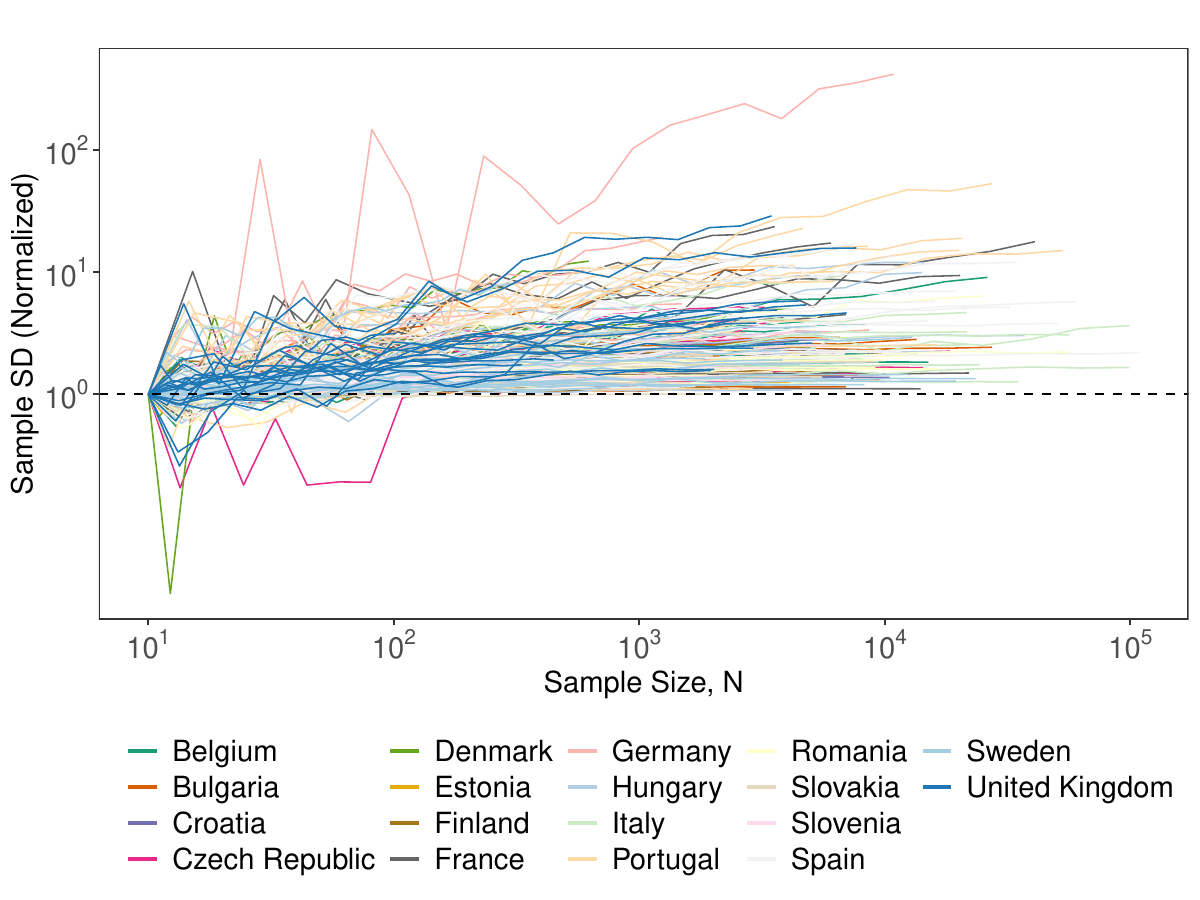}
	\end{minipage}
		\begin{minipage}{.49\textwidth}
		\includegraphics[width=\textwidth]{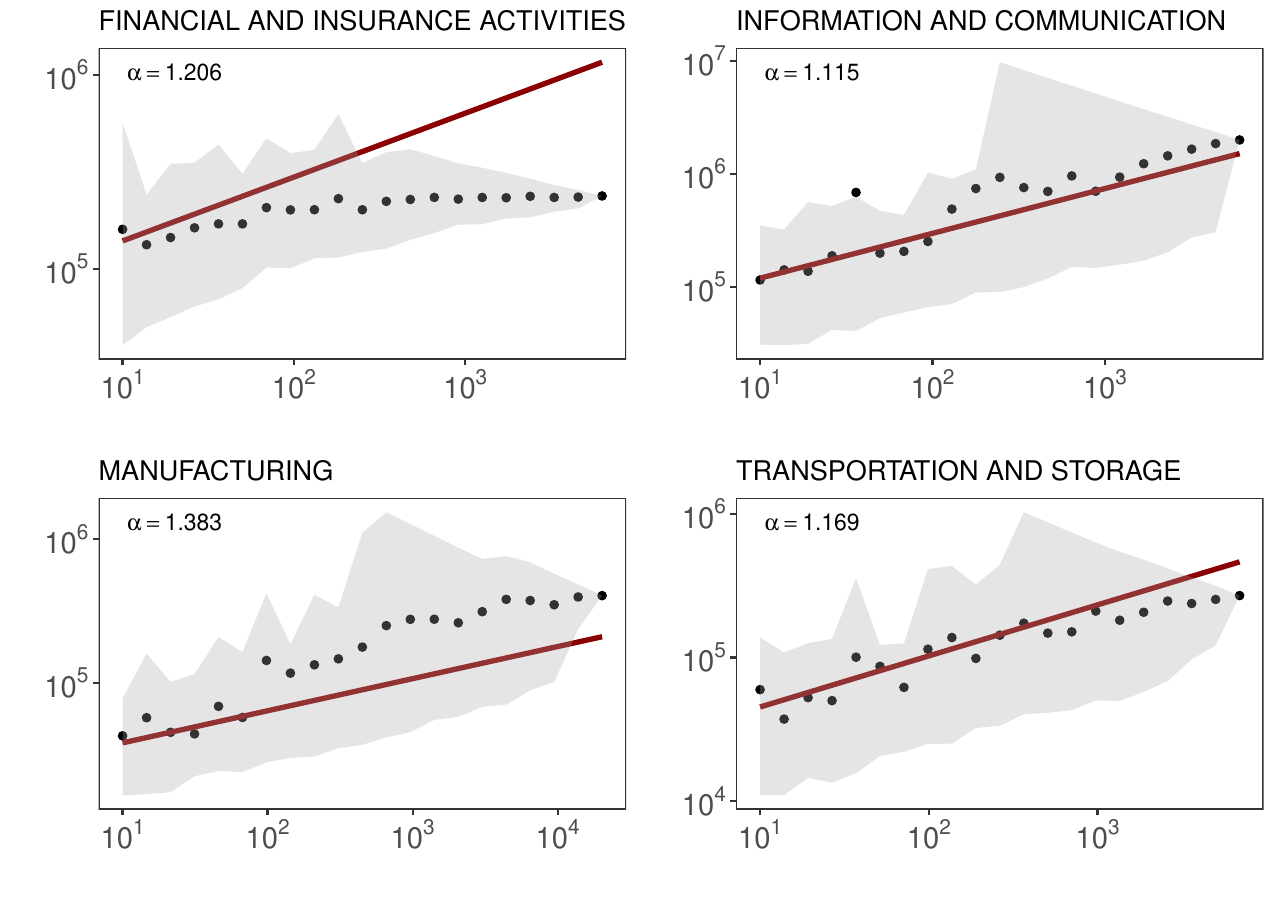}
	\end{minipage}
	\caption{\textbf{Measured standard deviation in sub-samples of firm labour productivity, country-industry sample for year 2015.} We construct a distribution by pooling together the productivity levels of firms in each of 19 countries for each industry. Then, for each subsample size $N$, we compute the standard deviation of each of 1,000 subsamples. The plot on the left shows the average of standard deviations (dots) for each of all country-industry samples at a 2-digit industry level for year 2015, while the plot on the right shows the same scaling for four selective industries for France, Year 2015. Similar pattern holds for all other years.}

		\label{fig:sd_scaling_industry}
	\end{center}
  \end{figure}

\section{Trapani's (2016) procedure for testing for infinite moments}
\label{app:finite_moment_test}

\citet{trapani2016testing} suggests a test for the divergence of arbitrary moments of order $p$, including fractional (non-integer) moments. As the moment of order $p$ may be infinite, it is unknown whether a limiting distribution of the moments exists. The test can therefore not be applied directly. Following a randomized testing approach, the test therefore adds artificial randomness to manipulate the quantity in question, the sample moment of order $p$, to yield a known distribution if the moment is infinite. The resulting test statistic can then be compared to this distribution to obtain a p-value for whether or not the null hypothesis, that the moment of order $p$ is infinite, is correct.

\subsection{Test procedure}
The test statistic is derived such that under the null hypothesis $H_0$, the $p$th moment of the distribution is infinite. For this, the approach starts with the absolute sample moment of order $p$, %$A_p$,
\begin{align*}
    A_p = \frac{1}{n} \sum^n_{k = 1} |X_k|^p.
\end{align*}
To make the resulting test statistic scale-invariant and comparable, the absolute moment must be rescaled,
\begin{align*}
    A_p^\ast = \frac{A_p}{(A_\psi)^{p/\psi}} \times \frac{(A_\psi^\mathcal{N})^{p/\psi}}{A_p^\mathcal{N}},
\end{align*}
with $\psi \in (0,p)$. $A_p^\mathcal{N}$ denotes the $p$th absolute fractional lower order moment of the standard normal distribution, $A_{\psi}^\mathcal{N}$ the $\psi$th absolute fractional lower order moment of the standard normal, etc. Next, an artificial random sample $\xi$ of size $r$\footnote{Trapani suggests $r=n^{0.8}$ where $n$ is the number of observations used to compute $A_p$.} is generated from a standard normal distribution and rescaled,
\begin{align*}
    \varphi_r = \sqrt{e^{A_p^\ast}} \times \xi_r.
\end{align*}
The intuition here is that $\varphi_r$ follows a normal distribution with mean zero and a finite variance, as $n \rightarrow \infty$, if $A_p^\ast$ is finite itself. Thus the problem has been reduced from testing for any moment $p$, to testing the existence of the variance of the transformed random variable $\varphi_r$.

The next step is to generate a sequence $\zeta_r$, given by
\begin{align*}
    \zeta_r(u) \equiv I\left[ \varphi_r \leq u \right],
\end{align*}
where $I[\cdot]$ is an indicator function, and $u \neq 0 $ is any real number. Under $H_0$, $\zeta_r(u)$ will have a Bernoulli distribution with mean $\frac{1}{2}$ and variance $\frac{1}{4}$. This is not the case under the alternative, where $A_p^\ast < \infty$, as $e^{A_p^\ast}$ converges to a finite value.

Values for $u$ are picked from some density, but for simplicity it can be taken from a uniform distribution $U(-1,1)$. As a result, the test statistic of interest is obtained as
\begin{align*}
    \Theta_{r} \equiv \int_{-1}^{1} \frac{1}{2} \vartheta_{r}^2(u) du, \\
    \vartheta_{r}(u) \equiv \frac{2}{\sqrt{r}} \sum^r_{j=1} \left[ \zeta_j(u) - \frac{1}{2} \right].
\end{align*}

Under the null hypothesis that moment $p$ is infinite, $\vartheta_{r}(u)$ should reduce to zero given that $\zeta_j(u)$ has a mean of two. $\Theta_{r}$ is thus shown to follow a $\chi^2$ distribution with $df=1$ if moment $p$ is infinite. 

\begin{figure}[H]
  \begin{center}
	\includegraphics[width=\textwidth]{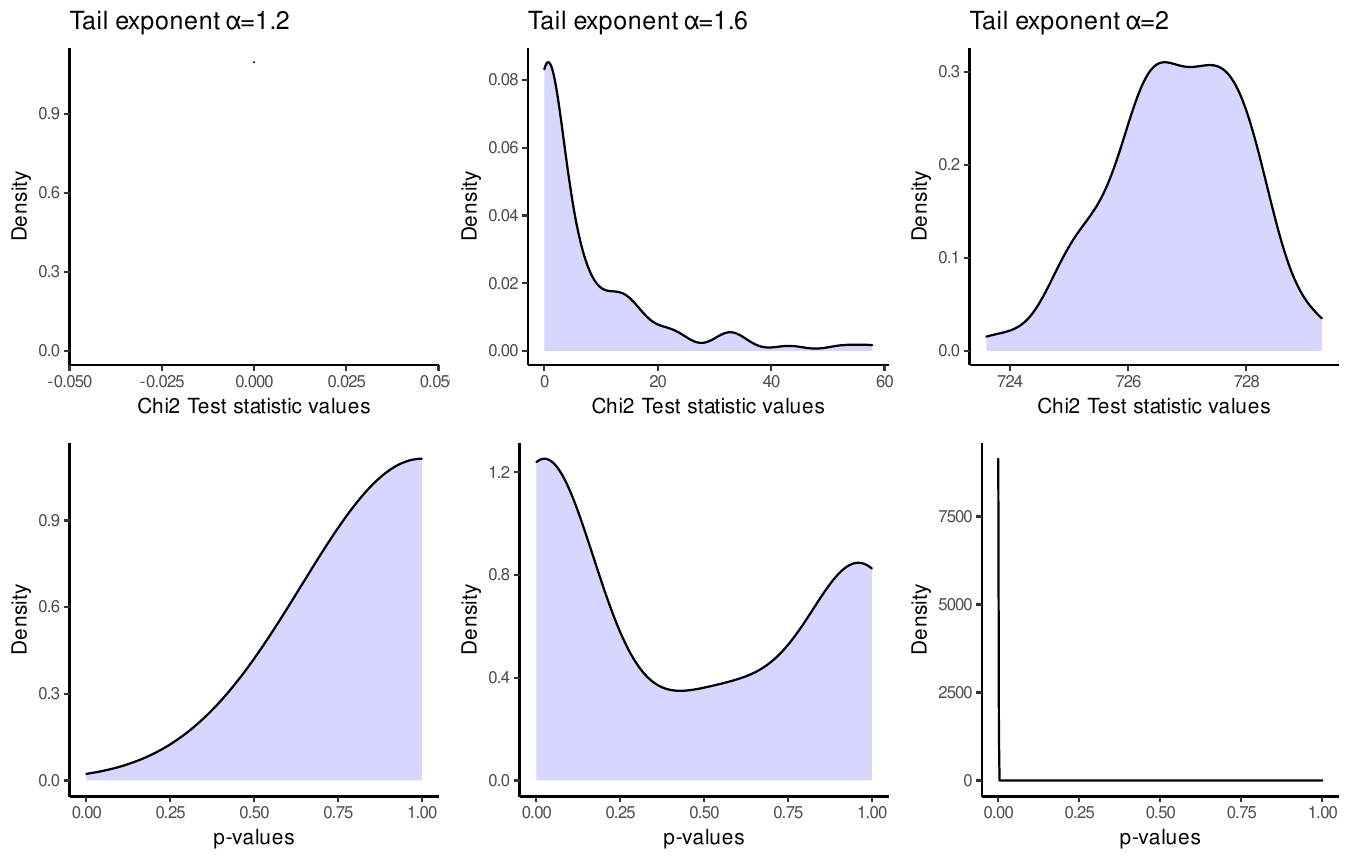}
  \end{center}
  \caption{\textbf{Performance test for Trapani's finite moment test conducted using the R package finity.} Distribution of $\chi^2$ values and p-values given by the test for the divergence of the second moment ($k=2$) for samples of size $N=100,000$ drawn from a L\'{e}vy alpha-stable distribution $S(\alpha, 0.5, 1.0, 0.0; 0)$ for three different values of $\alpha \in \{1.2, 1.6, 2.0\}$. }
  \label{fig:finity-performance}
\end{figure}

\subsection{Performance of the test}

Trapani's test depends on a number of parameters; the choice of $\psi$, $r$, and $u$ can influence the test results. We largely follow \citet{trapani2016testing} in our choices of the parameter values and choose 
$$u=1, r=n^{0.8}, \psi=\begin{cases}k-1 \quad \textup{if } k>1\\k/2 \quad \textup{if } k\leq 1\end{cases}.$$

The test results should be expected to be somewhat noisy. Especially for samples drawn from a L\'{e}vy alpha-stable distribution $S(\alpha, \dots)$ in the vicinity of the true tail parameter $k\approx \alpha$, the test loses accuracy. 

For the tests performed here, we use the R package \verb|finity| \citep{finity}. Fig.~\ref{fig:finity-performance} shows the distribution of $\chi^2$ statistics and p-values for a test of divergence of the second moment ($k=2$), for samples of size $N=100,000$ drawn from a L\'{e}vy alpha-stable distribution $S(\alpha, 0.5, 1.0, 0.0; 0)$ for three different values of $\alpha$. The results in Fig.~\ref{fig:finity-performance} reflect a typical pattern of performance of our implementation of the test at relatively large sample sizes ($N>1000$), where the test tends to seriously over-reject the null even for values of $2> \alpha > 1.6$, reinforcing the robustness of our conclusion that productivity has an infinite variance.

\subsection{Finite moment test results for country-year sample}

% latex table generated in R 3.5.1 by xtable 1.8-4 package
% Thu Apr 23 11:42:03 2020
\begin{table}[H]
\footnotesize
\centering
\begin{tabular}{|l|cccccccccccc|}
  \hline
Country & 2006 & 2007 & 2008 & 2009 & 2010 & 2011 & 2012 & 2013 & 2014 & 2015 & 2016 & 2017 \\ 
\hline
    Belgium & 1.00 & 0.30 & 0.72 & 0.00 & 0.00 & 0.00 & 0.27 & 0.08 & 0.95 & 0.98 & 0.99 & 0.98 \\ 
  Bulgaria &  & 0.98 & 0.86 & 0.91 & 1.00 & 1.00 & 1.00 & 0.99 & 0.93 & 0.99 & 0.92 & 1.00 \\ 
  Croatia & 1.00 & 1.00 & 1.00 & 1.00 & 0.99 & 0.99 & 0.98 & 0.99 & 0.92 & 0.00 & 0.32 &  \\ 
  Czech Republic & 0.99 & 0.97 & 0.99 & 1.00 & 1.00 & 0.99 & 0.83 & 1.00 & 0.65 & 0.52 & 1.00 & 0.99 \\ 
  Denmark &  &  &  &  &  &  & 1.00 & 1.00 & 1.00 & -- & 0.99 & 0.99 \\ 
  Estonia &  & 1.00 & 0.03 & 0.01 & 1.00 & 0.42 & 0.00 & 0.09 & 0.00 & 0.00 & 0.00 & 0.00 \\ 
  Finland & 0.10 & 0.08 & 0.63 & 0.08 & 0.28 & 0.59 & 0.63 & 0.82 & 0.92 & 0.62 & 0.76 & 1.00 \\ 
  France & 0.08 & 0.32 & 1.00 & 0.99 & 1.00 & 1.00 & 0.99 & 0.99 & 0.24 & 0.99 & 0.99 & 1.00 \\ 
  Germany & 1.00 & 0.99 & 1.00 & 1.00 & 1.00 & 0.99 & 0.99 & 1.00 & -- & -- & 1.00 &  \\ 
  Hungary &  & 1.00 & 0.98 & 1.00 & 0.99 & 0.99 & -- & -- & 0.99 & 0.99 & 1.00 & 0.99 \\ 
  Italy & 0.99 & 0.44 & 0.93 & 0.99 & 0.67 & 0.51 & 1.00 & 0.00 & 0.00 & 0.00 & 0.87 & 0.00 \\ 
  Poland & 1.00 & 1.00 & 0.59 & 0.99 & 0.99 & 0.98 & 0.94 & 1.00 &  &  &  & 0.96 \\ 
  Portugal & 0.95 & 0.95 & 1.00 & 1.00 & 0.99 & 0.99 & 1.00 & 1.00 & 1.00 & 1.00 & 0.99 & 0.99 \\ 
  Romania &  & 1.00 & 0.99 & 0.99 & 0.90 & 0.35 & 1.00 & 1.00 & 0.74 & 0.99 & 1.00 &  \\ 
  Slovakia &  & 0.98 & 0.99 & 1.00 & 0.16 & 0.37 & 1.00 & 0.43 & 0.78 & 0.00 & 0.86 &  \\ 
  Slovenia &  &  &  &  & 0.00 & 0.00 & 0.00 & 0.00 & 0.00 & 0.00 & 0.02 & 0.00 \\ 
  Spain & 0.28 & 1.00 & 1.00 & 1.00 & 1.00 & 1.00 & 1.00 & 1.00 & 1.00 & 1.00 & 1.00 & 1.00 \\ 
  Sweden & 0.00 & 0.95 & 1.00 & 1.00 & 1.00 & 0.00 & 0.00 & 0.00 & 0.00 & 0.93 & 0.00 & 0.00 \\ 
  United Kingdom & 0.98 & 0.99 & 0.99 & 0.99 & 0.99 & 0.99 & -- & 0.99 & -- & 1.00 & 0.99 & \\ 
   \hline
\end{tabular}
	\caption{\textbf{Testing for infinite moments of firm-level labour productivity.} P-values of Trapani's finite moment test for the nonexistence of the second moment of the distribution of labour productivity, for country-year samples. In the vast majority of cases, the p.value is above any sensible threshold (e.g., 1\%, 5\%, or 10\%), suggesting that the nonexistence of the second moment can not be rejected. For example, 84\% of the P-values are greater than 0.05, meaning that, for 84\% of country-year samples, the finite moment test failed to reject the null hypothesis of the infinite second moment at the 5\% confidence level. See Appendix \ref{app:finite_moment_test}.}
	\label{tab:finite_moment_test_year}
\end{table}

\subsection{Finite moment test results for country-year-industry sample}

Table~\ref{tab:finite_moment_test_ind_2} shows the proportion of the 2-digit industries in each country for which we cannot reject that the second moment diverges at the 5\% level. 

% latex table generated in R 3.5.1 by xtable 1.8-4 package
% Mon May 18 10:10:20 2020
\begin{table}[H]
\centering
\small
\begin{tabular}{ cc }
\begin{tabular}{lrr}
  \hline
Country & \# of Sectors & Proportion \\ 
  \hline
Belgium & 228 & 0.58 \\ 
  Bulgaria & 198 & 0.72 \\ 
  Croatia & 198 & 0.33 \\ 
  Czech Republic & 216 & 0.67 \\ 
  Denmark & 108 & 0.44 \\ 
  Estonia & 198 & 0.28 \\ 
  Finland & 216 & 0.28 \\ 
  France & 216 & 0.50 \\ 
  Germany & 209 & 0.83 \\ 
  Hungary & 209 & 0.61 \\ 
   \hline
\end{tabular}
\quad 
\begin{tabular}{lrr}
  \hline
Country & \# of Sectors & Proportion\\ 

  \hline
Italy & 216 & 0.56 \\ 
  Poland & 162 & 0.72 \\ 
  Portugal & 216 & 0.78 \\ 
  Romania & 190 & 0.53 \\ 
  Slovakia & 180 & 0.67 \\ 
  Slovenia & 162 & 0.17 \\ 
  Spain & 228 & 0.68 \\ 
  Sweden & 216 & 0.22 \\ 
  United Kingdom & 209 & 0.94 \\ 
   &  &  \\ 
   \hline
\end{tabular}
\end{tabular}
	\caption{\textbf{Proportion of the country-year-industry samples for which the hypothesis of infinite second moment of the distribution of LP is not rejected at the 5\% level.}  Across all country-year-industry samples, the proportion is 55\%. 
	}
		\label{tab:finite_moment_test_ind_2}
\end{table}

Even at a highly disaggregated country-year- industry level, 55\%  samples do not have the second moment in the distribution of labour productivity. For the largest 5 European countries (France, Germany, Italy, Spain, and the UK), the same result holds for nearly 70\% of samples.

\end{document}

%% file: Tables/intan_correlations.tex
% latex table generated in R 4.1.2 by xtable 1.8-4 package
% Thu Mar 24 14:58:16 2022
\begin{table}[H]
\centering
\caption{\textbf{Correlations between metrics of productivity dispersion computed at the industry-country-year level.}} 
\label{table:intan_correlations}
\begin{tabular}{r|p{25mm}p{25mm}p{25mm}p{25mm}}
  \hline
 & $\log(\text{IQR}_{90/10})$ & $\log(\text{IQR}_{95/50})$ & $\log(\gamma/\delta)$ & $-\alpha$ \\ 
  \hline
$\log(\text{IQR}_{90/10})$ & 1.00 & 0.72 & 0.69 & 0.46 \\ 
  $\log(\text{IQR}_{95/50})$ & 0.72 & 1.00 & 0.67 & 0.85 \\ 
  $\log(\gamma/\delta)$ & 0.69 & 0.67 & 1.00 & 0.37 \\ 
  $-\alpha$ & 0.46 & 0.85 & 0.37 & 1.00 \\ 
   \hline
\end{tabular}
\end{table}

%% file: Tables/intan_reg_iqr.tex
\begin{table}[!ht]
  \caption{\textbf{Intangible intensity and quantile-based metrics of productivity dispersion.}} 
  \label{table:intan_reg_iqr} 
\footnotesize 
\begin{center}
\begin{tabular}{@{\extracolsep{1pt}}lcccccc} 
\\[-1.8ex]\hline 
\hline \\[-1.8ex] 
 & \multicolumn{6}{c}{Dep. variable: Log intangible intensity of a country-industry-year triples} \\ 
\cline{2-7} 
\hline \\[-1.8ex] 
 log(q90/q10) & 0.123$^{*}$ &  & 0.045 & $-$0.029$^{*}$ &  & $-$0.030$^{*}$ \\ 
  & (0.075) &  & (0.094) & (0.016) &  & (0.016) \\ 
  & & & & & & \\ 
 log(q95/q50) &  & 0.431$^{***}$ & 0.221 &  & $-$0.005 & 0.005 \\ 
  &  & (0.104) & (0.181) &  & (0.034) & (0.056) \\ 
  & & & & & & \\ 
 log(q50) & 1.060$^{***}$ & 1.040$^{***}$ & 1.059$^{***}$ & 0.056 & 0.049 & 0.056 \\ 
  & (0.046) & (0.042) & (0.046) & (0.057) & (0.040) & (0.057) \\ 
  & & & & & & \\ 
\hline \\[-1.8ex] 
Fixed effects & None & None & None & CxJ, TxJ & CxJ, TxJ & CxJ, TxJ \\ 
Degrees of freedom & 2530 & 2835 & 2529 & 2064 & 2361 & 2063 \\ 
Observations & 2,533 & 2,838 & 2,533 & 2,533 & 2,838 & 2,533 \\ 
R$^{2}$ & 0.683 & 0.689 & 0.684 & 0.996 & 0.996 & 0.996 \\ 
Adjusted R$^{2}$ & 0.683 & 0.688 & 0.684 & 0.995 & 0.995 & 0.995 \\ 
Residual Std. Error & 1.114 & 1.109 & 1.112 & 0.135 & 0.141 & 0.135 \\ 
\hline 
\hline \\[-1.8ex] 
\end{tabular} 
\end{center}
\footnotesize{\textit{Notes:} This table regresses intangible capital intensity on various inter-quantile measures of labour productivity dispersion, by industry-country-year. The fourth, fifth and sixth columns include country-industry and year-industry fixed effects. Corresponding standard errors are in parentheses.

Significance level: $^{*}$p$<$0.1; $^{**}$p$<$0.05; $^{***}$p$<$0.01}
\end{table}

%% file: Tables/intan_reg_levy.tex
\begin{table}[!ht] 
  \caption{\textbf{Intangible intensity and L\'{e}vy-based metrics of productivity dispersion.}} 
  \label{table:intan_reg_levy} 
\footnotesize 
\begin{center}
\begin{tabular}{@{\extracolsep{1pt}}lcccccc} 
\\[-1.8ex]\hline 
\hline \\[-1.8ex] 
 & \multicolumn{6}{c}{Dep. variable: Log intangible intensity of a country-industry-year triples} \\ 
\cline{2-7} 
\hline \\[-1.8ex] 
 $\log(\gamma/\delta)$ & 0.754$^{***}$ &  & 0.678$^{***}$ & $-$0.008 &  & $-$0.007 \\ 
  & (0.115) &  & (0.128) & (0.038) &  & (0.038) \\ 
  & & & & & & \\ 
 $-\alpha$ &  & 1.043$^{***}$ & 0.506$^{*}$ &  & 0.010 & 0.008 \\ 
  &  & (0.277) & (0.283) &  & (0.068) & (0.069) \\ 
  & & & & & & \\ 
 $\log(\delta)$ & 1.048$^{***}$ & 1.022$^{***}$ & 1.046$^{***}$ & 0.011 & 0.018 & 0.012 \\ 
  & (0.043) & (0.044) & (0.042) & (0.041) & (0.018) & (0.042) \\ 
  & & & & & & \\ 
\hline \\[-1.8ex] 
Fixed effects & None & None & None & CxJ, TxJ & CxJ, TxJ & CxJ, TxJ \\ 
Degrees of freedom & 2826 & 2826 & 2825 & 2352 & 2352 & 2351 \\ 
Observations & 2,829 & 2,829 & 2,829 & 2,829 & 2,829 & 2,829 \\ 
R$^{2}$ & 0.688 & 0.667 & 0.691 & 0.996 & 0.996 & 0.996 \\ 
Adjusted R$^{2}$ & 0.688 & 0.666 & 0.691 & 0.995 & 0.995 & 0.995 \\ 
Residual Std. Error & 1.110 & 1.148 & 1.106 & 0.140 & 0.140 & 0.140 \\ 
\hline 
\hline \\[-1.8ex] 

\end{tabular} 
\end{center}
\footnotesize{\textit{Notes:} This table regresses intangible capital intensity on L\'evy-based measures of labour productivity dispersion, by industry-country-year. The fourth, fifth and sixth columns include country-industry and year-industry fixed effects. Corresponding standard errors are in parentheses.

Significance level: $^{*}$p$<$0.1; $^{**}$p$<$0.05; $^{***}$p$<$0.01}
\end{table}

%% file: Tables/intan_lasso.tex
% latex table generated in R 4.1.2 by xtable 1.8-4 package
% Thu Mar 24 15:02:54 2022
\begin{table}[!ht]
\begin{center}
\caption{\textbf{LASSO estimates for intangible capital intensity and labour productivity dispersion.}}
\label{table:intan_lasso}
\begin{tabular}{r|p{35mm}p{35mm}p{35mm}}
  \hline
\hline
 & Quantiles-only & L\'{e}vy-only & Both \\ 
  \hline
$\log(q90/q10)$ & -0.00 &  & -0.00 \\ 
  $\log(q95/q50)$ & 0.02 &  & 0.01 \\ 
  $\log(q50)$ & 0.45 &  & 0.43 \\ 
  $\log(\gamma/\delta)$ &  & 0.21 & 0.01 \\ 
  $-\alpha$ &  & 0.09 & -0.02 \\ 
  $\log(\delta)$ &  & 0.27 & 0.02 \\ 
   \hline
Observations & 2533 & 2829 & 2522 \\ 
  Non-zero coeffs & 470 & 476 & 474 \\ 
   \hline
\hline
\end{tabular}
\end{center}
\footnotesize{\textit{Notes:} This table regresses intangible capital intensity on various measures of labour productivity dispersion, by industry-country-year, using a LASSO algorithm. Dispersion measures in the first column are based on quantiles of the relevant labour productivity distribution, and are from L\'{e}vy-based measures in the second column. The third column includes all dispersion measures available. Each regression includes country-industry and year-industry fixed effects, although the number set to zero by the algorithm is reported in the final row.}
\end{table}

%% file: main.bib
@book{sornette2006critical,
  title={Critical phenomena in natural sciences: chaos, fractals, selforganization and disorder: concepts and tools},
  author={Sornette, Didier},
  year={2006},
  publisher={Springer Science \& Business Media}
}

@article{gouin2020productivity,
  title={Productivity Dispersion, Between-firm Competition and the Labor Share},
  author={Gouin-Bonenfant, Emilien},
  year={2020},
  journal={Draft}
}

@article{ilzetzki2017measuring,
  title={Measuring productivity dispersion: Lessons from counting one-hundred million ballots},
  author={Ilzetzki, Ethan and Simonelli, Saverio},
  year={2017},
  journal={CEPR Discussion Paper},
  volume={DP12273}
}

@article{aradanaz2017understanding,
  title={Understanding firms in the bottom 10\% of the labour productivity distribution in Great Britain:“the laggards”, 2003 to 2015},
  author={Aradanaz-Badia, A and Awano, Gaganan and Wales, Philip},
  journal={Office for National Statistics},
  notes={\url{https://www.beta.ons.gov.uk/economy/economicoutputandproductivity/productivitymeasures/articles/understandingfirmsinthebottom10ofthelabourproductivitydistributioningreatbritain/jantomar2017}},
  year={2017}
}

@article{souma2009distribution,
  title={Distribution of labour productivity in Japan over the period 1996--2006},
  author={Souma, Wataru and Ikeda, Yuichi and Iyetomi, Hiroshi and Fujiwara, Yoshi},
  journal={Economics: E-journal},
  volume={3},
  number={1},
  year={2009}
}

@article{campbell2019measuring,
  title={Measuring productivity dispersion in selected Australian industries},
  author={Campbell, Simon and Nguyen, Thai and Sibelle, Alexander and Soriano, Franklin},
  year={2019},
  journal={Treasury-ABS Working Paper}
}

@article{oulton2000tale,
  title={A Tale of Two Cycles: Closure, Downsizing and Productivity Growth in Uk Manufacturing, I973-89},
  author={Oulton, Nicholas},
  journal={National Institute Economic Review},
  volume={173},
  pages={66--79},
  year={2000},
  publisher={Cambridge University Press}
}

@article{baily1996downsizing,
  title={Downsizing and productivity growth: myth or reality?},
  author={Baily, Martin Neil and Bartelsman, Eric J and Haltiwanger, John},
  journal={Small Business Economics},
  volume={8},
  number={4},
  pages={259--278},
  year={1996},
  publisher={Springer}
}

@article{gu2019frontier,
  title={Frontier Firms, Productivity Dispersion and Aggregate Productivity Growth in Canada},
  author={Gu, Wulong},
  journal={International Productivity Monitor},
  number={37},
  volume={{}},
  pages={96--119},
  year={2019},
  publisher={Centre for the Study of Living Standards}
}

@article{clauset2009power,
  title={Power-law distributions in empirical data},
  author={Clauset, Aaron and Shalizi, Cosma Rohilla and Newman, Mark EJ},
  journal={SIAM review},
  volume={51},
  number={4},
  pages={661--703},
  year={2009},
  publisher={SIAM}
}

@article{voitalov2019scale,
  title={Scale-free networks well done},
  author={Voitalov, Ivan and van der Hoorn, Pim and van der Hofstad, Remco and Krioukov, Dmitri},
  journal={Physical Review Research},
  volume={1},
  number={3},
  pages={033034},
  year={2019},
  publisher={APS}
}

@article{friedman2017,
 title={Regularization Paths for Generalized Linear Models via Coordinate Descent},
 volume={33},
 number={1},
 journal={Journal of Statistical Software},
 author={Friedman, Jerome H. and Hastie, Trevor and Tibshirani, Rob},
 year={2010},
 pages={1–22}
}

@article{cunningham2021dispersion,
  title={Dispersion in dispersion: Measuring establishment-level differences in productivity},
  author={Cunningham, Cindy and Foster, Lucia and Grim, Cheryl and Haltiwanger, John and Pabilonia, Sabrina Wulff and Stewart, Jay and Wolf, Zoltan},
  year={2021},
  journal={IZA Discussion Paper},
  volume={14459}
}

@article{goldin2021productivity,
  title={Why is productivity slowing down?},
  author={Goldin, Ian and Koutroumpis, Pantelis and Lafond, Fran{\c{c}}ois and Winkler, Julian},
  year={2021},
  journal={INET Oxford Working Paper}
}

@article{berlingieri2017multiprod,
  title={The Multiprod project: A comprehensive overview},
  author={Berlingieri, Giuseppe and Blanchenay, Patrick and Calligaris, Sara and Criscuolo, Chiara},
  journal={OECD Science, Technology and Industry Working Papers},
  volume={2017},
  number={4},
  pages={1},
  year={2017},
  publisher={Organisation for Economic Cooperation and Development (OECD)}
}

@incollection{moran,
  author    = {Moran, Jos\'{e} and Secchi, Angelo and Bouchaud, Jean-Philippe},
  title     = {Statistical facts on growth rates},
  booktitle = {Statistical physics and anomalous macroeconomic fluctuations, PhD Thesis},
  publisher = {EHESS press},
  editor = {Jos\'{e} Moran},
  year      = {2020}
}

@article{rotemberg2021plant,
  title={Plant-to-Table (s and Figures): Processed Manufacturing Data and Measured Misallocation},
  author={Rotemberg, Martin and White, T Kirk},
  year={2021},
  journal={Working Paper},
  note={\url{https://wp.nyu.edu/mrotemberg/wp-content/uploads/sites/8049/2021/07/RW_plant_to_table_July_2021.pdf}}
}

@book{white2000asymptotic,
  title={Asymptotic theory for econometricians, revised edition},
  author={White, Halbert},
  year={2000},
  publisher={Academic press}
}

@article{bartkiewicz2011stable,
  title={Stable limits for sums of dependent infinite variance random variables},
  author={Bartkiewicz, Katarzyna and Jakubowski, Adam and Mikosch, Thomas and Wintenberger, Olivier},
  journal={Probability Theory and Related Fields},
  volume={150},
  number={3},
  pages={337--372},
  year={2011},
  publisher={Springer}
}

@book{kelejian2017spatial,
  title={Spatial Econometrics},
  author={Kelejian, Harry and Piras, Gianfranco},
  year={2017},
  publisher={Academic Press}
}

@article{shintani2018super,
  title={Super Generalized Central Limit Theorem—Limit Distributions for Sums of Non-identical Random Variables with Power Laws—},
  author={Shintani, Masaru and Umeno, Ken},
  journal={Journal of the Physical Society of Japan},
  volume={87},
  number={4},
  pages={043003},
  year={2018},
  publisher={The Physical Society of Japan}
}

@techreport{Jaeger17,
author={J{\"a}ger, Kirsten and {The Conference Board}},
title={{EU} {KLEMS} Growth and Productivity Accounts 2017 Release, Statistical Module 1},
institution={{EU} {KLEMS} {P}roject},
year={2018},
note={Revised 2018, Available online http://www.euklems.net/TCB/2018/Metholology\_EUKLEMS\_2017\_revised.pdf,  {http://www.euklems.net/} TCB/2018/EUKLEMS\_2018\_revision.pdf}
}

@article{bartelsman2000understanding,
  title={Understanding productivity: Lessons from longitudinal microdata},
  author={Bartelsman, Eric J and Doms, Mark},
  journal={Journal of Economic Literature},
  volume={38},
  number={3},
  pages={569--594},
  year={2000}
}

@article{bartelsman2013cross,
Author = {Bartelsman, Eric and Haltiwanger, John and Scarpetta, Stefano},
Title = {Cross-Country Differences in Productivity: The Role of Allocation and Selection},
Journal = {American Economic Review},
Volume = {103},
Number = {1},
Year = {2013},
Month = {February},
Pages = {305-34},
}

@article{gaffeo2008levy,
  title={L{\'e}vy-stable productivity shocks},
  author={Gaffeo, Edoardo},
  journal={Macroeconomic Dynamics},
  volume={12},
  number={3},
  pages={425--443},
  year={2008},
  publisher={Cambridge University Press}
}

@article{konig2016innovation,
  title={Innovation vs. imitation and the evolution of productivity distributions},
  author={K{\"o}nig, Michael D and Lorenz, Jan and Zilibotti, Fabrizio},
  journal={Theoretical Economics},
  volume={11},
  number={3},
  pages={1053--1102},
  year={2016},
  publisher={Wiley Online Library}
}

@article{perla2014equilibrium,
  title={Equilibrium imitation and growth},
  author={Perla, Jesse and Tonetti, Christopher},
  journal={Journal of Political Economy},
  volume={122},
  number={1},
  pages={52--76},
  year={2014},
  publisher={University of Chicago Press Chicago, IL}
}

@article{aoyama2010productivity,
  title={Productivity dispersion: facts, theory, and implications},
  author={Aoyama, Hideaki and Yoshikawa, Hiroshi and Iyetomi, Hiroshi and Fujiwara, Yoshi},
  journal={Journal of Economic Interaction and Coordination},
  volume={5},
  number={1},
  pages={27--54},
  year={2010},
  publisher={Springer}
}

@article{schwarzkopf2010explanation,
  title={An explanation of universality in growth fluctuations},
  author={Schwarzkopf, Yonathan and Axtell, Robert and Farmer, J Doyne},
  journal={SSRN},
  year={2010},
  volume={1597504}
}

@incollection{bartelsman2018measuring,
title = {Measuring Productivity Dispersion},
author = {Bartelsman, Eric J. and Wolf, Zoltan},
year = {2018},
editor = {Grifell-Tatj{\'e}, Emili and Lovell, C.A. Knox and Sickles, Robin C.},
booktitle = {Oxford Handbook of Productivity Analysis},
publisher = {Oxford University Press},
address = {United Kingdom},
}

@article{holly2013aggregate,
  title={Aggregate fluctuations and the cross-sectional dynamics of firm growth},
  author={Holly, Sean and Petrella, Ivan and Santoro, Emiliano},
  journal={Journal of the Royal Statistical Society: Series A (Statistics in Society)},
  volume={176},
  number={2},
  pages={459--479},
  year={2013},
  publisher={Wiley Online Library}
}

@article{gaffeo2011distribution,
  title={The distribution of sectoral {TFP} growth rates: International evidence},
  author={Gaffeo, Edoardo},
  journal={Economics Letters},
  volume={113},
  number={3},
  pages={252--255},
  year={2011},
  publisher={Elsevier}
}

@article{KalemliOzcanetal15,
 title = {How to Construct Nationally Representative Firm Level data from the ORBIS Global Database},
 author = {Kalemli-Ozcan, Sebnem and Sorensen, Bent and Villegas-Sanchez, Carolina and Volosovych, Vadym and Yesiltas, Sevcan},
 journal = {NBER Working Paper},
 volume = {21558},
 year = {2015}
}

@incollection{foster2018innovation,
  title={Innovation, Productivity Dispersion, and Productivity Growth},
  author={Foster, Lucia and Grim, Cheryl and Haltiwanger, John C and Wolf, Zoltan},
  booktitle={Measuring and Accounting for Innovation in the Twenty-First Century},
  pages={103},
  year={2021},
  editor={Carol Corrado and Jonathan Haskel and Javier Miranda and Daniel Sichel},
  publisher={University of Chicago Press}
}

@article{Bottazzi/Secchi11,
  title={A new class of asymmetric exponential power densities with applications to economics and finance},
  author={Bottazzi, Giulio and Secchi, Angelo},
  journal={Industrial and Corporate Change},
  volume={20},
  number={4},
  pages={991--1030},
  year={2011},
  publisher={Oxford University Press}
}

@article{McCulloch86,
  title={Simple consistent estimators of stable distribution parameters},
  author={McCulloch, J Huston},
  journal={Communications in Statistics-Simulation and Computation},
  volume={15},
  number={4},
  pages={1109--1136},
  year={1986},
  publisher={Taylor \& Francis}
}

@article{haltiwanger2018misallocation,
  title={Misallocation measures: The distortion that ate the residual},
  author={Haltiwanger, John and Kulick, Robert and Syverson, Chad},
  year={2018},
  journal={NBER Working Paper},
  volume={24199}
}

@article{syverson2011what,
Author = {Syverson, Chad},
Title = {What Determines Productivity?},
Journal = {Journal of Economic Literature},
Volume = {49},
Number = {2},
Year = {2011},
Month = {June},
Pages = {326-65},
}

@article{NOLAN1998187,
title = "Parameterizations and modes of stable distributions",
journal = "Statistics \& Probability Letters",
volume = "38",
number = "2",
pages = "187 - 195",
year = "1998",
author = "John P. Nolan"
}

@article{mitzenmacher2004brief,
  title={A brief history of generative models for power law and lognormal distributions},
  author={Mitzenmacher, Michael},
  journal={Internet Mathematics},
  volume={1},
  number={2},
  pages={226--251},
  year={2004},
  publisher={Taylor \& Francis}
}

@article{gabaix2009power,
  title={Power laws in economics and finance},
  author={Gabaix, Xavier},
  journal={Annual Review of Economics},
  volume={1},
  number={1},
  pages={255--294},
  year={2009},
  publisher={Annual Reviews}
}

@book{NOLAN2020,
title = {Stable Distributions - Models for Heavy Tailed Data},
publisher = {Springer},
author = {Nolan, John P.},
year = {2020}
}

@article{ASQUITH2014955,
title = "Parameter estimation for the 4-parameter {A}symmetric {E}xponential {P}ower distribution by the method of {L}-moments using {R}",
journal = "Computational Statistics \& Data Analysis",
volume = "71",
pages = "955 - 970",
year = "2014",
author = "William H. Asquith",
}

@Manual{lmonco,
    title = {lmomco - {L}-moments, censored {L}-moments, trimmed {L}-moments, {L}-comoments, and many distributions},
    author = {W.H. Asquith},
    year = {2018},
    note = {R package version 2.3.1},
  }

@article{Bottazzietal07,
year={2007},
journal={Small Business Economics},
volume={29},
number={1-2},
title={Invariances and Diversities in the Patterns of Industrial Evolution: Some Evidence from Italian Manufacturing Industries},
publisher={Kluwer Academic Publishers-Plenum Publishers},
author={Bottazzi, Giulio and Cefis, Elena and Dosi, Giovanni and Secchi, Angelo},
pages={137-159},
}

@book{resnick2007heavy,
  title={Heavy-Tail Phenomena: Probabilistic and Statistical Modeling},
  author={Resnick, Sidney I.},
  series = {Springer Series in Operations Research and Financial Engineering},
  year={2007},
  publisher={Springer}
}

@book{Ijiri/Simon77,
  title={Skew distributions and the sizes of business firms},
  author={Ijiri, Yuji and Simon, Herbert Alexander},
  volume={24},
  year={1977},
  publisher={North Holland},
  address={Amsterdam}
}

@article{Axtell01,
author = {Axtell, Robert L.}, 
title = {Zipf Distribution of {U}.{S}. Firm Sizes}, 
volume = {293}, 
number = {5536}, 
pages = {1818-1820}, 
year = {2001}, 
journal = {Science} 
}

@article{aoyama2009labour,
  title={Labour productivity superstatistics},
  author={Aoyama, Hideaki and Yoshikawa, Hiroshi and Iyetomi, Hiroshi and Fujiwara, Yoshi},
  journal={Progress of Theoretical Physics, Supplement},
  volume={179},
  pages={80--92},
  year={2009},
  publisher={Oxford Academic}
}

@article{mizuno2012power,
  title={Power laws in firm productivity},
  author={Mizuno, Takayuki and Ishikawa, Atushi and Fujimoto, Shouji and Watanabe, Tsutomu},
  journal={Progress of Theoretical Physics Supplement},
  volume={194},
  pages={122--134},
  year={2012},
  publisher={Oxford Academic}
}

@article{aoyama2015micro,
  title={Micro-macro relation of production: double scaling law for statistical physics of economy},
  author={Aoyama, Hideaki and Fujiwara, Yoshi and Gallegati, Mauro},
  journal={Journal of Economic Interaction and Coordination},
  volume={10},
  number={1},
  pages={67--78},
  year={2015},
  publisher={Springer}
}

@article{fujiwara2009distribution,
  title={Distribution of labour productivity in Japan over the period 1996-2006},
  author={Souma, Wataru and Ikeda, Yuichi and Iyetomi, Hiroshi  and Fujiwara, Yoshi},
  journal={Economics: The Open-Access, Open-Assessment E-Journal},
  volume={3},
  year={2009}
}

@article{Bontadini2021,
   author = {Bontadini, F. and Corrado, C. and Haskel, J. and Iommi, M. and Jona-Lasinio, C.},
   title = {EUKLEMS \& INTANProd: methods and data descriptions},
   journal = {EU KLEMS website},
   year = {2021},
   note={\url{https://euklems-intanprod-llee.luiss.it/wp-content/uploads/2022/02/EUKLEMSINTANProd_2021_Methods-and-data-description-Rev1.pdf}}
}

@article{ONS_ABS,
title = "Firm-level labour productivity estimates from the Annual Business Survey (ABS): summary statistics",
journal = "available online at",
year = "2022",
author = "{Office for National Statistics}",
note = {\url{https://www.ons.gov.uk/economy/economicoutputandproductivity/productivitymeasures/datasets/firmlevellabourproductivityestimatesfromtheannualbusinesssurveyabssummarystatistics}}
}

@article{cohen2020heavy,
  title={Heavy-tailed distributions, correlations, kurtosis and Taylor’s Law of fluctuation scaling},
  author={Cohen, Joel E and Davis, Richard A and Samorodnitsky, Gennady},
  journal={Proceedings of the Royal Society A},
  volume={476},
  number={2244},
  pages={20200610},
  year={2020},
  publisher={The Royal Society Publishing}
}

@incollection{de2022firms,
  author    = {De Loecker, Jan and Obermeier, Tim and Van Reenen, John},
  title     = {Firms and inequalities},
  booktitle = {The IFS Deaton Review of Inequalities},
  publisher = {IFS},
  editor = {IFS},
  year      = {2022}
}

@article{trapani2016testing,
title = "Testing for (in)finite moments",
journal = "Journal of Econometrics",
volume = "191",
number = "1",
pages = "57 - 68",
year = "2016",
author = "Lorenzo Trapani",
}

@article{andrews2016best,
   author = {Andrews, Dan and Criscuolo, Chiara and Gal, Peter N.},
   title = {The Best versus the Rest},
   journal = {OECD Productivity Working Papers},
   year = {2016},
   type = {Journal Article},
   volume={05}
}

@article{berlingieri2017great,
   author = {Berlingieri, Giuseppe and Blanchenay, Patrick and Criscuolo, Chiara},
   title = {The Great Divergence(s)},
   journal = {OECD STI Policy Papers},
   volume={39},
   year = {2017}
  }

@article{cette2018firm,
   author = {Cette, Gilbert and Corde, Simon and Lecat, R\'{e}my},
   title = {Firm-level Productivity Dispersion and Convergence},
   journal = {Economics Letters},
   volume = {166},
   pages = {76-78},
   year = {2018},
   type = {Journal Article}
}

@article{gabaix2011granular,
   author = {Gabaix, Xavier},
   title = {The Granular Origins of Aggregate Fluctuations},
   journal = {Econometrica},
   volume = {79},
   number = {3},
   pages = {733-772},
   year = {2011},
   type = {Journal Article}
}

@article{haldane2017productivity,
   author = {Haldane, Andy},
   title = {Productivity puzzles},
   journal = {Speech given by Andy Haldane, Chief Economist, Bank of England at the London School of Economics},
   year = {2017},
   type = {Journal Article}
}

@article{hsieh2009misallocation,
   author = {Hsieh, Chang-Tai and Klenow, Peter J.},
   title = {Misallocation and Manufacturing {TFP} in {C}hina and {I}ndia},
   journal = {The Quarterly Journal of Economics},
   volume = {124},
   number = {4},
   pages = {1403-1448},
   year = {2009},
   type = {Journal Article}
}

@article{lucas2014knowledge,
   author = {Lucas, Robert E. and Moll, Benjamin},
   title = {Knowledge Growth and the Allocation of Time},
   journal = {Journal of Political Economy},
   volume = {122},
   number = {1},
   pages = {1-51},
   year = {2014},
   type = {Journal Article}
}

@article{ghiglino2012random,
   author = {Ghiglino, Christian},
   title = {Random walk to innovation: Why productivity follows a power law},
   journal = {Journal of Economic Theory},
   volume = {147},
   number = {2},
   pages = {713-737},
   year = {2012},
   type = {Journal Article}
}

@article{Mandelbrot1963,
	author = {Mandelbrot, Benoit},
	title = {New Methods in Statistical Economics},
	journal = {Journal of Political Economy},
	volume = {71},
	number = {5},
	pages = {421-440},
	year = {1963},
	
}

@article{Mandelbrot1960,
	author = {Benoit Mandelbrot},
	journal = {International Economic Review},
	number = {2},
	pages = {79--106},
	publisher = {[Economics Department of the University of Pennsylvania, Wiley, Institute of Social and Economic Research, Osaka University]},
	title = {The {P}areto-{L}\'{e}vy Law and the Distribution of Income},
	volume = {1},
	year = {1960}
}

@article{Fama1965,
	author = {Eugene F. Fama},
	journal = {Management Science},
	number = {3},
	pages = {404--419},
	publisher = {INFORMS},
	title = {Portfolio Analysis in a Stable {P}aretian Market},
	volume = {11},
	year = {1965}
}

@article{samuelson_1967, 
title={Efficient Portfolio Selection for {P}areto-{L}\'{e}vy Investments}, 
volume={2}, DOI={10.2307/2329897}, 
number={2}, 
journal={Journal of Financial and Quantitative Analysis}, 
publisher={Cambridge University Press}, 
author={Samuelson, Paul A.}, 
year={1967}, 
pages={107–122}
}

@incollection{Emberchts,
	title={Time Series Analysis for Heavy-Tailed Processes},
	author={Embrechts, P. and Kluppelberg, C. and  Mikosch, T},
	booktitle={Modelling Extremal Events,},
	pages={371-412},
	year={1997},
	editor={},
	publisher={Springer},
	address={Berlin, Heidelberg}
}

@ARTICLE{Mitchell,
	title = {The Making and Using of Index Numbers},
	author = {Mitchell, Wesley C.},
	year = {1915},
	journal = {US BLS Bulletin},
	volume={284},
	note={Reprinted 1938, accessible at \url{https://fraser.stlouisfed.org/files/docs/publications/bls/bls_0656_1938.pdf}}
}

@book{bouchaud2003theory, 
place={Cambridge}, 
edition={2}, 
title={Theory of Financial Risk and Derivative Pricing: From Statistical Physics to Risk Management}, 
publisher={Cambridge University Press}, 
author={Bouchaud, Jean-Philippe and Potters, Marc}, 
year={2003}}

@article{newman2005power,
  title={Power laws, {P}areto distributions and {Z}ipf's law},
  author={Newman, Mark EJ},
  journal={Contemporary Physics},
  volume={46},
  number={5},
  pages={323--351},
  year={2005},
  publisher={Taylor \& Francis}
}

@article{gopinath2017capital,
  title={Capital allocation and productivity in South Europe},
  author={Gopinath, Gita and Kalemli-{\"O}zcan, {\c{S}}ebnem and Karabarbounis, Loukas and Villegas-Sanchez, Carolina},
  journal={The Quarterly Journal of Economics},
  volume={132},
  number={4},
  pages={1915--1967},
  year={2017},
  publisher={Oxford University Press}
}

@Manual{finity,
    title = {finity: Test for Finiteness of Moments in a Distribution},
    author = {Torsten Heinrich and Julian Winkler},
    year = {2020},
    note = {R package version 0.1.4.1},
    url = {https://CRAN.R-project.org/package=finity},
}

@article{fagiolo2008output,
  title={Are output growth-rate distributions fat-tailed? some evidence from OECD countries},
  author={Fagiolo, Giorgio and Napoletano, Mauro and Roventini, Andrea},
  journal={Journal of Applied Econometrics},
  volume={23},
  number={5},
  pages={639--669},
  year={2008},
  publisher={Wiley Online Library}
}

@article{acemoglu2017microeconomic,
  title={Microeconomic origins of macroeconomic tail risks},
  author={Acemoglu, Daron and Ozdaglar, Asuman and Tahbaz-Salehi, Alireza},
  journal={American Economic Review},
  volume={107},
  number={1},
  pages={54--108},
  year={2017}
}

@article{Ayebo/Kozubowski03,
  title={An asymmetric generalization of Gaussian and Laplace laws},
  author={Ayebo, Abraham and Kozubowski, Tomasz J},
  journal={Journal of Probability and Statistical Science},
  volume={1},
  number={2},
  pages={187--210},
  year={2003},
}

@article{Delicado/Goria08,
  title={A small sample comparison of maximum likelihood, moments and L-moments methods for the asymmetric exponential power distribution},
  author={Delicado, Pedro and Goria, MN},
  journal={Computational Statistics \& Data Analysis},
  volume={52},
  number={3},
  pages={1661--1673},
  year={2008},
  publisher={Elsevier},
  doi={10.1016/j.csda.2007.05.021}
}

@techreport{Bottazzi14,
  title={Subbotools user's manual: For version 0.9. 7.1, 8 September 2004},
  author={Bottazzi, Giulio},
  year={2014},
  institution={LEM Working Paper Series},
  note={Originally published 2004, updated 2014, http://www.lem.sssup.it/WPLem/files/2004-14.pdf},
}

@article{corrado2021new,
  title={New evidence on intangibles, diffusion and productivity},
  author={Corrado, Carol and Criscuolo, Chiara and Haskel, Jonathan and Himbert, Alexander and Jona-Lasinio, Cecilia},
  year={2021},
  journal={OECD STI Working Papers},
 volume={2021/10},
  publisher={OECD}
}

@misc{axtell2019dynamics,
  title={Dynamics of Firms from the Bottom Up: Data, Theories and Models},
  author={Axtell, RL and Guerrero, O},
  year={forthcoming},
  publisher={MIT Press Cambridge, MA}
}

@article{bottazzi2003common,
  title={Common properties and sectoral specificities in the dynamics of {US} manufacturing companies},
  author={Bottazzi, Giulio and Secchi, Angelo},
  journal={Review of Industrial Organization},
  volume={23},
  number={3},
  pages={217--232},
  year={2003},
  publisher={Springer}
}

@Manual{boot,
    title = {boot: Bootstrap R (S-Plus) Functions},
    author = {Angelo Canty and B. D. Ripley},
    year = {2021},
    note = {R package version 1.3-28},
  }

@techreport{oliveira2021business,
  title={Business Time: How Ready Are UK Firms for the Decisive Decade?},
  author={Oliveira-Cunha, J and Kozler, J and Shah, P and Thwaites, G and Valero, A},
  institution={The Resolution Foundation},
  year={2021}
}
